\documentstyle[aps]{revtex}

\begin{document}
\title{Renormalization of the SU(2)-symmetric model of hadrodynamics}
\author{Jun-Chen Su and Hai-Jun Wang }
\address{Center for Theoretical Physics, School of Physics, Changchun 130023,
People's Republic of China}
\maketitle

\begin{abstract}
It is proved that the SU(2)-symmetric model of hadrodynamics can well be set
up on the gauge-invariance principle. The quantization of the model can
readily be performed in the Lagrangian path-integral formalisms by using the
Lagrangian undetermined multiplier method. Furthermore, it is shown that the
quantum theory is invariant with respect to a kind of BRST-transformations.
From the BRST-symmetry of the theory, the Ward-Takahashi identities
satisfied by the generating functionals of full Green functions, connected
Green functions and proper vertex functions are successively derived. As an
application of the above Ward-Takahashi identities, the Ward-Takahashi
identities obeyed by the propagators and various proper vertices are
derived. Based on these identities, the propagators and vertices are
perfectly renormalized. Especially, as a result of the renormalization, the
Slavnov-Taylor identity satisfied by renormalization constants is natually
deduced. To demonstrate the renormalizability of the theory, the one-loop
renormalization of the theory is carried out by means of the mass-dependent
momentum space subtraction and the renormalization group approach, giving an
exact one-loop effective coupling constant and one-loop effective nucleon,
pion and $\rho -$meson masses.

\end{abstract}

In the early time, a $SU(2)$-symmetric model of hadrodynamics was proposed
by Sakurai [1,2] based on the non-Abelian gauge-field theory which was first
initiated by C. N. Yang and R. L. Mills [3]. A tempting feature of this
model is that the model gives a complete description of the interactions
among nucleons, pions and $\rho $-mesons. However, the model was beset with
two serious difficulties: one is the non-gauge-invariance of the $\rho $%
-meson mass term because according to the prevailing viewpoint, the
requirement of gauge-invariance does not admit the $\rho $-meson mass term
to enter the Lagrangian; another is the unrenormalizability of the model as
argued in the previous literature [4-9]. Due to these difficulties, the
model was eventually relinquished even though a part of the interactions
between nucleons and $\rho $-mesons and the interaction between nucleons and
pions which are all included in the model have been widely applied in
nuclear physics.

Against the difficulties mentioned above, this paper attempts to answer the
questions: whether the $SU(2)$-symmetric model of hadrodynamics could be set
up on the gauge-invariance principle and whether the model could be
renormalizable? The first question has been answered in our published papers
[10-12]. In the papers, it is argued that a non-Abelian massive gauge field
theory in which the masses of all gauge fields are the same can actually be
set up on the principle of gauge-invariance without need of introducing the
Higgs mechanism. This means that the model under consideration can exactly
be made up of the massive gauge field theory with SU(2) gauge symmetry. The
essential points to achieve this conclusion are as follows. (a) The gauge
boson fields such as the $\rho $-meson fields must be viewed as a
constrained system in the whole space of vector potentials and the Lorentz
condition, as a necessary constraint, must be introduced from the beginning
and imposed on the massive Yang-Mills Lagrangian; (b) The gauge-invariance\
of a gauge field theory should be generally examined from its action other
than from the Lagrangian because action is of more fundamental dynamical
meaning than Lagrangian. Particularly, for a constrained system such as the $%
\rho $-meson field, the gauge-invariance should be seen from its action
given in the physical subspace defined by the Lorentz condition because the
field exists and moves only in the physical subspace; (c) In the physical
subspace, only infinitesimal gauge transformations are possibly allowed and
necessary to be considered in examination of whether the theory is
gauge-invariant or not; This fact was clarified originally in Ref. [13]; (d)
To construct a correct gauge field theory, the residual gauge degrees of
freedom existing in the physical subspace must be eliminated by a constraint
condition on the gauge group. This constraint condition may be determined by
requiring the action to be gauge-invariant [10]. Based on these points of
view, as will be shown in this paper, the $SU(2$)-symmetric model of
hadrodynamics can exactly be set up on the basis of gauge-invariance and the
quantization of the model can readily be performed in the path-integral
formalism by means of the Lagrange undetermined multiplier method.

The main purpose of this paper is to show that the quantum theory of the $%
SU(2)$-symmetric hadrodynamics built up on the gauge-invariance principle,
as the $U(1)$-symmetric hadrodynamics [14], can perfectly be renormalized.
Since a correct renormalization should be performed by exactly respecting
Ward-Takahashi (W-T) identities [15-17] which follow from the gauge-symmetry
of the theory, we first show that the quantum theory established here has an
important property that the effective action appearing in the generating
functional of Green functions is invariant with respect to a kind of
BRST-transformations [18]. From the BRST-symmetry of the theory, we will
derive various W-T identities satisfied by the generating functionals of
Green functions and proper vertex functions. Furthermore, from the W-T
identities obeyed by the generating functionals, we will derive W-T
identities satisfied by the $\rho -$meson and ghost particle propagators and
various proper vertices which appear in the perturbative expansion of
S-matrix elements. Based on these W-T identities, the propagators and
vertices will be perfectly renormalized. As a result of the
renormalizations, the Slavnov-Taylor (S-T) identity satisfied by the
renormalization constants [19, 20] will be derived. This identity is much
useful for practical calculations of the renormalization carried out by the
approach of renormalization group equation [21-23]. It should be mentioned
that the previous conclusion for the unrenormalizability of the model was
drawn from the quantum theory which was not established correctly [4-9]
because the unphysical degrees of freedom involved in the theory are not
eliminated by introducing appropriate constraint conditions. In our theory,
the unphysical degrees of freedom appearing in the theory, i.e., the
unphysical longitudinal components of vector potentials for the $\rho $%
-meson fields and the residual gauge degrees of freedom existing in the
subspace defined by the Lorentz condition are respectively eliminated by the
introduced Lorentz condition and the ghost equation which acts as the
constraint condition on the gauge group. This guarantees that the quantum
theory set up in this paper must be renormalizable. To demonstrate further
the renormalizability of the theory, in this paper, the one loop
renormalization will specifically be carried out by means of the
mass-dependent momentum space subtraction scheme and the renormalization
group equation (RGE). In this renormalization, we derive an exact one-loop
effective coupling constant and one-loop effective nucleon, $\rho $-meson
and pion masses without any ambiguity.

The arrangement of this paper is as follows. In Sec. 2, we will formulate
the quantization of the model in the path-integral formalism and derive the
BRST-transformations under which the effective action of the model is
invariant. In Sec. 3, we will derive the W-T identities satisfied by various
generating functionals. In Sec. 4, the W-T identity obeyed by the $\rho $%
-meson propagator and ghost propagator will be derived and the
renormalization of these propagators will be discussed. In Sec. 5, the W-T
identity obeyed by the $\rho $-meson three-line proper vertex will be
derived and the renormalization of the vertex will be discussed. In Sec. 6,
the same thing will be done for the $\rho $-meson four-line proper vertex.
In section 7, the W-T identity satisfied by the nucleon- $\rho $-meson
vertex will be derived and the renormalization of the vertex will be
discussed. In Sec. 8, the W-T identity obeyed by the pion-$\rho $-meson
three-line proper vertex will be derived and the renormalization of the
vertex will be discussed. In Sec. 9, the same thing will be done for the
pion-$\rho $-meson four-line proper vertex. Section 10 serves derive the
one-loop effective coupling constant. Section 11 is used to discuss the
renormalization of pion propagator and derive the one-loop effective boson ($%
\rho $-meson and pion) masses. In Sec. 12, we will discuss the
renormalization of nucleon propagator and derive the one-loop effective
nucleon mass. In the last section, some conclusions and discussions are
made. In Appendix, the Feynman rules derived from the model action will be
listed for convenience of perturbative calculations.

\section{Quantization and BRST-transformations}

The $SU(2)$-symmetric model of hadrodynamics is described by the following
Lagrangian [1,2]: 
\begin{equation}
{\cal L=L}_N+{\cal L}_\rho +{\cal L}_\pi +{\cal L}_{N\pi }  \eqnum{2.1}
\end{equation}
where 
\begin{equation}
{\cal L}_N=\bar \psi \{i\gamma ^\mu D_\mu -M\}\psi ,  \eqnum{2.2}
\end{equation}

\begin{equation}
{\cal L}_\rho =-\frac 14F^{a\mu \nu }F_{\mu \nu }^a+\frac 12m_\rho ^2A^{a\mu
}A_\mu ^a,  \eqnum{2.3}
\end{equation}
\begin{equation}
{\cal L}_\pi =\frac 12(D^\mu \pi )^{+}(D_\mu \pi )-\frac 12m_\pi ^2\pi ^2 
\eqnum{2.4}
\end{equation}
and

\begin{equation}
{\cal L}_{N\pi }=ig\bar \psi \gamma _5\tau ^a\psi \pi ^a.  \eqnum{2.5}
\end{equation}
In Eq. (2.2), 
\begin{equation}
\psi =\left( 
\begin{array}{c}
\psi _p \\ 
\psi _n
\end{array}
\right)  \eqnum{2.6}
\end{equation}
is the nucleon isospin doublet in which $\psi _p$ and $\psi _n$ denote the
proton and neutron field functions respectively, $M$ is the nucleon mass and 
\begin{equation}
D_\mu =\partial _\mu -igT^aA_\mu ^a  \eqnum{2.7}
\end{equation}
is the covariant derivative in which $T^a=\frac{\tau ^a}2$ ($a=1,2,3$) are
the generators of $SU(2$) gauge group with $\tau ^a$ being the isospin Pauli
matrices, $g$ is the coupling constant and $A_\mu ^a(x)$ represent the
vector potentials of $\rho $-meson fields. In Eq. (2.3), 
\begin{equation}
F_{\mu \nu }^a=\partial _\mu A_\nu ^a-\partial _\nu A_\mu ^a+g\varepsilon
^{abc}A_\mu ^bA_\nu ^c  \eqnum{2.8}
\end{equation}
stand for the strength tensors in which $\varepsilon ^{abc}$ are the
structure constants of $SU(2)$ group (the $3$-rank Levi-Civita tensor) and $%
m_\rho $ is the $\rho $-meson mass. In Eq. (2.4), 
\begin{equation}
\pi =\left( 
\begin{array}{c}
\pi _1 \\ 
\pi _2 \\ 
\pi _3
\end{array}
\right)  \eqnum{2.9}
\end{equation}
designates the pion isospin triplet and $m_\pi $ denotes the pion mass. It
is noted that the covariant derivative $D_\mu $ in Eq. (2.4) is still
represented in Eq. (2.7), but the matrix $T^a$ is now given in the adjoint
representation with matrix elements $(T^a)_{bc}=-i\varepsilon _{abc}$.

The Lagrangian described above, except for the $\rho $-meson mass term, was
derived from the requirement of $SU(2)$ gauge symmetry. The $\rho $-meson
mass term is added from the physical requirement and is obviously not
gauge-invariant. Therefore, the model described by the Lagrangian was
considered to be gauge-non-invariance in the past. In the model, the nucleon
fields are spinor fields and the pion fields are pseudoscalar fields which
are all independent physical field variables, while, the $\rho $-meson
fields, acting as the $SU(2)$ massive gauge fields, are vector fields for
which the four Lorentz components of a vector potential $A^{a\mu }$ (for a
certain $a$ ) are not all independent physical variables. Since a massive
vector field has only three degrees of freedom of polarization which can
completely be described by the transverse part $A_T^{a\mu }$ of the vector
potential $A^{a\mu }$, the longitudinal part $A_L^{a\mu }$ of the vector
potential appears to be a redundant degree of freedom which must be
eliminated by introducing the Lorentz condition ( the constraint condition
on the gauge field) 
\begin{equation}
\partial ^\mu A_\mu ^a=0  \eqnum{2.10}
\end{equation}
whose solution is

\begin{equation}
A_L^{a\mu }=0.  \eqnum{2.11}
\end{equation}
Therefore, the $\rho $-meson fields must be viewed as a constrained system
and the Lorentz condition must be introduced from the beginning and imposed
on the Lagrangian so as to eliminate the redundant unphysical variables. As
mentioned in Introduction, the gauge-invariance\ of a system should be
generally examined from its action other than from its Lagrangian because
action is of more fundamental dynamical meaning than Lagrangian.
Particularly, for a constrained system, the gauge-invariance should be seen
from its action given in the physical subspace defined by the constraint
condition because the fields exist and move only in the physical subspace.
And, as pointed out originally in Ref. [13], in the physical subspace, only
infinitesimal gauge transformations of gauge fields are possibly allowed and
necessary to be considered in examination of whether a field theory is
gauge-invariant or not. In accordance with these point of view, it is easy
to prove that the action given by the Lagrangian in Eqs. (2.1)-(2.9) is
locally gauge-invariant under the Lorentz condition. In fact, when we
perform the following infinitesimal $SU(2)$ local gauge transformations in
the action given by the Lagrangian in Eqs. (2.1)-(2.9): 
\begin{equation}
\begin{array}{c}
\delta A_\mu ^a(x)=D_\mu ^{ab}(x)\theta ^b(x), \\ 
\delta \psi (x)=igT^a\theta ^a(x)\psi (x), \\ 
\delta \bar \psi (x)=-ig\bar \psi (x)T^a\theta ^a(x), \\ 
\delta \pi ^a(x)=g\varepsilon ^{abc}\pi ^b(x)\theta ^c(x)
\end{array}
\eqnum{2.12}
\end{equation}
where $\theta ^a(x)$ ($a=1,2,3$) are the parametric functions of the local
gauge group $U(x)=\exp \{igT^a\theta ^a(x)\}$ and 
\begin{equation}
D_\mu ^{ab}=\delta ^{ab}\partial _\mu -g\varepsilon ^{abc}A_\mu ^c, 
\eqnum{2.13}
\end{equation}
and apply the Lorentz condition to the action, it is found that 
\begin{equation}
\delta S=-m_\rho ^2\int d^4x\theta ^a\partial ^\mu A_\mu ^a=0  \eqnum{2.14}
\end{equation}
which implies that the model constructed by the Lagrangian in Eqs.
(2.1)-(2.9) and the Lorentz condition in Eq. (2.10) is gauge-invariant.

According to the general procedure of dealing with the constrained system,
the Lorentz condition may be incorporated into the Lagrangian by means of
the Lagrange undetermined multiplier method. In doing this, it is convenient
to generalize the\ model Lagrangian and the Lorentz condition to the
following forms: 
\begin{equation}
{\cal L}_\lambda ={\cal L}-\frac 12\alpha (\lambda ^a)^2  \eqnum{2.15}
\end{equation}
and

\begin{equation}
\partial ^\mu A_\mu ^a+\alpha \lambda ^a=0  \eqnum{2.16}
\end{equation}
where ${\cal L}$ was given in Eqs. (2.1)-(2.9). When the constraint in Eq.
(2.16) is incorporated into the Lagrangian in Eq. (2.15) by the Lagrange
multiplier method, we obtain 
\begin{equation}
\begin{array}{c}
{\cal L}_\lambda ={\cal L}-\frac 12\alpha (\lambda ^a)^2+\lambda ^a(\partial
^\mu A_\mu ^a+\alpha \lambda ^a) \\ 
={\cal L}+\lambda ^a\partial ^\mu A_\mu ^a+\frac 12\alpha (\lambda ^a)^2.
\end{array}
\eqnum{2.17}
\end{equation}
This Lagrangian is obviously not gauge-invariant. However, to build up a
correct gauge field theory, it is necessary to require the action given by
the Lagrangian (2.17) to be invariant under the gauge transformations\
denoted in Eqs. (2.12). By this requirement, noticing the identity $%
\varepsilon ^{abc}A^{a\mu }A_\mu ^b=0$ and applying the constraint condition
in Eq. (2.16), we find

\begin{equation}
\delta S_\lambda =-\frac 1\alpha \int d^4x\partial ^\nu A_\nu ^a(x)\partial
^\mu ({\cal D}_\mu ^{ab}(x)\theta ^b(x))=0  \eqnum{2.18}
\end{equation}
where 
\begin{equation}
{\cal D}_\mu ^{ab}(x)=\delta ^{ab}\frac{\sigma ^2}{\Box _x}\partial _\mu
^x+D_\mu ^{ab}(x)  \eqnum{2.19}
\end{equation}
in which $\sigma ^2=\alpha m_\rho ^2$ and $D_\mu ^{ab}(x)$ was defined in
Eq. (2.13). From Eq. (2.16), we see $\frac 1\alpha \partial ^\nu A_\nu
^a=-\lambda ^a\neq 0$. Therefore, to ensure the action to be
gauge-invariant, the following constraint condition on the gauge group is
necessary to be required 
\begin{equation}
\partial _x^\mu ({\cal D}_\mu ^{ab}(x)\theta ^b(x))=0.  \eqnum{2.20}
\end{equation}
When we introduce, as usual, the ghost field variables $C^a(x)$ in such a
manner 
\begin{equation}
\theta ^a(x)=\xi C^a(x),  \eqnum{2.21}
\end{equation}
where $\xi $ is an infinitesimal Grassmann's number, the constraint
condition in Eq. (2.20) can be rewritten as 
\begin{equation}
\partial ^\mu ({\cal D}_\mu ^{ab}C^b)=0,  \eqnum{2.22}
\end{equation}
where the number $\xi $ has been dropped. This constraint condition usually
is called ghost equation. Certainly, the condition in Eq. (2.22) may also be
incorporated into the Lagrangian in Eq. (2.17) by the Lagrange multiplier
method to give a more generalized Lagrangian as follows 
\begin{equation}
{\cal L}_\lambda ={\cal L}+\lambda ^a\partial ^\mu A_\mu ^a+\frac 12\alpha
(\lambda ^a)^2+\bar C^a\partial ^\mu ({\cal D}_\mu ^{ab}C^b)  \eqnum{2.23}
\end{equation}
where $\bar C^a(x)$, acting as Lagrange undetermined multipliers, are the
new scalar variables conjugate to the ghost variables $C^a(x).$

At present, we are in a position to proceed the quantization of the model.
As we learn from the Lagrange undetermined multiplier method, the dynamical
and constrained variables as well as the Lagrange multipliers in the
Lagrangian (2.23) can all be treated as free ones, varying independently.
Therefore, in the Lagrangian path-integral formalism of quantization [10,
16, 17], we are allowed to use this kind of Lagrangian to construct the
generating functional of Green functions 
\begin{equation}
\begin{array}{c}
\ Z[\overline{\eta },\eta ,J^{a\mu },K^a,\overline{\xi }^a,\xi ^a]=\frac 1N%
\int D(\bar \psi ,\psi ,A_\mu ^a,\pi ^a,\bar C^a,C^a,\lambda ^a)exp\{i\int
d^4x[{\cal L}_\lambda (x) \\ 
+\bar \psi (x)\eta (x)+\overline{\eta }(x)\psi (x)+J^{a\mu }(x)A_\mu
^a(x)+K^a\pi ^a+\overline{\xi }^a(x)C^a(x)+\bar C^a(x)\xi ^a(x)]\}
\end{array}
\eqnum{2.24}
\end{equation}
where $D(A_\mu ^a,\cdots ,\lambda ^a)$ denotes the functional integration
measure, $\overline{\eta },\eta ,J_\mu ^a,K^a,\overline{\xi }^a$ and $\xi ^a$
are the external sources coupled to the nucleon, $\rho $-meson, pion and
ghost fields respectively and $N$ is the normalization constant. The
integral over $\lambda ^a(x)$ in Eq. (2.24), as seen from Eq. (2.23), is of
Gaussian-type. After calculating this integral, we arrive at 
\begin{equation}
\begin{array}{c}
Z[\overline{\eta },\eta ,J^{a\mu },K^a,\overline{\xi }^a,\xi ^a]=\frac 1N%
\int D(\bar \psi ,\psi ,A_\mu ^a,\pi ^a,\bar C^a,C^a)exp\{i\int d^4x[{\cal L}%
_{eff}(x) \\ 
+\bar \psi (x)\eta (x)+\overline{\eta }(x)\psi (x)+J^{a\mu }(x)A_\mu
^a(x)+K^a\pi ^a+\overline{\xi }^a(x)C^a(x)+\bar C^a(x)\xi ^a(x)]\}
\end{array}
\eqnum{2.25}
\end{equation}
where 
\begin{equation}
{\cal L}_{eff}={\cal L}-\frac 1{2\alpha }(\partial ^\mu A_\mu ^a)^2-\partial
^\mu \bar C^a{\cal D}_\mu ^{ab}C^b  \eqnum{2.26}
\end{equation}
is the effective Lagrangian given in arbitrary gauges in which ${\cal L}$ is
the Lagrangian written in Eqs. (2.1)-(2.9), and the second and third terms
usually are named as gauge-fixing term and ghost term, respectively.

Similar to other gauge field theories such as the standard model, for the
quantum hadrodynamics described above, there are a set of
BRST-transformations including the infinitesimal gauge transformations shown
in Eq. (2.12) and the transformations for the ghost fields under which the
effective action is invariant. The transformations for the ghost fields may
be found from the stationary condition of the effective action under the
BRST-transformations. By applying the transformations in Eq. (2.12) to the
action given by the Lagrangian in Eq. (2.26), one can derive 
\begin{equation}
\delta S=\int d^4x\{[\delta \bar C^a-\frac \xi \alpha \partial ^\nu A_\nu
^a]\partial ^\mu ({\cal D}_\mu ^{ab}C^b)+\bar C^a\partial ^\mu \delta ({\cal %
D}_\mu ^{ab}C^b)\}=0.  \eqnum{2.27}
\end{equation}
This expression suggests that if we set 
\begin{equation}
\delta \bar C^a=\frac \xi \alpha \partial ^\nu A_\nu ^a  \eqnum{2.28}
\end{equation}
and 
\begin{equation}
\partial ^\mu \delta ({\cal D}_\mu ^{ab}C^b)=0.  \eqnum{2.29}
\end{equation}
The action will be invariant. Eq. (2.28) gives the transformation law of the
ghost field variable $\bar C^a(x)$ which is similar to the one in quantum
chromodynamics (QCD). From Eq. (2.29), we may derive a transformation law of
the ghost field variables $C^a(x)$. Noticing the relation in Eq. (2.19), we
can write 
\begin{equation}
\delta ({\cal D}_\mu ^{ab}(x)C^b(x))=\frac{\sigma ^2}{\Box _x}\partial _\mu
^x\delta C^a(x)+\delta (D_\mu ^{ab}(x)C^b(x)).  \eqnum{2.30}
\end{equation}
Paralleling to the proof in QCD [9, 18], it can be found that

\begin{equation}
\delta (D_\mu ^{ab}(x)C^b(x))=D_\mu ^{ab}(x)[\delta C^b(x)+\frac \xi 2%
g\varepsilon ^{bcd}C^c(x)C^d(x)].  \eqnum{2.31}
\end{equation}
With this result, Eq. (2.30) can be represented as 
\begin{equation}
\delta ({\cal D}_\mu ^{ab}(x)C^b(x))={\cal D}_\mu ^{ab}(x)\delta
C^b(x)-D_\mu ^{ab}(x)\delta C_0^b(x)  \eqnum{2.32}
\end{equation}
where 
\begin{equation}
\delta C_0^a(x)\equiv -\frac{\xi g}2\varepsilon ^{abc}C^b(x)C^c(x). 
\eqnum{2.33}
\end{equation}
On substituting Eq. (2.32) into Eq. (2.29), we have 
\begin{equation}
M^{ab}(x)\delta C^b(x)=M_0^{ab}(x)\delta C_0^b(x)  \eqnum{2.34}
\end{equation}
where we have defined 
\begin{equation}
M^{ab}(x)\equiv \partial _x^\mu {\cal D}_\mu ^{ab}(x)=\delta ^{ab}(\Box
_x+\sigma ^2)-g\varepsilon ^{abc}A_\mu ^c(x)\partial _x^\mu  \eqnum{2.35}
\end{equation}
and

\begin{equation}
M_0^{ab}(x)\equiv \partial _x^\mu D_\mu ^{ab}(x)=M^{ab}(x)-\sigma ^2\delta
^{ab}.  \eqnum{2.36}
\end{equation}
It is noted that the operator in Eq. (2.35 ) is just the operator appearing
in Eq. (2.22). Corresponding to Eq. (2.22), we may write an equation
satisfied by the Green function $\Delta ^{ab}(x-y)$, 
\begin{equation}
M^{ac}(x)\Delta ^{cb}(x-y)=\delta ^{ab}\delta ^4(x-y).  \eqnum{2.37}
\end{equation}
The function $\Delta ^{ab}(x-y)$ just is the exact propagator of the ghost
field which is the inverse of the operator $M^{ab}(x)$. With the help of Eq.
(2.37) and noticing Eq. (2.36), we may solve out the $\delta C^a(x)$ from
Eq. (2.34) 
\begin{equation}
\begin{array}{c}
\delta C^a(x)=(M^{-1}M_0\delta C_0)^a(x)=\{M^{-1}(M-\sigma ^2)\delta
C_0\}^a(x) \\ 
=\delta C_0^a(x)-\sigma ^2\int d^4y\Delta ^{ab}(x-y)\delta C_0^b(y).
\end{array}
\eqnum{2.38}
\end{equation}
This just is the transformation law for the ghost field variables $C^a(x)$.
It is interesting to note that in the Landau gauge ($\alpha =0),$ due to $%
\sigma =0$, the above transformation will reduce to the form similar to the
one given in QCD. This result is natural since in the Landau gauge, the $%
\rho $-meson field mass term in the action is gauge-invariant. However, in
general gauges, the mass term is no longer gauge-invariant. In this case, to
maintain the action to be gauge-invariant, it is necessary to give the ghost
field a mass $\sigma $ so as to counteract the gauge-non-invariance of the $%
\rho $-meson field mass term. As a result, in the transformation given in
Eq. (2.38) appears a term proportional to $\sigma ^2$.

\section{Ward-Takahashi identities}

This section is devoted to deriving the W-T identities satisfied by
generating functionals based on the BRST-symmetry of the theory. When we
make the BRST-transformations shown in Eqs. (2.12), (2.28) and (2.38) to the
generating functional in Eq. (2.25) and consider the invariance of the
generating functional, the action and the integration measure under the
transformations (the invariance of the integration measure is easy to
check), we obtain an identity such that [9, 17] 
\begin{equation}
\begin{array}{c}
\frac 1N\int {\cal D}(A_\mu ^a,\bar C^a,C^a,\pi ^a,\bar \psi ,\psi )\int
d^4x\{J^{a\mu }(x)\delta A_\mu ^a(x)+\bar \eta (x)\delta \psi (x)+\delta 
\bar \psi (x)\eta (x) \\ 
+K^a(x)\delta \pi ^a(x)+\delta \overline{C}^a(x)\xi ^a(x)+\overline{\xi }%
^a(x)\delta C^a(x)\}e^{iS+EST} \\ 
=0
\end{array}
\eqnum{3.1}
\end{equation}
where $EST$ is an abbreviation of the external source terms appearing in Eq.
(2.25). The Grassmann number $\xi $ contained in the BRST-transformations in
Eq. (3.1) may be eliminated by performing a partial differentiation of Eq.
(3.1) with respect to $\xi $. As a result, we get a W-T identity as follows 
\begin{equation}
\begin{array}{c}
\frac 1N\int {\cal D}(A_\mu ^a,\bar C^a,C^a,\bar \psi ,\psi )\int
d^4x\{J^{a\mu }(x)\Delta A_\mu ^a(x)-\bar \eta (x)\Delta \psi (x)+\Delta 
\bar \psi (x)\eta (x) \\ 
+K^a(x)\Delta \pi ^a(x)+\triangle \overline{C}^a(x)\xi ^a(x)-\overline{\xi }%
^a(x)\Delta C^a(x)\}e^{iS+EST} \\ 
=0
\end{array}
\eqnum{3.2}
\end{equation}
where 
\begin{equation}
\begin{array}{c}
{\Delta }A_\mu ^a(x)=D_\mu ^{ab}(x)C^b(x), \\ 
{\Delta }\psi (x)=igT^aC^a(x)\psi (x), \\ 
{\Delta }\bar \psi (x)=ig\bar \psi (x)T^aC^a(x), \\ 
\Delta \pi ^a(x)=g\varepsilon ^{abc}\pi ^b(x)C^c(x), \\ 
{\Delta }\bar C^a(x)=\frac 1\alpha \partial ^\mu A_\mu ^a(x), \\ 
{\Delta }C^a(x)=\int d^4y[\delta ^{ab}\delta ^4(x-y)-\sigma ^2\Delta
^{ab}(x-y)]{\triangle }C_0^b(y), \\ 
{\Delta }C_0^b(y)=-\frac 12g\varepsilon ^{bcd}C^c(y)C^d(y).
\end{array}
\eqnum{3.3}
\end{equation}
The functions defined above are finite. Each of them differs from the
corresponding BRST-transformation written in Eqs. (2.12), (2.28) and (2.38)
by an infinitesimal Grassmann parameter $\xi .$

In order to represent the composite field functions $\Delta A_\mu ^a,\Delta 
\bar \psi ,\Delta \psi ,\Delta \pi ^a$ and $\Delta C^a$ in Eq. (3.2) in
terms of differentials of the functional $Z$ with respect to external
sources, we may, as usual, construct a generalized generating functional by
introducing new external sources (called BRST-sources later on) into the
generating functional written in Eq. (2.25), as shown in the following [9,
17] 
\begin{equation}
\begin{array}{c}
Z[J_\mu ^a,\overline{\xi }^a,\xi ^a,K^a,\bar \eta ,\eta ;u^{a\mu },v^a,\chi
^a,\bar \zeta ,\zeta ] \\ 
=\frac 1N\int {\cal D}[\bar \psi ,\psi ,A_\mu ^a,\pi ^a,\bar C%
^a,C^a,]exp\{iS+i\int d^4x[u^{a\mu }\Delta A_\mu ^a+\Delta \bar \psi \zeta +%
\bar \zeta \Delta \psi \\ 
+\chi ^a\Delta \pi ^a+v^a\Delta C^a+J^{a\mu }A_\mu ^a+\overline{\xi }^aC^a+%
\bar C^a\xi ^a+K^a\pi ^a+\bar \eta \psi +\bar \psi \eta ]\}
\end{array}
\eqnum{3.4}
\end{equation}
where $u^{a\mu }$, $\overline{v}^a$, $\chi ^a$ $\bar \zeta $ and $\zeta $
are the sources of the functions $\Delta A_{\mu \text{, }}^a$ $\Delta C^a$, $%
\Delta \pi ^a$, $\Delta \Psi $ and $\Delta \overline{\psi }$, respectively.
Obviously, the $u^{a\mu }$, $\Delta A_\mu ^a$, $\chi ^a$ and $\Delta \pi ^a$
are anticommuting quantities, while, the $v^a$, $\bar \zeta $, $\zeta $, $%
\Delta C^a$, $\Delta \bar \psi $ and $\Delta \psi $ are commuting ones. We
may start from the above generating functional to re-derive the W-T
identity. In order that the identity thus derived is identical to that as
given in Eq. (3.2), it is necessary to require the BRST-source terms $%
u_i\Delta \Phi _i$, where $u_i=u^{a\mu }$, $v^a$, $\chi ^a$, $\overline{%
\zeta }$ or $\zeta $ and $\Delta \Phi _i=\Delta A_\mu ^a$, $\Delta C^a$, $%
\Delta \pi ^a$, $\Delta \Psi $ or $\Delta \overline{\Psi }$ to be invariant
under the BRST-transformations. How to ensure the BRST-invariance of the
source terms? For illustration, let us introduce the source terms in such a
fashion 
\begin{equation}
\begin{array}{c}
\int d^4x[\widetilde{u}^{a\mu }\delta A_\mu ^a+\widetilde{v}^a\delta C^a+%
\widetilde{\chi }^a\delta \pi ^a+\overline{\widetilde{\zeta }}\delta \psi
+\delta \overline{\psi }\widetilde{\zeta }] \\ 
=\int d^4x[u^{a\mu }\triangle A_\mu ^a+v^a\triangle C^a+\chi ^a\Delta \pi ^a+%
\overline{\zeta }\triangle \psi +\triangle \overline{\psi }\zeta ]
\end{array}
\eqnum{3.5}
\end{equation}
where 
\begin{equation}
u^{a\mu }=\tilde u^{a\mu }\xi ,\;\;v^a=\tilde v^a\xi ,\;\;\chi ^a=\widetilde{%
\chi }^a\xi ,\;\;\bar \varsigma =\overline{\widetilde{\varsigma }}\xi
,\;\;\varsigma =-\tilde \varsigma \xi .  \eqnum{3.6}
\end{equation}
These external sources are defined by including the Grassmann number $\xi $
and hence products of them with $\xi $ vanish. This suggests that we may
generally define the sources by the following condition 
\begin{equation}
u_i\xi =0.  \eqnum{3.7}
\end{equation}
Considering that under the BRST-transformation, the variation of the
composite field functions given in arbitrary gauges can be represented in
the form $\delta \Delta \Phi _i=\xi \widetilde{\Phi }_i$ where $\widetilde{%
\Phi }_i$ are functions without including the parameter $\xi $. Clearly, the
definition in Eq. (3.7) for the sources would guarantee the BRST- invariance
of the BRST-source terms. When the BRST-transformations in Eqs. (2.12),
(2.28) and (2.38) are made to the generating functional in Eq. (3.4), due to
the definition in Eq. (3.7) for the sources, we have $u_i\delta \Delta \Phi
_i=0$ which means that the BRST-source terms give a vanishing contribution
to the identity in Eq. (3.1). Therefore, we still obtain the identity as
shown in Eq. (3.1) except that the external source terms is now extended to
include the BRST-external source terms. This fact indicates that we may
directly insert the BRST-source terms into the exponent in Eq. (3.1) without
changing the identity itself. When performing a partial differentiation of
the identity with respect to $\xi $, we obtain a W-T identity which is the
same as written in Eq. (3.2) except that the BRST-source terms are now
included in the identity. Therefore, Eq. (3.2) may be expressed as 
\begin{equation}
\begin{array}{c}
\int d^4x[J^{a\mu }(x)\frac \delta {\delta u^{a\mu }(x)}-\bar \eta (x)\frac 
\delta {\delta \bar \zeta (x)}+\eta (x)\frac \delta {\delta \zeta (x)}+K^a(x)%
\frac \delta {\delta \chi ^a(x)} \\ 
-\overline{\xi }^a(x)\frac \delta {\delta v^a(x)}+\frac 1\alpha \xi
^a(x)\partial _x^\mu \frac \delta {\delta J^{a\mu }(x)}]Z[J_\mu ^a,\cdots
,\zeta ] \\ 
=0.
\end{array}
\eqnum{3.8}
\end{equation}
This is the W-T identity satisfied by the generating functional of full
Green functions.

On substituting in Eq. (3.8) the relation [9, 17] 
\begin{equation}
Z=e^{iW}  \eqnum{3.9}
\end{equation}
where $W$ denotes the generating functional of connected Green functions,
one may obtain a W-T identity expressed by the functional $W$ 
\begin{equation}
\begin{array}{c}
\int d^4x[J^{a\mu }(x)\frac \delta {\delta u^{a\mu }(x)}-\bar \eta (x)\frac 
\delta {\delta \bar \zeta (x)}+\eta (x)\frac \delta {\delta \zeta (x)}+K^a(x)%
\frac \delta {\delta \chi ^a(x)} \\ 
-\overline{\xi }^a(x)\frac \delta {\delta v^a(x)}+\frac 1\alpha \xi
^a(x)\partial _x^\mu \frac \delta {\delta J^{a\mu }(x)}]W[J_u^a,\cdots
,\zeta ] \\ 
=0.
\end{array}
\eqnum{3.10}
\end{equation}
From this identity, one may get another W-T identity satisfied by the
generating functional $\Gamma $ of proper (one-particle-irreducible) vertex
functions. The functional $\Gamma $ is usually defined by the following
Legendre transformation [9, 17]

\begin{equation}
\begin{array}{c}
\Gamma [A^{a\mu },\bar C^a,C^a,\pi ^a,\bar \psi ,\psi ;u_\mu ^a,v^a,\chi ^a,%
\bar \zeta ,\zeta ]=W[J_\mu ^a,\overline{\xi }^a,\xi ^a,K^a,\bar \eta ,\eta
;u_\mu ^a,v^a,\chi ^a,\bar \zeta ,\zeta ] \\ 
-\int d^4x[J_\mu ^aA^{a\mu }+\overline{\xi }^aC^a+\bar C^a\xi ^a+K^a\pi ^a+%
\bar \eta \psi +\bar \psi \eta ]
\end{array}
\eqnum{3.11}
\end{equation}
where $A_\mu ^a,\bar C^a,C^a,\pi ^a,\bar \psi $ and $\psi $ are the field
variables defined by the following functional derivatives 
\begin{equation}
\begin{array}{c}
A_\mu ^a(x)=\frac{\delta W}{\delta J^{a\mu }(x)},\;\;\bar C^a(x)=-\frac{%
\delta W}{\delta \xi ^a(x)},C^a(x)=\frac{\delta W}{\delta \overline{\xi }%
^a(x)}, \\ 
\pi ^a(x)=\frac{\delta W}{\delta K^a(x)},\;\;\bar \psi (x)=-\frac{\delta W}{%
\delta \eta (x)},\;\;\psi (x)=\frac{\delta W}{\delta \bar \eta (x)}.
\end{array}
\eqnum{3.12}
\end{equation}
From Eq. (3.11), it is not difficult to get the inverse transformations [9,
17] 
\begin{equation}
\begin{array}{c}
J^{a\mu }(x)=-\frac{\delta \Gamma }{\delta A_\mu ^a(x)},\;\;\overline{\xi }%
^a(x)=\frac{\delta \Gamma }{\delta C^a(x)},\xi ^a(x)=-\frac{\delta \Gamma }{%
\delta \bar C^a(x)}, \\ 
K^a(x)=-\frac{\delta \Gamma }{\delta \pi ^a(x)},\;\;\bar \eta (x)=\frac{%
\delta \Gamma }{\delta \psi (x)},\;\;\eta (x)=-\frac{\delta \Gamma }{\delta 
\bar \psi (x)}.
\end{array}
\eqnum{3.13}
\end{equation}
It is obvious that 
\begin{eqnarray}
\frac{\delta W}{\delta u_\mu ^a}=\frac{\delta \Gamma }{\delta u_\mu ^a},\;\;%
\frac{\delta W}{\delta v^a}=\frac{\delta \Gamma }{\delta v^a},\;\;\frac{%
\delta \Gamma }{\delta \chi ^a}=\frac{\delta W}{\delta \chi ^a},\;\;\frac{%
\delta W}{\delta \zeta }=\frac{\delta \Gamma }{\delta \zeta },\;\;\frac{%
\delta W}{\delta \bar \zeta }=\frac{\delta \Gamma }{\delta \bar \zeta }. 
\eqnum{3.14}
\end{eqnarray}
Employing Eqs. (3.13) and (3.14), the W-T identity in Eq. (3.10) will be
written as [9, 17] 
\begin{equation}
\begin{array}{c}
\int d^4x\{\frac{\delta \Gamma }{\delta A_\mu ^a(x)}\frac{\delta \Gamma }{%
\delta u^{a\mu }(x)}+\frac{\delta \Gamma }{\delta C^a(x)}\frac{\delta \Gamma 
}{\delta v^a(x)}+\frac{\delta \Gamma }{\delta \psi (x)}\frac{\delta \Gamma }{%
\delta \bar \zeta (x)} \\ 
+\frac{\delta \Gamma }{\delta \bar \psi (x)}\frac{\delta \Gamma }{\delta
\zeta (x)}+\frac{\delta \Gamma }{\delta \pi ^a(x)}\frac{\delta \Gamma }{%
\delta \chi ^a(x)}+\frac 1\alpha \partial _x^\mu A_\mu ^a(x)\frac{\delta
\Gamma }{\delta \overline{C}^a(x)}\} \\ 
=0.
\end{array}
\eqnum{3.15}
\end{equation}
This is the W-T identity satisfied by the generating functional of proper
vertex functions.

The above identity may be represented in another form with the aid of the
so-called ghost equation of motion. The ghost equation may easily be derived
by first making the translation transformation: $\bar C^a\rightarrow \bar C%
^a+\bar \lambda ^a$ in Eq. (2.25) where $\bar \lambda ^a$ is an arbitrary
Grassmann variable, then differentiating Eq. (2.25) with respect to the $%
\bar \lambda ^a$ and finally setting $\overline{\lambda }^a=0$. The result
is [9, 17] 
\begin{equation}
\frac 1N\int D(A_\mu ^a,\bar C^a,C^a,K^a,\bar \psi ,\psi )\{\xi
^a(x)+\partial _x^\mu ({\cal D}_\mu ^{ab}(x)C^b(x))\}e^{iS+EST}=0. 
\eqnum{3.16}
\end{equation}
When we use the generating functional defined in Eq. (3.4) and notice the
relation in Eq. (2.19), the above equation may be represented as [9, 17] 
\begin{equation}
\lbrack \xi ^a(x)-i\partial _x^\mu \frac \delta {\delta u^{a\mu }(x)}-i{%
\sigma }^2\frac \delta {\delta \overline{\xi }^a(x)}]Z[J_\mu ^a,\cdots
,\zeta ]=0.  \eqnum{3.17}
\end{equation}
On substituting the relation in Eq. (3.9) into the above equation, we may
write a ghost equation satisfied by the functional $W$ such that 
\begin{equation}
\xi ^a(x)+\partial _x^\mu \frac{\delta W}{\delta u^{a\mu }(x)}+{\sigma }^2%
\frac{\delta W}{\delta \overline{\xi }^a(x)}=0.  \eqnum{3.18}
\end{equation}
From this equation, the ghost equation obeyed by the functional $\Gamma $ is
easy to be derived by virtue of Eqs. (3.12)-(3.14) [9, 17] 
\begin{equation}
\frac{\delta \Gamma }{\delta \bar C^a(x)}-\partial _x^\mu \frac{\delta
\Gamma }{\delta u^{a\mu }(x)}-\sigma ^2C^a(x)=0.  \eqnum{3.19}
\end{equation}
Upon applying the above equation to the last term in Eq. (3.15). the
identity in Eq. (3.15) will be rewritten as 
\begin{equation}
\begin{array}{c}
\int d^4x\{\frac{\delta \Gamma }{\delta A_\mu ^a}\frac{\delta \Gamma }{%
\delta u^{a\mu }}+\frac{\delta \Gamma }{\delta C^a}\frac{\delta \Gamma }{%
\delta v^a}+\frac{\delta \Gamma }{\delta \psi }\frac{\delta \Gamma }{\delta 
\bar \zeta }+\frac{\delta \Gamma }{\delta \bar \psi }\frac{\delta \Gamma }{%
\delta \zeta } \\ 
+\frac{\delta \Gamma }{\delta \pi ^a(x)}\frac{\delta \Gamma }{\delta \chi
^a(x)}+m_\rho ^2\partial ^\nu A_\nu ^aC^a-\frac 1\alpha \partial ^\mu
\partial ^\nu A_\nu ^a\frac{\delta \Gamma }{\delta u^{a\mu }}\} \\ 
=0.
\end{array}
\eqnum{3.20}
\end{equation}

Now, let us define a new functional $\hat \Gamma $ in such a manner 
\begin{equation}
\hat \Gamma =\Gamma +\frac 1{2\alpha }\int d^4x(\partial ^\mu A_\mu ^a)^2. 
\eqnum{3.21}
\end{equation}
From this definition, it follows that 
\begin{equation}
\frac{\delta \Gamma }{\delta A_\mu ^a}=\frac{\delta \hat \Gamma }{\delta
A_\mu ^a}+\frac 1\alpha \partial ^\mu \partial ^\nu A_\nu ^a.  \eqnum{3.22}
\end{equation}
When inserting Eq. (3.21) into Eq. (3.20) and considering the relation in
Eq. (3.22), we arrive at 
\begin{equation}
\int d^4x\{\frac{\delta \hat \Gamma }{\delta A_\mu ^a}\frac{\delta \hat 
\Gamma }{\delta u^{a\mu }}+\frac{\delta \hat \Gamma }{\delta C^a}\frac{%
\delta \hat \Gamma }{\delta v^a}+\frac{\delta \hat \Gamma }{\delta \psi }%
\frac{\delta \hat \Gamma }{\delta \bar \zeta }+\frac{\delta \hat \Gamma }{%
\delta \bar \psi }\frac{\delta \hat \Gamma }{\delta \zeta }+\frac{\delta
\Gamma }{\delta \pi ^a}\frac{\delta \Gamma }{\delta \chi ^a}+m_\rho
^2\partial ^\nu A_\nu ^aC^a\}=0.  \eqnum{3.23}
\end{equation}
The ghost equation represented through the functional ${\hat \Gamma }$ is of
the same form as that in Eq. (3.19) 
\begin{equation}
\frac{\delta \hat \Gamma }{\delta \bar C^a(x)}-\partial _x^\mu \frac{\delta 
\hat \Gamma }{\delta u^{a\mu }(x)}-{\sigma }^2C^a(x)=0.  \eqnum{3.24}
\end{equation}
In the Landau gauge, since $\sigma =0$ and ${\partial ^\nu A}_\nu ^a{=0}$,
Eqs. (3.23) and (3.24) respectively reduce to [9, 17] 
\begin{equation}
\int d^4x\{\frac{\delta \hat \Gamma }{\delta A_\mu ^a}\frac{\delta \hat 
\Gamma }{\delta u^{a\mu }}+\frac{\delta \hat \Gamma }{\delta C^a}\frac{%
\delta \hat \Gamma }{\delta v^a}+\frac{\delta \hat \Gamma }{\delta \psi }%
\frac{\delta \hat \Gamma }{\delta \bar \zeta }+\frac{\delta \hat \Gamma }{%
\delta \bar \psi }\frac{\delta \hat \Gamma }{\delta \zeta }+\frac{\delta
\Gamma }{\delta \pi ^a}\frac{\delta \Gamma }{\delta \chi ^a}\}=0 
\eqnum{3.25}
\end{equation}
and 
\begin{equation}
\frac{\delta \hat \Gamma }{\delta \bar C^a}-\partial ^\mu \frac{\delta \hat 
\Gamma }{\delta u^{a\mu }}=0.  \eqnum{3.26}
\end{equation}
which are homogeneous equations

From the W-T identities formulated above, we may derive various W-T
identities obeyed by Green functions and vertices, as will be illustrated in
the next sections.

\section{$\rho $-meson and ghost particle propagators}

In this section, we plan to derive the W-T identities satisfied by the $\rho 
$-meson and ghost particle propagators by starting from the W-T identity
represented in Eq. (3.8) and the ghost equation shown in Eq. (3.17) and then
discuss their renormalization. Let us perform differentiations of the
identities in Eqs. (3.8) and (3.17) with respect to the external sources $%
\xi ^a(x)$ and $\xi ^b(y)$ respectively and then set all the sources except
for the source $J_\mu ^a(x)$ to be zero. In this way, we obtain the
following identities 
\begin{eqnarray}
\frac 1\alpha \partial _x^\mu \frac{\delta Z[J]}{\delta J^{a\mu }(x)}+\int
d^4yJ^{b\nu }(y)\frac{\delta ^2Z[J,\xi ,u]}{\delta \xi ^a(x)\delta u^{b\nu
}(y)}|_{\xi =u=0}=0  \eqnum{4.1}
\end{eqnarray}
and 
\begin{equation}
\begin{array}{c}
i\partial _\mu ^x\frac{\delta ^2Z[J.\xi .u]}{\delta u_\mu ^a(x)\delta \xi
^b(y)}|_{\xi =u=0}+i{\sigma }^2\frac{\delta ^2Z[J,\overline{\xi },\xi ]}{%
\delta \overline{\xi }^a(x)\delta \xi ^b(y)}|_{\overline{\xi }=\xi =0} \\ 
+\delta ^{ab}\delta ^4(x-y)Z[J]=0.
\end{array}
\eqnum{4.2}
\end{equation}
Furthermore, on differentiating Eq. (4.1) with respect to $J_\nu ^b(y)$ and
then letting the source $J$ vanish, we may get an identity which is, in
operator representation, of the form [9, 17] $\ $%
\begin{equation}
\frac 1\alpha \partial _x^\mu <0^{+}|T[\hat A_\mu ^a(x)\hat A_\nu
^b(y)]|0^{-}>=<0^{+}|T^{*}[\hat {\bar C^a}(x)\hat D_\nu ^{bd}(y)\hat C%
^d(y)]|0^{-}>  \eqnum{4.3}
\end{equation}
where $\hat A_\nu ^a(x)$, $\hat C^a(x)$ and $\hat {\bar C^a}(x)$ stand for
the gluon field and ghost field operators and $T^{*}$ symbolizes the
covariant time-ordering product. When the source $J$ is set to vanish, Eq.
(4.2) gives such an equation [9, 17] 
\begin{equation}
\begin{array}{c}
i\partial _y^\nu <0^{+}|T^{*}\{\hat {\bar C^a}(x)\hat D_\nu ^{bd}(y)\hat C%
^d(y)\}|0^{-}> \\ 
+i{\sigma }^2<0^{+}|T[\hat {\bar C^a}(x)\hat C^b(y)]|0^{-}>=\delta
^{ab}\delta ^4(x-y).
\end{array}
\eqnum{4.4}
\end{equation}
Upon inserting Eq. (4.3) into Eq. (4.4), we have 
\begin{equation}
\partial _x^\mu \partial _y^\nu D_{\mu \nu }^{ab}(x-y)-\alpha \sigma
^2\Delta ^{ab}(x-y)=-\alpha \delta ^{ab}\delta ^4(x-y)  \eqnum{4.5}
\end{equation}
where 
\begin{equation}
iD_{\mu \nu }^{ab}(x-y)=<0^{+}|T\{\hat A_\mu ^a(x)\hat A_\nu ^b(y)\}|0^{-}> 
\eqnum{4.6}
\end{equation}
which is the full $\rho $-meson propagator and 
\begin{equation}
i\Delta ^{ab}(x-y)=<0^{+}|T\{\hat C^a(x)\hat {\bar C^b}(y)\}|0^{-}> 
\eqnum{4.7}
\end{equation}
which is the full ghost particle propagator. Eq. (4.5) just is the W-T
identity respected by the $\rho $-meson propagator which establishes a
relation between the longitudinal part of $\rho $-meson propagator and the
ghost particle propagator. Particularly, in the Landau gauge ($\alpha =0$),
as we see, Eq. (4.5) reduces to the form which exhibits the transversity of
the $\rho $-meson propagator. By the Fourier transformation, Eq. (4.5) will
be converted to the form given in the momentum space as follows 
\begin{equation}
k^\mu k^\nu D_{\mu \nu }^{ab}(k)-\alpha \sigma ^2\Delta ^{ab}(k)=-\alpha
\delta ^{ab}.  \eqnum{4.8}
\end{equation}
The ghost particle propagator may be determined by the ghost equation shown
in Eq. (4.4). However, we would rather here to derive its expression from
the Dyson-Schwinger equation [24] satisfied by the propagator which may be
established by the perturbation method. 
\begin{equation}
\Delta ^{ab}(k)=\Delta _0^{ab}(k)+\Delta _0^{aa^{\prime }}(k){\Omega }%
^{a^{\prime }b^{\prime }}(k)\Delta ^{b^{\prime }b}(k)  \eqnum{4.9}
\end{equation}
where 
\begin{equation}
i\Delta _0^{ab}(k)=i\delta ^{ab}\Delta _0(k)=\frac{-i\delta ^{ab}}{%
k^2-\sigma ^2+i\varepsilon }  \eqnum{4.10}
\end{equation}
is the free ghost particle propagator which can be derived from the
generating functional in Eq. (2.25) by a perturbative calculation and $%
-i\Omega ^{ab}(k)=-i\delta ^{ab}\Omega (k)$ denotes the proper self-energy
operator of ghost particle. From Eq. (4.9), it is easy to solve that 
\begin{equation}
i\Delta ^{ab}(k)=\frac{-i\delta ^{ab}}{k^2[1+\hat \Omega (k^2)]-\sigma
^2+i\varepsilon }  \eqnum{4.11}
\end{equation}
where the self-energy has properly been expressed as 
\begin{equation}
\Omega (k)=k^2\hat \Omega (k^2).  \eqnum{4.12}
\end{equation}
Similarly, we may write a Dyson-Schwinger equation for the $\rho $-meson
propagator by the perturbation procedure$\ $[24] 
\begin{equation}
D_{\mu \nu }(k)=D_{\mu \nu }^0(k)+D_{\mu \lambda }^0(k)\Pi ^{\lambda \rho
}(k)D_{\rho \nu }(k)  \eqnum{4.13}
\end{equation}
where the color indices are suppressed for simplicity and 
\begin{equation}
iD_{\mu \nu }^{(0)ab}(k)=i\delta ^{ab}D_{\mu \nu }^{(0)}(k)=-i\delta ^{ab}[%
\frac{g_{\mu \nu }-k_\mu k_\nu /k^2}{k^2-m_\rho ^2+i\varepsilon }+\frac{%
\alpha k_\mu k_\nu /k^2}{k^2-\sigma ^2+i\varepsilon }]  \eqnum{4.14}
\end{equation}
is the free $\rho $-meson propagator which can easily be derived from the
perturbative expansion of the generating functional in Eq. (2.25) and $-i\Pi
_{\mu \nu }^{ab}(k)=-i\delta ^{ab}\Pi _{\mu \nu }(k)$ stands for the $\rho $%
-meson proper self-energy operator. Let us decompose the propagator and the
self-energy operator into a transverse part and a longitudinal part: 
\begin{equation}
D^{\mu \nu }(k)=D_T^{\mu \nu }(k)+D_L^{\mu \nu }(k),\;\Pi ^{\mu \nu }(k)=\Pi
_T^{\mu \nu }(k)+\Pi _L^{\mu \nu }(k)  \eqnum{4.15}
\end{equation}
where 
\begin{equation}
\begin{array}{c}
D_T^{\mu \nu }(k)={\cal P}_T^{\mu \nu }(k)D_T(k^2),\;D_L^{\mu \nu }(k)={\cal %
P}_L^{\mu \nu }(k)D_L(k^2), \\ 
\Pi _T^{\mu \nu }(k)={\cal P}_T^{\mu \nu }(k)\Pi _T(k^2),\;\Pi _L^{\mu \nu
}(k)={\cal P}_L^{\mu \nu }(k)\Pi _L(k^2)
\end{array}
\eqnum{4.16}
\end{equation}
here ${\cal P}_T^{\mu \nu }(k)=(g^{\mu \nu }-\frac{k^\mu k^\nu }{k^2})$ and $%
{\cal P}_L^{\mu \nu }(k)=\frac{k^\mu k^\nu }{k^2}$ are the transverse and
longitudinal projectors respectively. Considering these decompositions and
the orthogonality between the transverse and longitudinal parts, Eq. (4.13)
will be split into two equations 
\begin{equation}
D_{T\mu \nu }(k)=D_{T\mu \nu }^0(k)+D_{T\mu \lambda }^0(k)\Pi _T^{\lambda
\rho }(k)D_{T\rho \nu }(k)  \eqnum{4.17}
\end{equation}
and 
\begin{equation}
D_{L\mu \nu }(k)=D_{L\mu \nu }^0(k)+D_{L\mu \lambda }^0(k)\Pi _L^{\lambda
\rho }(k)D_{L\rho \nu }(k).  \eqnum{4.18}
\end{equation}
Solving the equations (4.17) and (4.18), one can get 
\begin{equation}
iD_{\mu \nu }^{ab}(k)=-i\delta ^{ab}\{\frac{g_{\mu \nu }-k_\mu k_\nu /k^2}{%
k^2+\Pi _T(k^2)-m_\rho ^2+i\varepsilon }+\frac{\alpha k_\mu k_\nu /k^2}{%
k^2+\alpha \Pi _L(k^2)-\sigma ^2+i\varepsilon }\}.  \eqnum{4.19}
\end{equation}
With setting

\begin{equation}
\Pi _T(k^2)=k^2\Pi _1(k^2)+m_\rho ^2\Pi _2(k^2)  \eqnum{4.20}
\end{equation}
which follows from the Lorentz-covariance of the operator $\Pi _T(k^2)$ and

\begin{equation}
\alpha \Pi _L(k^2)=k^2\hat \Pi _L(k^2),  \eqnum{4.21}
\end{equation}
Eq. (4.19) will be written as 
\begin{equation}
iD_{\mu \nu }^{ab}(k)=-i\delta ^{ab}\{\frac{g_{\mu \nu }-k_\mu k_\nu /k^2}{%
k^2[1+\Pi _1(k^2)]-m_\rho ^2[1-\Pi _2(k^2)]+i\varepsilon }+\frac{\alpha
k_\mu k_\nu /k^2}{k^2[1+\hat \Pi _L(k^2)]-\sigma ^2+i\varepsilon }. 
\eqnum{4.22}
\end{equation}
We would like to note that the expressions given in Eqs. (4.12), (4.20) and
(4.21) can be verified by practical calculations and are important for the
renormalization of the propagators and the $\rho $-meson mass.

Substitution of Eqs. (4.11) and (4.22) into Eq. (4.8) yields 
\begin{equation}
\hat \Pi _L(k^2)=\frac{\sigma ^2\hat \Omega (k^2)}{k^2[1+\hat \Omega (k^2)]}.
\eqnum{4.23}
\end{equation}
From this relation, we see, either in the Landau gauge or in the zero-mass
limit, the $\hat \Pi _L(k^2)$ vanishes.

Now let us discuss the renormalization. The function $\hat \Omega (k^2)$ in
Eq. (4.11) and the functions $\Pi _1(k^2)$, $\Pi _2(k^2)$ and $\hat \Pi
_L(k^2)$ in Eq. (4.22) are generally divergent in higher order perturbative
calculations. According to the conventional procedure of renormalization,
the divergences included in the functions $\hat \Omega (k^2),$ $\Pi _1(k^2),$
$\Pi _2(k^2)$ and $\hat \Pi _L(k^2)$ may be subtracted at a renormalization
point, say, $k^2=\mu ^2$. Thus, we can write [9, 16, 17] 
\begin{equation}
\begin{array}{c}
\hat \Omega (k^2)=\hat \Omega (\mu ^2)+\hat \Omega ^c(k^2),\;\;\Pi
_1(k^2)=\Pi _1(\mu ^2)+\Pi _1^c(k^2), \\ 
\Pi _2(k^2)=\Pi _2(\mu ^2)+\Pi _2^c(k^2),\text{ }\hat \Pi _L(k^2)=\hat \Pi
_L(\mu ^2)+\hat \Pi _L^c(k^2)
\end{array}
\eqnum{4.24}
\end{equation}
where $\hat \Omega (\mu ^2)$, $\Pi _1(\mu ^2),$ $\Pi _2(\mu ^2),$ $\hat \Pi
_L(\mu ^2)$ and $\Omega ^c(k^2)$, $\Pi _1^c(k^2)$, $\Pi _2^c(k^2),$ $\hat \Pi
_L^c(k^2)$ are respectively the divergent parts and the finite parts of the
functions $\Omega (k^2)$, $\Pi _1(k^2)$, $\Pi _2(k^2)$ and $\hat \Pi _L(k^2)$%
. The divergent parts can be absorbed in the following renormalization
constants defined by [9, 16, 17] 
\begin{equation}
\begin{array}{c}
\tilde Z_3^{-1}=1+\hat \Omega (\mu ^2),\;\;Z_3^{-1}=1+\Pi _1(\mu ^2),\text{ }%
Z_3^{\prime -1}=1+\hat \Pi _L(\mu ^2), \\ 
Z_{m_\rho }^{-1}=\sqrt{Z_3[1-\Pi _2(\mu ^2)]}=\sqrt{[1-\Pi _1(\mu ^2)][1-\Pi
_2(\mu ^2)]}
\end{array}
\eqnum{4.25}
\end{equation}
where $Z_3$ and $\tilde Z_3$ are the renormalization constants of $\rho $%
-meson and ghost particle propagators respectively, $Z_3^{\prime }$ is the
additional renormalization constant of the longitudinal part of gluon
propagator and $Z_{m_\rho }$ is the renormalization constant of gluon mass.
With the above definitions of the renormalization constants, on inserting
Eq. (4.24) into Eqs. (4.11) and (4.22) , the ghost particle propagator and
the gluon propagator can be renormalized, respectively, in such a manner 
\begin{equation}
i\Delta ^{ab}(k)=\tilde Z_3i\Delta _R^{ab}(k)  \eqnum{4.26}
\end{equation}
and 
\begin{equation}
iD_{\mu \nu }^{ab}(k)=Z_3iD_{R\mu \nu }^{~~ab}(k)  \eqnum{4.27}
\end{equation}
where 
\begin{equation}
i\Delta _R^{ab}(k)=\frac{-i\delta ^{ab}}{k^2[1+\Omega _R(k^2)]-\sigma
_R^2+i\varepsilon }  \eqnum{4.28}
\end{equation}
and 
\begin{equation}
iD_{R\mu \nu }^{ab}(k)=-i\delta ^{ab}\{\frac{g_{\mu \nu }-k_\mu k_\nu /k^2}{%
k^2-m_\rho ^{R2}+\Pi _R^T(k^2)+i\varepsilon }+\frac{Z_3^{\prime }\alpha
_Rk_\mu k_\nu /k^2}{k^2[1+\Pi _R^L(k^2)]-\overline{\sigma }_R^2+i\varepsilon 
}\}  \eqnum{4.29}
\end{equation}
are the renormalized propagators in which $m_\rho ^R,$ $\sigma _R$ and $\bar 
\sigma _R$ are the renormalized masses, $\alpha _R$ is the renormalized
gauge parameter, $\Omega _R(k^2),\Pi _R^T(k^2)$ and $\Pi _R^L(k^2)$ denote
the finite corrections coming from the loop diagrams. They are defined as 
\begin{equation}
\begin{array}{c}
m_\rho ^R=Z_{m_\rho }^{-1}m_\rho ,\;\alpha _R=Z_3^{-1}\alpha ,\;\overline{%
\sigma }_R=\sqrt{Z_3^{^{\prime }}}\sigma ,\text{ }\sigma _R=\sqrt{\widetilde{%
Z}_3}\sigma , \\ 
\Omega _R(k^2)=\tilde Z_3\hat \Omega ^c(k^2),\text{ }\Pi
_R^T(k^2)=Z_3[k^2\Pi _1^c(k^2)+m_\rho ^2\Pi _2^c(k^2)],\;\Pi
_R^L(k^2)=Z_3^{\prime }\hat \Pi _L^c(k^2).
\end{array}
\eqnum{4.30}
\end{equation}
From the definitions in Eqs. (4.24) and (4.30), it is clearly seen that at
the renormalization point $\mu ,$ the finite corrections $\Omega _R(k^2),\Pi
_R^T(k^2)$ and $\Pi _R^L(k^2)$ vanish. In this case, the propagators reduce
to the form of free ones. As we see from Eq. (4.29), the longitudinal part
of the gluon propagator, except for in the Landau gauge, needs to be
renormalized and has an extra renormalization constant ${Z}_3^{\prime }$.
This fact coincides with the general property of the massive vector boson
propagator (see Ref. (9), Chap.V). From Eqs. (4.23)-(4.25) , it is easy to
find that the longitudinal part in Eq. ( 4.22) can be renormalized as 
\begin{equation}
\frac \alpha {k^2[1+\hat \Pi _L(k^2)]-\sigma ^2+i\varepsilon }=Z_3\alpha
_R[1+\Omega _R(k^2)]\Delta _R(k^2)  \eqnum{4.31}
\end{equation}
where 
\begin{equation}
\Delta _R(k^2)=\frac 1{k^2[1+\Omega _R(k^2)]-\sigma _R^2+i\varepsilon } 
\eqnum{4.32}
\end{equation}
which appears in Eq. (4.28) and the renormalization constant $Z_3^{\prime }$
can be expressed as 
\begin{equation}
Z_3^{\prime }=[1+\frac{\sigma _R^2}{\mu ^2}\frac{(1-\tilde Z_3)}{\tilde Z_3}%
]^{-1}.  \eqnum{4.33}
\end{equation}
If choosing $\mu =\sigma _R$, we have

\begin{equation}
Z_3^{\prime }=\tilde Z_3.  \eqnum{4.34}
\end{equation}

\section{$\rho $-meson three-line proper vertex}

The aim of this section is to derive the W-T identity satisfied by the $\rho 
$-meson three-line proper vertex and discuss the renormalization of the
vertex. For this purpose, we first derive a W-T identity satisfied by the $%
\rho $-meson three-point Green function. Let us begin with the derivation
from the W-T identity in Eq. (4.1) and the ghost equation in Eq. (4.2). By
taking successive differentiations of Eq. (4.1) with respect to the sources $%
J_\nu ^b(y)$ and $J_\lambda ^c(z)$ and then setting the sources to vanish,
one may obtain the W-T identity obeyed by the $\rho $-meson three-point
Green function which is written in the operator form as follows 
\begin{eqnarray}
\frac 1\alpha \partial _x^\mu G_{\mu \nu \lambda }^{abc}(x,y,z)
&=&<0^{+}|T^{*}[\hat {\bar C^a}(x)\hat D_\nu ^{bd}(y)\hat C^d(y)\hat A%
_\lambda ^c(z)]|0^{-}>  \nonumber \\
+ &<&0^{+}|T^{*}[\hat {\bar C^a}(x)\hat A_\nu ^b(y)\hat D_\lambda ^{cd}(z)%
\hat C^d(z)]|0^{-}>  \eqnum{5.1}
\end{eqnarray}
where 
\begin{equation}
G_{\mu \nu \lambda }^{abc}(x,y,z)=<0^{+}|T[\hat A_\mu ^a(x)\hat A_\nu ^b(y)%
\hat A_\lambda ^c(z)]|0^{-}>  \eqnum{5.2}
\end{equation}
is the three-point Green function mentioned above. The identity in Eq. (5.1)
will be simplified by a ghost equation which may be derived by
differentiating Eq. (4.2) with respect to the source $J_\lambda ^c(z)$ 
\begin{equation}
\begin{array}{c}
\partial _x^\mu <0^{+}|T^{*}\{\hat D_\mu ^{ad}(x)\hat C^d(x)\hat {\bar C^b}%
(y)\hat A_\lambda ^c(z)\}|0^{-}> \\ 
+{\sigma }^2<0^{+}|T[\hat C^a(x)\hat {\bar C^b}(y)\hat A_\lambda
^c(z)]|0^{-}]>=0.
\end{array}
\eqnum{5.3}
\end{equation}
Taking derivatives of Eq. (5.1) with respect to $y$ and $z$ and employing
Eq. (5.3), we get 
\begin{equation}
\partial _x^\mu \partial _y^\nu \partial _z^\lambda G_{\mu \nu \lambda
}^{abc}(x,y,z)=\alpha \sigma ^2\{\partial _y^\nu G_{~~\nu
}^{cab}(z,x,y)+\partial _z^\lambda G_{~~\lambda }^{bac}(y,x,z)\}  \eqnum{5.4}
\end{equation}
where 
\begin{equation}
G_{~~\mu }^{abc}(x,y,z)=<0^{+}|T\{\hat C^a(x)\hat {\bar C^b}(y)\hat A_\mu
^c(z)\}|0^{-}>.  \eqnum{5.5}
\end{equation}
is the three-point $\rho $-meson-ghost particle Green function. In the
Landau gauge, Eq. (5.4) reduces to 
\begin{equation}
\partial _x^\mu \partial _y^\nu \partial _z^\lambda G_{\mu \nu \lambda
}^{abc}(x,y,z)=0  \eqnum{5.6}
\end{equation}
which shows the transversity of the Green function. From Eq. (5.4), we may
derive a W-T identity for the $\rho $-meson three-line vertex. For this
purpose, it is necessary to use the following one-particle-irreducible
decompositions of the Green functions which can easily be obtained by the
well-known procedure [9, 17] 
\begin{equation}
\begin{array}{c}
G_{\mu \nu \lambda }^{abc}(x,y,z)=\int d^4x^{\prime }d^4y^{\prime
}d^4z^{\prime }iD_{\mu \mu ^{\prime }}^{aa^{\prime }}(x-x^{\prime }) \\ 
\times iD_{\nu \nu ^{\prime }}^{bb^{\prime }}(y-y^{\prime })iD_{\lambda
\lambda ^{\prime }}^{cc^{\prime }}(z-z^{\prime })\Gamma _{a^{\prime
}b^{\prime }c^{\prime }}^{\mu ^{\prime }\nu ^{\prime }\lambda ^{\prime
}}(x^{\prime },y^{\prime },z^{\prime })
\end{array}
\eqnum{5.7}
\end{equation}
and 
\begin{equation}
\begin{array}{c}
G_{~~\,\nu }^{abc}(x,y,z)=\int d^4x^{\prime }d^4y^{\prime }d^4z^{\prime
}i\Delta ^{aa^{\prime }}(x-x^{\prime })\Gamma ^{a^{\prime }b^{\prime
}c^{\prime },\nu ^{\prime }}(x^{\prime },y^{\prime },z^{\prime }) \\ 
\times i\Delta ^{b^{\prime }b}(y^{\prime }-y)iD_{\nu ^{\prime }\nu
}^{c^{\prime }c}(z^{\prime }-z)
\end{array}
\eqnum{5.8}
\end{equation}
where $iD_{\mu \mu ^{\prime }}^{aa^{\prime }}(x-x^{\prime })$ and $i\Delta
^{aa^{\prime }}(x-x^{\prime })$ are respectively the $\rho $-meson and the
ghost particle propagators discussed in the preceding section, $\Gamma
_{abc}^{\mu \nu \lambda }(x,y,z)$ and $\Gamma _{~~\lambda }^{abc}(x,y,z)$
are the three-line $\rho $-meson proper vertex and the three-line $\rho -$%
meson-ghost particle proper vertex respectively. They are defined as [9, 17] 
\begin{equation}
\Gamma _{abc}^{\mu \nu \lambda }(x,y,z)=i\frac{\delta ^3\Gamma }{\delta
A_\mu ^a(x)\delta A_\nu ^b(y)\delta A_\lambda ^c(z)}|_{J=0}  \eqnum{5.9}
\end{equation}
and 
\begin{equation}
\Gamma _{~~\,\lambda }^{abc}(x,y,z)=\frac{\delta ^3\Gamma }{i\delta \bar C%
^a(x)\delta C^b(y)\delta A^{c\lambda }(z)}|_{J=0}  \eqnum{5.10}
\end{equation}
where $J$ stands for all the external sources. Substituting Eqs. (5.7) and
(5.8) into Eq. (5.4) and transforming Eq. (5.4) into the momentum space, one
can derive an identity which establishes the relation between the
longitudinal part of the three-line $\rho $-meson vertex and the three-line
ghost vertex as follows 
\begin{equation}
\begin{array}{c}
p^\mu q^\nu k^\lambda \Lambda _{\mu \nu \lambda }^{abc}(p,q,k)=-\frac{\sigma
^2}\alpha \chi (p^2)[\chi (k^2)q^\nu \Lambda _{~~\nu }^{cab}(k,p,q) \\ 
+\chi (q^2)k^\lambda \Lambda _{~~\,\lambda }^{bac}(q,p,k)]
\end{array}
\eqnum{5.11}
\end{equation}
where we have defined 
\begin{equation}
\begin{array}{c}
\Gamma _{\mu \nu \lambda }^{abc}(p,q,k)=(2\pi )^4\delta ^4(p+q+k)\Lambda
_{\mu \nu \lambda }^{abc}(p,q,k), \\ 
\Gamma _{~~\,\lambda }^{abc}(p,q,k)=(2\pi )^4\delta ^4(p+q+k)\Lambda
_{~~\lambda }^{abc}(p,q,k)
\end{array}
\eqnum{5.12}
\end{equation}
and 
\begin{equation}
\begin{array}{c}
\chi (p^2)=\{k^2[1+\hat \Pi _L(p^2)]-\sigma ^2+i\varepsilon \}\{k^2[1+\hat 
\Omega (p^2)]-\sigma ^2+i\varepsilon \}^{-1} \\ 
=[1+\hat \Omega (k^2)]^{-1}
\end{array}
\eqnum{5.13}
\end{equation}
here $\hat \Pi _L(p^2)$ and $\hat \Omega (p^2)$ are the self-energies
defined in Eqs. (4.12) and (4.21). The second equality in Eq. (5.13) is
obtained by inserting the relation in Eq. (4.23) into the first equality.

Obviously, in the Landau gauge, Eq. (5.11) reduces to 
\begin{equation}
p^\mu q^\nu k^\lambda \Lambda _{\mu \nu \lambda }^{abc}(p,q,k)=0 
\eqnum{5.14}
\end{equation}
which implies that the vertex is transverse in this case. In the lowest
order approximation, owing to 
\begin{equation}
\chi (p^2)=1  \eqnum{5.15}
\end{equation}
and 
\begin{equation}
\Lambda _{~~~~~\;\mu }^{(0)abc}(p,q,k)=g\varepsilon ^{abc}p_\mu , 
\eqnum{5.16}
\end{equation}
the right hand side (RHS) of Eq. (5.11) vanishes, therefore, we have 
\begin{equation}
p^\mu q^\nu k^\lambda \Lambda _{~~~\mu \nu \lambda }^{(0)abc}(p,q,k)=0. 
\eqnum{5.17}
\end{equation}
This result is consistent with that for the bare three-line $\rho $-meson
vertex given by the Feynman rule.

Now, let us discuss the renormalization of the three-line gluon vertex. From
the renormalization of the $\rho $-meson and ghost particle propagators
described in Eqs. (4.26) and (4.27) and the definitions of the propagators
written in Eqs. (4.6) and (4.7), one can see 
\begin{equation}
\begin{array}{c}
A^{a\mu }(x)=\sqrt{Z_3}A_R^{a\mu }(x), \\ 
C^a(x)=\sqrt{\tilde Z_3}C_R^a(x),\text{ }\bar C^a(x)=\sqrt{\tilde Z_3}\bar C%
_R^a(x)
\end{array}
\eqnum{5.18}
\end{equation}
(hereafter the subscript $R$ marks renormalized quantities). According to
above relations and the definitions given in Eqs. (5.9), (5.10) and (5.12),
we find 
\begin{equation}
\begin{array}{c}
\Lambda _{\mu \nu \lambda }^{abc}(p,q,k)=Z_3^{-3/2}\Lambda _{R\mu \nu
\lambda }^{~~abc}(p,q,k), \\ 
\Lambda _{~~\ \lambda }^{abc}(p,q,k)=\tilde Z_3^{-1}Z_3^{-1/2}\Lambda
_{R~~\lambda }^{~~abc}(p,q,k).
\end{array}
\eqnum{5.19}
\end{equation}
Applying these relations, the renormalized version of the identity written
in Eq. (5.11) will be 
\begin{equation}
\begin{array}{c}
p^\mu q^\nu k^\lambda \Lambda _{R\mu \nu \lambda }^{~~abc}(p,q,k)=-\frac{%
\sigma _R^2}{\alpha _R}{\chi }_R(p^2)[{\chi _R}(k^2)q^\nu \Lambda _{R~~\nu
}^{~~cab}(k,p,q) \\ 
+\chi _R(q^2)k^\lambda \Lambda _{R~~\lambda }^{~~bac}(q,p,k)]
\end{array}
\eqnum{5.20}
\end{equation}
where $\alpha _R$ and $\sigma _R$ were defined in Eq. (4.30) and 
\begin{equation}
\chi _R(k^2)=\frac 1{1+\Omega _R(k^2)]}  \eqnum{5.21}
\end{equation}
is the renormalized expression of the function $\chi (k^2)$. In the above,
we have considered 
\begin{equation}
\chi (k^2)=\widetilde{Z}_3\chi _R(k^2)  \eqnum{5.22}
\end{equation}
which follows from $\hat \Omega (k^2)=\hat \Omega (\mu ^2)+\hat \Omega
^c(k^2),$ $\tilde Z_3^{-1}=1+\hat \Omega (\mu ^2)$ and$\;\Omega _R(k^2)=%
\tilde Z_3\hat \Omega ^c(k^2)$ defined in the preceding section. At the
renormalization point chosen to be $p^2=q^2=k^2=\mu ^2$, we see, $\chi
_R(\mu ^2)=1$. In this case, the renormalized ghost vertex takes the form of
the bare vertex so that the RHS of Eq. (5.20) vanishes, therefore, we have 
\begin{equation}
p^\mu q^\nu k^\lambda \Lambda _{R\mu \nu \lambda
}^{~~abc}(p,q,k)|_{p^2=q^2=k^2=\mu ^2}=0.  \eqnum{5.23}
\end{equation}

Ordinarily, one is interested in discussing the renormalization of such
three-line vertices that they are defined from the vertices defined in Eqs.
(5.9) and (5.10) by extracting a coupling constant $g$. These vertices are
denoted by $\tilde \Lambda _{\mu \nu \lambda }^{abc}(p,q,k)$ and $\tilde 
\Lambda _{~~\lambda }^{abc}(p,q,k)$. Commonly, they are renormalized in such
a fashion [9, 17] 
\begin{equation}
\begin{array}{c}
\tilde \Lambda _{\mu \nu \lambda }^{abc}(p,q,k)=Z_1^{-1}\widetilde{\Lambda }%
_{R\mu \nu \lambda }^{abc}(p,q,k), \\ 
\tilde \Lambda _{~~\lambda }^{abc}(p,q,k)=\widetilde{Z}_1^{-1}\widetilde{%
\Lambda }_{R~~\lambda }^{abc}(p,q,k)
\end{array}
\eqnum{5.24}
\end{equation}
where $Z_1$ and $\tilde Z_1$ are referred to as the renormalization
constants of the $\rho $-meson three-line vertex and the $\rho -$meson-ghost
particle vertex, respectively. It is clear that the W-T identity shown in
Eq. (5.11) also holds for the vertices $\tilde \Lambda _{\mu \nu \lambda
}^{abc}(p,q,k)$ and $\tilde \Lambda _{~~\lambda }^{abc}(p,q,k)$. So, when
the vertices $\Lambda _{\mu \nu \lambda }^{abc}(p,q,k)$ and $\Lambda
_{~~\lambda }^{abc}(p,q,k)$ in Eqs. (5.11) are replaced by $\tilde \Lambda
_{\mu \nu \lambda }^{abc}(p,q,k)$ and $\tilde \Lambda _{~~\lambda
}^{abc}(p,q,k)$ respectively and then Eq. (5.24) is inserted to such an
identity, we obtain a renormalized version of the identity as follows 
\begin{equation}
\begin{array}{c}
p^\mu q^\nu k^\lambda \tilde \Lambda _{R\mu \nu \lambda }^{~~abc}(p,q,k)=-%
\frac{Z_1\tilde Z_3}{Z_3\tilde Z_1}\frac{\sigma _R^2}{\alpha _R}\chi
_R(p^2)[\chi _R(k^2) \\ 
\times q^\nu \tilde \Lambda _{R~~\nu }^{~~cab}(k,p,q)+\chi _R(q^2)k^\lambda 
\tilde \Lambda _{R~~\lambda }^{~~bac}(q,p,k)].
\end{array}
\eqnum{5.25}
\end{equation}
When multiplying the both sides of Eq. (5.25) with a renormalized coupling
constant $g_R$ and absorbing it into the vertices, noticing 
\begin{equation}
\begin{array}{c}
\Lambda _{R\mu \nu \lambda }^{abc}(p,q,k)=g_R\widetilde{\Lambda }_{R\mu \nu
\lambda }^{abc}(p,q,k), \\ 
\Lambda _{R~~\lambda }^{abc}(p,q,k)=g_R\widetilde{\Lambda }_{R~~\lambda
}^{abc}(p,q,k),
\end{array}
\eqnum{5.26}
\end{equation}
we have 
\begin{equation}
\begin{array}{c}
p^\mu q^\nu k^\lambda \Lambda _{R\mu \nu \lambda }^{~~abc}(p,q,k)=-\frac{Z_1%
\tilde Z_3}{Z_3\tilde Z_1}\frac{\sigma _R^2}{\alpha _R}\chi _R(p^2)[\chi
_R(k^2) \\ 
\times q^\nu \Lambda _{R~~\nu }^{~~cab}(k,p,q)+\chi _R(q^2)k^\lambda \Lambda
_{R~~\lambda }^{~~bac}(q,p,k)].
\end{array}
\eqnum{5.27}
\end{equation}
In comparison of Eq. (5.27) with Eq. (5.20), we see, except for the factor $%
Z_1\tilde Z_3Z_3^{-1}\tilde Z_1^{-1}$, the both identities are identical to
each other. From this observation, we deduce 
\begin{equation}
\frac{Z_1}{Z_3}=\frac{\tilde Z_1}{\tilde Z_3}.  \eqnum{5.28}
\end{equation}
This is the S-T identity similar to that given in QCD [9, 16, 17, 19].

\section{$\rho $-meson four-line proper vertex}

By the similar procedure as deriving Eqs. (5.1) and (5.3), the W-T identity
obeyed by the $\rho $-meson four-point Green function may be derived by
differentiating Eq. (4.1) with respect to the sources $J_\mu ^b(y),J_\lambda
^c(z)$ and $J_\tau ^d(u)$. The result represented in the operator form is as
follows 
\begin{equation}
\begin{array}{c}
\frac 1\alpha \partial _x^\mu G_{\mu \nu \lambda \tau }^{abcd}(x,y,z,u) \\ 
=<0^{+}|T^{*}[\hat {\bar C^a}(x)\hat D_\nu ^{be}(y)\hat C^e(y)\hat A_\lambda
^c(z)\hat A_\tau ^d(u)]|0^{-}> \\ 
+<0^{+}|T^{*}[\hat {\bar C^a}(x)\hat A_\nu ^b(y)\hat D_\lambda ^{ce}(z)\hat C%
^e(z)\hat A_\tau ^d(u)]|0^{-}> \\ 
+<0^{+}|T^{*}[\hat {\bar C^a}(x)\hat A_\nu ^b(y)\hat A_\lambda ^c(z)\hat D%
_\tau ^{de}(u)\hat C^e(u)]|0^{-}>
\end{array}
\eqnum{6.1}
\end{equation}
where 
\begin{equation}
G_{\mu \nu \lambda \tau }^{abcd}(x,y,z,u)=<0^{+}|T[\hat A_\mu ^a(x)\hat A%
_\nu ^b(y)\hat A_\lambda ^c(z)\hat A_\tau ^d(u)]|0^{-}>  \eqnum{6.2}
\end{equation}
is the $\rho $-meson four-point Green function. The accompanying ghost
equation may be obtained by differentiating Eq. (4.2) with respect to the
sources $J_\lambda ^c(z)$ and $J_\tau ^d(u)$. The result is 
\begin{equation}
\begin{array}{c}
\partial _x^\mu <0^{+}|T^{*}[\hat D_\mu ^{ae}(x)\hat C^e(x)\hat {\bar C^b}(y)%
\hat A_\lambda ^c(z)\hat A_\tau ^d(u)]|0^{-}> \\ 
+{\sigma }^2G_{~~\lambda \tau }^{abcd}(x,y,z,u)=-\delta ^{ab}\delta
^4(x-y)D_{\lambda \tau }^{cd}(z-u)
\end{array}
\eqnum{6.3}
\end{equation}
where 
\begin{equation}
G_{~~\lambda \tau }^{abcd}(x,y,z,u)=<0^{+}|T[\hat C^a(x)\hat {\bar C^b}(y)%
\hat A_\lambda ^c(z)\hat A_\tau ^d(u)]|0^{-}>  \eqnum{6.4}
\end{equation}
is the four-point $\rho $-meson-ghost particle Green function.
Differentiation of Eq. (6.1) with respect to the coordinates $y$, $z$ and $u$
and use of Eq. (6.3) lead to 
\begin{equation}
\begin{array}{c}
\frac 1\alpha \partial _x^\mu \partial _y^\nu \partial _z^\lambda \partial
_u^\tau G_{\mu \nu \lambda \tau }^{abcd}(x,y,z,u)=\delta ^{ab}\delta
^4(x-y)\partial _z^\lambda \partial _u^\tau D_{\lambda \tau }^{cd}(z-u) \\ 
+\delta ^{ac}\delta ^4(x-z)\partial _y^\nu \partial _u^\tau D_{\nu \tau
}^{bd}(y-u)+\delta ^{ad}\delta ^4(x-u)\partial _y^\nu \partial _z^\lambda
D_{\nu \lambda }^{bc}(y-z) \\ 
+\sigma ^2\{\partial _z^\lambda \partial _u^\tau G_{~~\lambda \tau
}^{bacd}(y,x,z,u)+\partial _y^\nu \partial _u^\tau G_{~~\nu \tau
}^{cabd}(z,x,y,u) \\ 
+\partial _y^\nu \partial _z^\lambda G_{~~\nu \lambda }^{dabc}(u,x,y,z)\}.
\end{array}
\eqnum{6.5}
\end{equation}

It is noted that the four-point Green functions appearing in the above
equations are unconnected. Their decompositions to connected Green functions
are not difficult to be found by making use of the relation between the
generating functionals $Z$ for the full Green functions and $W$ for the
connected Green functions as written in Eq. (3.9). The result is 
\begin{equation}
\begin{array}{c}
G_{\mu \nu \lambda \tau }^{abcd}(x,y,z,u)=G_{\mu \nu \lambda \tau
}^{abcd}(x,y,z,u)_c-D_{\mu \nu }^{ab}(x-y)D_{\lambda \tau }^{cd}(z-u) \\ 
-D_{\mu \lambda }^{ac}(x-z)D_{\nu \tau }^{bd}(y-u)-D_{\mu \tau
}^{ad}(x-u)D_{\nu \lambda }^{bc}(y-z)
\end{array}
\eqnum{6.6}
\end{equation}
and 
\begin{equation}
G_{~~\lambda \tau }^{abcd}(x,y,z,u)=G_{~~\,\lambda \tau
}^{abcd}(x,y,z,u)_c-\Delta ^{ab}(x-y)D_{\lambda \tau }^{cd}(z-u). 
\eqnum{6.7}
\end{equation}
The first terms marked by the subscript ''$c$'' in Eqs. ( 6.6) and (6.7) are
connected Green functions. When inserting Eqs. (6.6) and (6.7) into Eq.
(6.5) and using the W-T identity in Eq. (4.5), one may find 
\begin{equation}
\begin{array}{c}
\partial _x^\mu \partial _y^\nu \partial _z^\lambda \partial _u^\tau G_{\mu
\nu \lambda \tau }^{abcd}(x,y,z,u)_c=\alpha \sigma ^2\{\partial _y^\nu
\partial _z^\lambda G_{~~\nu \lambda }^{dabc}(u,x,y,z)_c \\ 
+\partial _y^\nu \partial _u^\tau G_{~~\nu \tau }^{cabd}(z,x,y,u)_c+\partial
_z^\lambda \partial _u^\tau G_{~~\lambda \tau }^{bacd}(y,x,z,u)_c\}.
\end{array}
\eqnum{6.8}
\end{equation}
This is the W-T identity satisfied by the connected four-point Green
functions. In the Landau gauge, we have 
\begin{equation}
\partial _x^\mu \partial _y^\nu \partial _z^\lambda \partial _u^\tau G_{\mu
\nu \lambda \tau }^{abcd}(x,y,z,u)_c=0  \eqnum{6.9}
\end{equation}
which shows the transversity of the Green function.

The W-T identity for the four-line proper $\rho $-meson vertex may be
derived from Eq. (6.8) with the help of the following
one-particle-irreducible decompositions of the connected Green functions
which can easily be found by the standard procedure [9, 17], 
\begin{equation}
\begin{array}{c}
~G_{\mu \nu \lambda \tau }^{abcd}(x_1,x_2,x_3,x_4)_c \\ 
=\int \prod\limits_{i=1}^4d^4y_iD_{\mu \mu ^{\prime }}^{aa^{\prime
}}(x_1-y_1)D_{\nu \nu ^{\prime }}^{bb^{\prime }}(x_2-y_2)\Gamma _{a^{\prime
}b^{\prime }c^{\prime }d^{\prime }}^{\mu ^{\prime }\nu ^{\prime }\lambda
^{\prime }\tau ^{\prime }}(y_1,y_2,y_3,y_4) \\ 
\times D_{\lambda ^{\prime }\lambda }^{c^{\prime }c}(y_3-x_3)D_{\tau
^{\prime }\tau }^{d^{\prime }d}(y_4-x_4) \\ 
+i\int \prod\limits_{i=1}^4d^4y_id^4z_i\{D_{\mu \mu ^{\prime }}^{aa^{\prime
}}(x_1-y_1)D_{\nu \nu ^{\prime }}^{bb^{\prime }}(x_2-y_2)\Gamma _{a^{\prime
}b^{\prime }e}^{\mu ^{\prime }\nu ^{\prime }\rho }(y_1,y_2,y_3) \\ 
\times D_{\rho \rho ^{\prime }}^{ee^{\prime }}(y_3-z_1)\Gamma _{e^{\prime
}c^{\prime }d^{\prime }}^{\rho ^{\prime }\lambda ^{\prime }\tau ^{\prime
}}(z_1,z_2,z_3)D_{\lambda ^{\prime }\lambda }^{c^{\prime }c}(z_2-x_3)D_{\tau
^{\prime }\tau }^{d^{\prime }d}(z_3-x_4) \\ 
+D_{\mu \mu ^{\prime }}^{aa^{\prime }}(x_1-y_1)D_{\lambda \lambda ^{\prime
}}^{cc^{\prime }}(x_3-y_2)\Gamma _{a^{\prime }c^{\prime }e}^{\mu ^{\prime
}\lambda ^{\prime }\rho }(y_1,y_2,y_3)D_{\rho \rho ^{\prime }}^{ee^{\prime
}}(y_3-z_1) \\ 
\times \Gamma _{e^{\prime }b^{\prime }d^{\prime }}^{\rho ^{\prime }\nu
^{\prime }\tau ^{\prime }}(z_1,z_2,z_3)D_{\nu ^{\prime }\nu }^{b^{\prime
}b}(z_2-x_2)D_{\tau ^{\prime }\tau }^{d^{\prime }d}(z_3-x_4) \\ 
+D_{\nu \nu ^{\prime }}^{bb^{\prime }}(x_2-y_1)D_{\lambda \lambda ^{\prime
}}^{cc^{\prime }}(x_3-y_2)\Gamma _{b^{\prime }c^{\prime }e}^{\nu ^{\prime
}\lambda ^{\prime }\rho }(y_1,y_2,y_3)D_{\rho \rho ^{\prime }}^{ee^{\prime
}}(y_3-z_1) \\ 
\times \Gamma _{e^{\prime }a^{\prime }d^{\prime }}^{\rho ^{\prime }\mu
^{\prime }\tau ^{\prime }}(z_1,z_2,z_3)D_{\mu ^{\prime }\mu }^{a^{\prime
}a}(z_2-x_1)D_{\tau ^{\prime }\tau }^{d^{\prime }d}(z_3-x_4)\}
\end{array}
\eqnum{6.10}
\end{equation}
and 
\begin{equation}
\begin{array}{c}
G_{~~\lambda \tau }^{abcd}(x_1,x_2,x_3,x_4)_c \\ 
=\int \prod\limits_{i=1}^4d^4y_i\Delta ^{aa^{\prime }}(x_1-y_1)\Gamma
_{a^{\prime }b^{\prime }c^{\prime }d^{\prime }}^{~~~~\lambda ^{\prime }\tau
^{\prime }}(y_1,y_2,y_3,y_4)\Delta ^{b^{\prime }b}(y_2-x_2) \\ 
\times D_{\lambda ^{\prime }\lambda }^{c^{\prime }c}(y_3-x_3)D_{\tau
^{\prime }\tau }^{d^{\prime }d}(y_4-x_4) \\ 
+i\int \prod\limits_{i=1}^4d^4y_id^4z_i\{\Delta ^{aa^{\prime
}}(x_1-y_1)\Gamma _{a^{\prime }ed^{\prime }}^{~~~\tau ^{\prime
}}(y_1,y_2,y_3)\Delta ^{ee^{\prime }}(y_2-z_1) \\ 
\times D_{\tau ^{\prime }\tau }^{d^{\prime }d}(y_3-x_4)\Gamma _{e^{\prime
}b^{\prime }c^{\prime }}^{~~~~\lambda ^{\prime }}(z_1,z_2,z_3)\Delta
^{b^{\prime }b}(z_2-x_2)D_{\lambda ^{\prime }\lambda }^{c^{\prime
}c}(z_3-x_3) \\ 
+\Delta ^{aa^{\prime }}(x_1-y_1)\Gamma _{a^{\prime }ec^{\prime
}}^{~~~\lambda ^{\prime }}(y_1,y_2,y_3)\Delta ^{ee^{\prime
}}(y_2-z_1)D_{\lambda ^{\prime }\lambda }^{c^{\prime }c}(y_3-x_3) \\ 
\times \Gamma _{e^{\prime }b^{\prime }d^{\prime }}^{~~~\tau ^{\prime
}}(z_1,z_2,z_3)\Delta ^{b^{\prime }b}(z_2-x_2)D_{\tau ^{\prime }\tau
}^{d^{\prime }d}(z_3-x_4) \\ 
+\Delta ^{aa^{\prime }}(x_1-y_1)\Gamma _{a^{\prime }b^{\prime }e}^{~~~\rho
}(y_1,y_2,y_3)\Delta ^{b^{\prime }b}(y_2-x_2)D_{\rho \rho ^{\prime
}}^{ee^{\prime }}(y_3-z_1) \\ 
\times \Gamma _{e^{\prime }c^{\prime }d^{\prime }}^{\rho ^{\prime }\lambda
^{\prime }\tau ^{\prime }}(z_1,z_2,z_3)D_{\lambda ^{\prime }\lambda
}^{c^{\prime }c}(z_2-x_3)D_{\tau ^{\prime }\tau }^{d^{\prime }d}(z_3-x_4)\}
\end{array}
\eqnum{6.11}
\end{equation}
where $\Gamma _{\mu \nu \lambda \tau }^{abcd}(x_1,x_2,x_3,x_4)$ is the
four-line $\rho $-meson proper vertex and $\Gamma _{~~\lambda \tau
}^{abcd}(x_1,x_2,x_3,x_4)$ is the four-line $\rho -$meson- ghost particle
proper vertex. They are defined as [9, 16, 17] 
\begin{equation}
\begin{array}{c}
\Gamma _{\mu \nu \lambda \tau }^{abcd}(x_1,x_2,x_3,x_4)=i\frac{\delta
^4\Gamma }{\delta A^{a\mu }(x_1)\delta A^{b\nu }(x_2)\delta A^{c\lambda
}(x_3)\delta A^{d\tau }(x_4)}|_{J=0}, \\ 
\Gamma _{~~\lambda \tau }^{abcd}(x_1,x_2,x_3,x_4)=\frac{\delta ^4\Gamma }{%
i\delta \bar C^a(x_1)\delta C^b(x_2)\delta A^{c\lambda }(x_3)\delta A^{d\tau
}(x_4)}|_{J=0}.
\end{array}
\eqnum{6.12}
\end{equation}
When substituting Eqs. (6.10) and (6.11) into Eq. (6.8) and transforming Eq.
(6.8) into the momentum space, one can find the following identity satisfied
by the four-line $\rho -$meson proper vertex 
\begin{equation}
\begin{array}{c}
k_1^\mu k_2^\nu k_3^\lambda k_4^\tau \Lambda _{\mu \nu \lambda \tau
}^{abcd}(k_1,k_2,k_3,k_4)=\Psi \left( 
\begin{array}{cccc}
a & b & c & d \\ 
k_1 & k_2 & k_3 & k_4
\end{array}
\right) \\ 
+\Psi \left( 
\begin{array}{cccc}
a & c & d & b \\ 
k_1 & k_3 & k_4 & k_2
\end{array}
\right) +\Psi \left( 
\begin{array}{cccc}
a & d & b & c \\ 
k_1 & k_4 & k_2 & k_3
\end{array}
\right)
\end{array}
\eqnum{6.13}
\end{equation}
where 
\begin{equation}
\begin{array}{c}
\Psi \left( 
\begin{array}{cccc}
a & b & c & d \\ 
k_1 & k_2 & k_3 & k_4
\end{array}
\right) \\ 
=-ik_1^\mu k_2^\nu \Lambda _{\mu \nu \sigma
}^{abe}(k_1,k_2,-(k_1+k_2))D_{ef}^{\sigma \rho }(k_1+k_2)k_3^\lambda
k_4^\tau \Lambda _{\rho \lambda \tau }^{fcd}(-(k_3+k_4),k_3,k_4) \\ 
+\frac{i\sigma ^2}\alpha \chi (k_1^2)\chi (k_2^2)[ik_3^\lambda k_4^\tau
\Lambda _{~~\lambda \tau }^{bacd}(k_2,k_1,k_3,k_4) \\ 
-\Lambda _{~~\sigma }^{bae}(k_2,k_1,-(k_1+k_2))D_{ef}^{\sigma \rho
}(k_1+k_2)k_3^\lambda k_4^\tau \Lambda _{\rho \lambda \tau
}^{fcd}(-(k_3+k_4),k_3,k_4) \\ 
-k_4^\tau \Lambda _{~~\tau }^{bed}(k_2,-(k_2+k_4),k_4)\Delta
^{ef}(k_2+k_4)k_3^\lambda \Lambda _{~~\lambda }^{fac}(-(k_1+k_3),k_1,k_3) \\ 
-k_3^\lambda \Lambda _{~~\lambda }^{bec}(k_2,-(k_2+k_3),k_3)\Delta
^{ef}(k_2+k_3)k_4^\tau \Lambda _{~~~\tau }^{fad}(-(k_1+k_4),k_1,k_4)].
\end{array}
\eqnum{6.14}
\end{equation}
The second and third terms in Eq .(6.13) can be written out from Eq. (6.14)
through cyclic permutations. In the above, we have defined 
\begin{equation}
\begin{array}{c}
\Gamma _{\mu \nu \lambda \tau }^{abcd}(k_1,k_2,k_3,k_4)=(2\pi )^4\delta
^4(\sum\limits_{i=1}^4k_i)\Lambda _{\mu \nu \lambda \tau
}^{abcd}(k_1,k_2,k_3,k_4), \\ 
\Gamma _{~~\lambda \tau }^{abcd}(k_1,k_2,k_3,k_4)=(2\pi )^4\delta
^4(\sum\limits_{i=1}^4k_i)\Lambda _{~~\lambda \tau }^{abcd}(k_1,k_2,k_3,k_4).
\end{array}
\eqnum{6.15}
\end{equation}
In the lowest order approximation, we have checked that except for the first
term in Eq. (6.14) which was encountered in the massless theory, the
remaining mass-dependent terms are cancelled out with the corresponding
terms contained in the second and third terms in Eq. (6.13). Therefore, the
identity in Eq. (6.13) leads to a result in the lowest order approximation
which is consistent with the Feynman rule.

The renormalization of the four-line vertices is similar to that for the
three-line ones. From the definitions given in Eqs. (6.12), (6.15) and
(5.18), it is clearly seen that the four-line vertices should be
renormalized in such a manner 
\begin{equation}
\begin{array}{c}
\Lambda _{\mu \nu \lambda \tau }^{abcd}(k_1,k_2,k_3,k_4)=Z_3^{-2}\Lambda
_{R\mu \nu \lambda \tau }^{~~abcd}(k_1,k_2,k_3,k_4), \\ 
\Lambda _{~~\;\lambda \tau }^{abcd}(k_1,k_2,k_3,k_4)=\tilde Z%
_3^{-1}Z_3^{-1}\Lambda _{R\text{ }~~\lambda \tau }^{~~abcd}(k_1,k_2,k_3,k_4).
\end{array}
\eqnum{6.16}
\end{equation}
On inserting these relations into Eqs. (6.13) and (6.14), one can obtain a
renormalized identity similar to Eq. (5.20), that is 
\begin{equation}
\begin{array}{c}
k_1^\mu k_2^\nu k_3^\lambda k_4^\tau \Lambda _{R\mu \nu \lambda \tau }^{%
\text{ }abcd}(k_1,k_2,k_3,k_4)=\Psi _R\left( 
\begin{array}{cccc}
a & b & c & d \\ 
k_1 & k_2 & k_3 & k_4
\end{array}
\right) \\ 
+\Psi _R\left( 
\begin{array}{cccc}
a & c & d & b \\ 
k_1 & k_3 & k_4 & k_2
\end{array}
\right) +\Psi _R\left( 
\begin{array}{cccc}
a & d & b & c \\ 
k_1 & k_4 & k_2 & k_3
\end{array}
\right)
\end{array}
\eqnum{6.17}
\end{equation}
where 
\begin{equation}
\begin{array}{c}
\Psi _R\left( 
\begin{array}{cccc}
a & b & c & d \\ 
k_1 & k_2 & k_3 & k_4
\end{array}
\right) \\ 
=-ik_1^\mu k_2^\nu \Lambda _{R\mu \nu \sigma }^{\text{ }%
abe}(k_1,k_2,-(k_1+k_2))D_{Ref}^{\text{ }\sigma \rho }(k_1+k_2)k_3^\lambda
k_4^\tau \Lambda _{R\rho \lambda \tau }^{fcd}(-(k_3+k_4),k_3,k_4) \\ 
+\frac{i\sigma _R^2}{\alpha _R}\chi _R(k_1^2)\chi _R(k_2^2)[ik_3^\lambda
k_4^\tau \Lambda _{R~~\lambda \tau }^{bacd}(k_2,k_1,k_3,k_4) \\ 
-\Lambda _{R~\sigma }^{bae}(k_2,k_1,-(k_1+k_2))D_{Ref}^{\text{ }\sigma \rho
}(k_1+k_2)k_3^\lambda k_4^\tau \Lambda _{R\rho \lambda \tau }^{\text{ }%
fcd}(-(k_3+k_4),k_3,k_4) \\ 
-k_4^\tau \Lambda _{R~~\tau }^{bed}(k_2,-(k_2+k_4),k_4)\Delta
_R^{ef}(k_2+k_4)k_3^\lambda \Lambda _{R~~\lambda }^{\text{ }%
fac}(-(k_1+k_3),k_1,k_3) \\ 
-k_3^\lambda \Lambda _{R~\lambda }^{bec}(k_2,-(k_2+k_3),k_3)\Delta
_R^{ef}(k_2+k_3)k_4^\tau \Lambda _{R~\tau }^{fad}(-(k_1+k_4),k_1,k_4)].
\end{array}
\eqnum{6.18}
\end{equation}
We can also define vertices $\tilde \Lambda _{\mu \nu \lambda \tau
}^{abcd}(k_1,k_2,k_3,k_4)$ and $\tilde \Lambda _{~~\text{ }\lambda \tau
}^{abcd}(k_1,k_2,k_3,k_4)$ from the vertices $\Lambda _{\mu \nu \lambda \tau
}^{abcd}(k_1,k_2,k_3,k_4)$ and $\Lambda _{~~\lambda \tau
}^{abcd}(k_1,k_2,k_3,k_4)$ by taking out the coupling constant squared,
respectively. The renormalization of these vertices are usually defined by
[9, 17] 
\begin{equation}
\begin{array}{c}
\tilde \Lambda _{\mu \nu \lambda \tau }^{abcd}(k_1,k_2,k_3,k_4)=Z_4^{-1}%
\tilde \Lambda _{R\mu \nu \lambda \tau }^{~~abcd}(k_1,k_2,k_3,k_4), \\ 
\tilde \Lambda _{~~\lambda \tau }^{abcd}(k_1,k_2,k_3,k_4)=\tilde Z_4^{-1}%
\tilde \Lambda _{R~~\lambda \tau }^{~~abcd}(k_1,k_2,k_3,k_4).
\end{array}
\eqnum{6.19}
\end{equation}
where $Z_4$ and $\tilde Z_4$ are the renormalization constants of the
four-line $\rho $-meson vertex and the four-line $\rho -$meson-ghost
particle one respectively. Obviously, the identity in Eqs. (6.13) and (6.14)
remains formally unchanged if we replace all the vertices $\Lambda _i$ in
the identity with the ones $\widetilde{\Lambda }_i$. Substituting Eqs.
(5.24), (6.19), (4.26) and (4.27) into such an identity, one may write a
renormalized identity similar to Eq. (5.25), that is 
\begin{equation}
\begin{array}{c}
k_1^\mu k_2^\nu k_3^\lambda k_4^\tau \widetilde{\Lambda }_{R\mu \nu \lambda
\tau }^{\text{ }abcd}(k_1,k_2,k_3,k_4)=\widetilde{\Psi }_R\left( 
\begin{array}{cccc}
a & b & c & d \\ 
k_1 & k_2 & k_3 & k_4
\end{array}
\right) \\ 
+\widetilde{\Psi }_R\left( 
\begin{array}{cccc}
a & c & d & b \\ 
k_1 & k_3 & k_4 & k_2
\end{array}
\right) +\widetilde{\Psi }_R\left( 
\begin{array}{cccc}
a & d & b & c \\ 
k_1 & k_4 & k_2 & k_3
\end{array}
\right)
\end{array}
\eqnum{6.20}
\end{equation}
where 
\begin{equation}
\begin{array}{c}
\widetilde{\Psi }_R\left( 
\begin{array}{cccc}
a & b & c & d \\ 
k_1 & k_2 & k_3 & k_4
\end{array}
\right) \\ 
=\frac{Z_4Z_3}{Z_1^2}\{-ik_1^\mu k_2^\nu \widetilde{\Lambda }_{R\mu \nu
\sigma }^{\text{ }abe}(k_1,k_2,-(k_1+k_2))D_{Ref}^{\text{ }\sigma \rho
}(k_1+k_2)k_3^\lambda k_4^\tau \widetilde{\Lambda }_{R\rho \lambda \tau }^{%
\text{ }fcd}(-(k_3+k_4),k_3,k_4)\} \\ 
+\frac{i\sigma _R^2}{\alpha _R}\chi _R(k_1^2)\chi _R(k_2^2)\{\frac{%
\widetilde{Z}_3Z_4}{Z_3\widetilde{Z_4}}ik_3^\lambda k_4^\tau \widetilde{%
\Lambda }_{R~~\lambda \tau }^{\text{ }bacd}(k_2,k_1,k_3,k_4) \\ 
-\frac{Z_4\widetilde{Z}_3}{Z_1\widetilde{Z_1}}\widetilde{\Lambda }%
_{R~~\sigma }^{\text{ }bae}(k_2,k_1,-(k_1+k_2))D_{Ref}^{\text{ }\sigma \rho
}(k_1+k_2)k_3^\lambda k_4^\tau \widetilde{\Lambda }_{R\rho \lambda \tau }^{%
\text{ }fcd}(-(k_3+k_4),k_3,k_4) \\ 
-\frac{Z_4\widetilde{Z}_3^2}{Z_3\widetilde{Z}_1^2}[k_4^\tau \widetilde{%
\Lambda }_{R~\tau }^{bed}(k_2,-(k_2+k_4),k_4)\Delta
_R^{ef}(k_2+k_4)k_3^\lambda \widetilde{\Lambda }_{R~~\lambda }^{\text{ }%
fac}(-(k_1+k_3),k_1,k_3) \\ 
+k_3^\lambda \widetilde{\Lambda }_{R~\lambda
}^{bec}(k_2,-(k_2+k_3),k_3)\Delta _R^{ef}(k_2+k_3)k_4^\tau \widetilde{%
\Lambda }_{R~\tau }^{fad}(-(k_1+k_4),k_1,k_4)]\}.
\end{array}
\eqnum{6.21}
\end{equation}
Multiplying the both sides of Eqs. (6.20) and (6.21) by $g_R^2$, according
to the relations given in Eqs. (5.26) and in the following 
\begin{equation}
\begin{array}{c}
\Lambda _{R\mu \nu \lambda \tau }^{abcd}(k_{1,}k_2,k_3,k_4)=g_R^2\tilde 
\Lambda _{R\mu \nu \lambda \tau }^{~~abcd}(k_1,k_2,k_3,k_4), \\ 
\Lambda _{R~~\lambda \tau }^{abcd}(k_1,k_2,k_3,k_4)=g_R^2\widetilde{\Lambda }%
_{R~~\lambda \tau }^{~~abcd}(k_1,k_2,k_3,k_4),
\end{array}
\eqnum{6.22}
\end{equation}
we have an identity which is of the same form as the identity in Eqs. (6.20)
and (6.21) except that the vertices $\widetilde{\Lambda }_R^i$ in Eqs.
(6.20) and (6.21) are all replaced by the vertices $\Lambda _R^i$. Comparing
this identity with that written in Eqs. (6.17) and (6.18), one may find 
\begin{equation}
\frac{Z_3Z_4}{Z_1^2}=1,\frac{\widetilde{Z}_3Z_4}{Z_3\widetilde{Z}_4}=1,\frac{%
Z_4\widetilde{Z}_3}{Z_1\widetilde{Z}_1}=1,\frac{Z_4\widetilde{Z}_3^2}{Z_3%
\widetilde{Z}_1^2}=1  \eqnum{6.23}
\end{equation}
which lead to 
\begin{equation}
\frac{Z_1}{Z_3}=\frac{\widetilde{Z}_1}{\widetilde{Z}_3}=\frac{Z_4}{Z_1},%
\frac{Z_1}{\widetilde{Z}_1}=\frac{Z_3}{\widetilde{Z}_3}=\frac{Z_4}{%
\widetilde{Z}_4}  \eqnum{6.24}
\end{equation}
This just is the S-T identity analogous to that for QCD [9, 16, 17, 19].

\section{Nucleon-$\rho $-meson vertex and nucleon propagator}

This section is used to derive the W-T identity for nucleon-$\rho $-meson
vertex and discuss its renormalization. First we derive a W-T identity
satisfied by the nucleon-$\rho $-meson three-point Green function. This
identity can easily be derived by differentiating the W-T identity in Eq.
(3.8) or (3.10) with respect to the sources $\xi ^b(z)$, $\eta (y)$ and $%
\overline{\eta }(x)$ and then setting all the sources to be zero. The result
written in the operator form is as follows 
\begin{equation}
\partial _z^\mu G_\mu ^a(x,y,z)=i\alpha
g[G_1^{ba}(x,y,z)T^b-T^bG_2^{ba}(x,y,z)]  \eqnum{7.1}
\end{equation}
where 
\begin{equation}
G_\mu ^a(x,y,z)=\left\langle 0^{+}\mid \widehat{\psi }(x)\widehat{\overline{%
\psi }}(y)\widehat{A}(z)\mid 0^{-}\right\rangle  \eqnum{7.2}
\end{equation}
is the nucleon-$\rho $-meson three-point Green function, 
\begin{equation}
G_1^{ba}(x,y,z)=\left\langle 0^{+}\mid \widehat{\psi }(x)\widehat{\overline{%
\psi }}(y)\widehat{C}^b(y)\widehat{\overline{C}}^a(z)\mid 0^{-}\right\rangle
\eqnum{7.3}
\end{equation}
and 
\begin{equation}
G_2^{ba}(x,y,z)=\left\langle 0^{+}\mid \widehat{\psi }(x)\widehat{\overline{%
\psi }}(y)\widehat{C}^b(x)\widehat{\overline{C}}^a(z)\mid 0^{-}\right\rangle
\eqnum{7.4}
\end{equation}
are the nucleon-ghost particle mixed Green functions. The Green functions in
Eqs. (7.3) and (7.4) are connected because a nucleon field and a ghost field
are of a common coordinate.

The W-T identity for nucleon-$\rho $-meson vertex can be derived from Eq.
(7.1) with the help of one-particle irreducible decompositions of the Green
functions shown in Eqs. (7.2)-(7.4). The decompositions can easily be
obtained by the standard procedure [9, 17]. The results are given in the
following. 
\begin{equation}
G_\mu ^a(x,y,z)=\int d^4x^{\prime }d^4y^{\prime }d^4z^{\prime
}iS_F(x-x^{\prime })\Gamma ^{b\nu }(x^{\prime },y^{\prime },z^{\prime
})iS_F(y^{\prime }-y)iD_{\nu \mu }^{ba}(z^{\prime }-z)  \eqnum{7.5}
\end{equation}
where $D_{\nu \mu }^{ba}(z^{\prime }-z)$ is the $\rho $-meson propagator
defined in Eq. (4.6), 
\begin{equation}
iS_F(x-x^{\prime })=\left\langle 0^{+}\mid \widehat{\psi }(x)\widehat{%
\overline{\psi }}(x^{\prime })\mid 0^{-}\right\rangle  \eqnum{7.6}
\end{equation}
is the nucleon propagator and 
\begin{equation}
\Gamma ^{b\nu }(x^{\prime },y^{\prime },z^{\prime })=\frac{\delta ^3\Gamma }{%
i\delta \overline{\psi }(x^{\prime })\delta \psi (y^{\prime })\delta A_\nu
^b(z^{\prime })}\mid _{J=0}  \eqnum{7.7}
\end{equation}
is the nucleon-$\rho $-meson proper vertex. 
\begin{equation}
G_1^{ba}(x,y,z)=\int d^4x^{\prime }d^4z^{\prime }S_F(x-x^{\prime })\gamma
^{bc}(x^{\prime },y,z^{\prime })\Delta ^{ca}(z^{\prime }-z)  \eqnum{7.8}
\end{equation}
where $\Delta ^{ca}(z^{\prime }-z)$ is the ghost particle propagator defined
in Eq. (4.7) and 
\begin{equation}
\gamma _1^{bc}(x^{\prime },y,z^{\prime })=\int d^4ud^4v\Delta
^{bd}(y-u)\Gamma ^{cd}(x^{\prime },v,u,z^{\prime })S_F(v-y)  \eqnum{7.9}
\end{equation}
in which 
\begin{equation}
\Gamma ^{cd}(x^{\prime },v,u,z^{\prime })=i\frac{\delta ^4\Gamma }{\delta 
\overline{\psi }(x^{\prime })\delta \psi (v)\delta \overline{C}^c(u)\delta
C^d(z^{\prime })}\mid _{J=0}  \eqnum{7.10}
\end{equation}
is the nucleon-ghost particle vertex. Similarly, 
\begin{equation}
G_2^{ba}(x,y,z)=\int d^4y^{\prime }d^4z^{\prime }\gamma ^{bc}(x,y^{\prime
},z^{\prime })S_F(y^{\prime }-y)\Delta ^{ca}(z^{\prime }-z)  \eqnum{7.11}
\end{equation}
where 
\begin{equation}
\gamma _2^{bc}(x,y^{\prime },z^{\prime })=\int d^4ud^4vS_F(x-u)\Delta
^{bd}(x-v)\Gamma ^{dc}(u,y^{\prime },v,z^{\prime }).  \eqnum{7.12}
\end{equation}
On substituting Eqs. (7.5), (7.8) and (7.11) into Eq. (7.1) and then
transforming Eq. (7.1) into the momentum space, we have 
\begin{equation}
\begin{array}{c}
S_F(p)\Gamma ^{b\nu }(p,q,k)S_F(q)k^\mu D_{\mu \nu }^{ab}(k) \\ 
=-i\alpha g[S_F(p)\gamma _1^b(p,q,k)-\gamma _2^b(p,q,k)S_F(q)]\Delta
^{ab}(k)]
\end{array}
\eqnum{7.13}
\end{equation}
where we have defined 
\begin{equation}
\begin{array}{c}
\gamma _1^a(p,q,k)=\gamma _1^{ab}(p,q,k)T^b, \\ 
\gamma _2^a(p,q,k)=T^b\gamma _2^{ba}(p,q,k).
\end{array}
\eqnum{7.14}
\end{equation}
Considering that the vertex functions $\Gamma ^{b\nu }(p,q,k)$ and $\gamma
_i^a(p,q,k)$ ($i=1,2$) contain a common delta-function representing the
energy-momentum conservation, we may set

\begin{equation}
\begin{array}{c}
\Gamma ^{a\mu }(p,q,k)=(2\pi )^4\delta ^4(p-q+k)\Lambda ^{a\mu }(p,q,k), \\ 
\gamma _i^a(p,q,k)=(2\pi )^4\delta ^4(p-q+k)\widetilde{\gamma }_i^a(p,q,k)
\end{array}
\eqnum{7.15}
\end{equation}
where $\Lambda ^{a\mu }(p,q,k)$ and $\widetilde{\gamma }_i^a(p,q,k)$ are the
new vertex functions in which $k=q-p$. Noticing the above relations and the
expressions of $\rho $-meson and ghost particle propagators as given in Eqs.
(4.11) and (4.22), the W-T identity in Eq. (7.13) can be rewritten via the
functions $\Lambda ^{a\mu }(p,q,k)$ and $\widetilde{\gamma }_i^a(p,q,k)$ in
the form 
\begin{eqnarray}
k_\mu \Lambda ^{a\mu }(p,q,k)=ig\chi (k^2)[S_F^{-1}(p)\widetilde{\gamma }%
_2^a(p,q,k)-\widetilde{\gamma }_1^a(p,q,k)S_F^{-1}(q)]  \eqnum{7.16}
\end{eqnarray}
where $\chi (k^2)$ was defined in Eq. (5.13).

Let us turn to discuss the renormalized form of the above W-T identity. It
is well-known that the nucleon propagator can be expressed in the form 
\begin{equation}
S_F(p)=\frac 1{{\bf p-}M-\Sigma (p)+i\varepsilon }  \eqnum{7.17}
\end{equation}
where ${\bf p=\gamma }^\mu p_\mu $ and $\Sigma (p)$ denotes the nucleon
self-energy. The above expression can easily be derived from the Dyson
equation [24]. Usually, the nucleon propagator is renormalized in such a
fashion 
\begin{equation}
S_F(p)=Z_2S_F^R(p)  \eqnum{7.18}
\end{equation}
which implies 
\begin{equation}
\psi (x)=\sqrt{Z_2}\psi _R(x),\text{ }\overline{\psi }(x)=\sqrt{Z_2}%
\overline{\psi }_R(x).  \eqnum{7.19}
\end{equation}
From the relations in Eqs. (7.19) and (5.18), it is clearly seen that the
vertex defined in Eq. (7.7) is renormalized as 
\begin{equation}
\Gamma ^{a\mu }(x,y,z)=Z_2^{-1}Z_3^{-\frac 12}\Gamma _R^{a\mu }(x,y,z) 
\eqnum{7.20}
\end{equation}
which leads to 
\begin{equation}
\Lambda ^{a\mu }(p,q,k)=Z_2^{-1}Z_3^{-\frac 12}\Lambda _R^{a\mu }(p,q,k). 
\eqnum{7.21}
\end{equation}
While, for the functions $\widetilde{\gamma }_i^a(p,q,k)$, according to the
relations in Eqs. (5.18) and (7.19), they seem to be
renormalization-invariant. But, these functions actually are divergent in
the perturbative calculation. We assume that they are renormalized in such a
manner 
\begin{equation}
\widetilde{\gamma }_i^a(p,q,k)=Z_\gamma ^{-1}\widetilde{\gamma }%
_{iR}^a(p,q,k).  \eqnum{7.22}
\end{equation}
where $Z_\gamma $ is the renormalization constant of the functions $%
\widetilde{\gamma }_i^a(p,q,k)$. Based on the relations in Eqs. (5.22),
(7.18), (7.21), and (7.22), Eq. (7.16) can be represented in terms of the
renormalized quantities 
\begin{eqnarray}
k_\mu \Lambda _R^{a\mu }(p,q,k)=ig_R\chi _R(k^2)[S_F^{R-1}(p)\widetilde{%
\gamma }_{2R}^a(p,q,k)-\widetilde{\gamma }_{1R}^a(p,q,k)S_F^{R-1}(q)] 
\eqnum{7.23}
\end{eqnarray}
where $g_R$ is the renormalized coupling constant defined by 
\begin{equation}
g_R=\tilde Z_3Z_3^{\frac 12}\widetilde{Z}_\gamma ^{-1}g  \eqnum{7.24}
\end{equation}
It is well-known that 
\begin{equation}
g_R=\tilde Z_3Z_3^{\frac 12}\widetilde{Z}_1^{-1}g  \eqnum{7.25}
\end{equation}
where $\widetilde{Z}_1$ is the ghost vertex renormalization constant as
defined in Eq. (5.24). The relation in Eq. (7.25) ordinarily is determined
from the renormalization of S-matrix elements. In comparison of Eq. (7.24)
with Eq. (7.25), we see 
\begin{equation}
\widetilde{Z}_\gamma =\widetilde{Z}_1  \eqnum{7.26}
\end{equation}
which means that the functions $\widetilde{\gamma }_i^a(p,q,k)$ are
renormalized in the same way as for the ghost vertex (the three-line $\rho -$%
meson-ghost particle vertex). This result arises from the fact that the
functions $\widetilde{\gamma }_i^a(p,q,k)$ contains a ghost vertex as easily
seen from perturbative calculations.

In the conventional discussion of the vertex renormalization, one considers
such a vertex denoted by $\widetilde{\Lambda }^{a\mu }(p,q,k)$ that it is
defined from $\Lambda ^{a\mu }(p,q,k)$ by taking out a coupling constant.
Obviously, the W-T identity obeyed by the $\widetilde{\Lambda }^{a\mu
}(p,q,k)$ can be written out from (7.16) by taking away the coupling
constant on the RHS of Eq. (7.16), that is 
\begin{eqnarray}
k_\mu \widetilde{\Lambda }^{a\mu }(p,q,k)=i\chi (k^2)[S_F^{-1}(p)\widetilde{%
\gamma }_2^a(p,q,k)-\widetilde{\gamma }_1^a(p,q,k)S_F^{-1}(q)].  \eqnum{7.27}
\end{eqnarray}
The renormalization of the vertex $\widetilde{\Lambda }^{a\mu }(p,q,k)$
usually is defined by 
\begin{equation}
\widetilde{\Lambda }^{a\mu }(p,q,k)=Z_F^{-1}\widetilde{\Lambda }_R^{a\mu
}(p,q,k)  \eqnum{7.28}
\end{equation}
where $Z_F$ is the nucleon-$\rho $-meson vertex renormalization constant.
When Eqs. (5.22), (7.18), (7.22) and (7.28) are inserted into Eq. (7.27) and
then multiplying the both sides of Eq. (7.27) with a renormalized coupling
constant, we arrive at 
\begin{eqnarray}
k_\mu \Lambda _R^{a\mu }(p,q,k)=iZ_F\widetilde{Z}_3Z_2^{-1}Z_\gamma
^{-1}g_R\chi _R(k^2)[S_F^{R-1}(p)\widetilde{\gamma }_{2R}^a(p,q,k)-%
\widetilde{\gamma }_{1R}^a(p,q,k)S_F^{R-1}(q)]  \eqnum{7.29}
\end{eqnarray}
where 
\begin{equation}
\Lambda _R^{a\mu }(p,q,k)=g_R\widetilde{\Lambda }_R^{a\mu }(p,q,k). 
\eqnum{7.30}
\end{equation}
In comparison of Eq. (7.29) with Eq. (7.23) and considering the equality in
Eq. (7.26), we find, the following identity must hold 
\begin{equation}
\frac{Z_F}{Z_2}=\frac{\widetilde{Z}_1}{\widetilde{Z}_3}.  \eqnum{7.31}
\end{equation}
Combining the relations in Eqs. (5.28), (6.24) and (7.31), we have 
\begin{equation}
\frac{Z_F}{Z_2}=\frac{\widetilde{Z}_1}{\widetilde{Z}_3}=\frac{Z_1}{Z_3}=%
\frac{Z_4}{Z_1}.  \eqnum{7.32}
\end{equation}
This just is the well-known S-T identity which is formally identical to that
given in QCD [9, 17, 19].

\section{Pion-$\rho -$meson three-line vertex and pion propagator}

In this section, we are devoted to deriving the W-T identity obeyed by the
pion-$\rho -$meson three-line vertex and discuss its renormalization. Taking
derivative of Eq. (3.10) with respect to the source $\xi ^c(z)$ and then
setting $\bar \eta =\eta =\bar \xi ^a=0$, we have 
\begin{equation}
\frac 1\alpha \partial _z^\mu \frac{\delta W}{\delta J^{c\mu }(z)}-\int
d^4x[K^b(x)\frac{\delta ^2W}{\delta \chi ^b(x)\delta \xi ^c(z)}+J^{a\mu }(x)%
\frac{\delta ^2W}{\delta u^{a\mu }(x)\delta \xi ^c(z)}]=0  \eqnum{8.1}
\end{equation}
Furthermore, differentiating the above identity with respect to $K^b(y)$ and 
$K^a(x)$ and then turning off all the sources, one may obtain an identity
satisfied by the $\pi -\rho $ three-point Green function. Witten in the
operator formulation, it is represented as 
\begin{equation}
\partial _z^\mu \bar G_{\;\;\mu }^{abc}(x,y,z)=-\alpha g\varepsilon ^{bde}%
\bar G^{adec}(x,y,z)  \eqnum{8.2}
\end{equation}
where 
\begin{equation}
\bar G_{\;\;\mu }^{abc}(x,y,z)=\left\langle 0^{+}\left| T[\hat \pi ^a(x)\hat 
\pi ^b(y)\hat A_\mu ^c(z)]\right| 0^{-}\right\rangle  \eqnum{8.3}
\end{equation}
is the $\pi -\rho $ three-point Green function and 
\begin{equation}
\bar G^{adec}(x,y,z)=\left\langle 0^{+}\left| T[\hat \pi ^a(x)\hat \pi ^d(y)%
\hat C^e(y)\widehat{\overline{C}}^c(z)]\right| 0^{-}\right\rangle 
\eqnum{8.4}
\end{equation}
is the pion-ghost particle three-point Green function with a pion operator
and a ghost particle operator being put on the same position. Since the
identity in Eq. (8.2) is derived from the W-T identity in Eq. (3.10), the
Green functions in Eq. (8.2) are connected. In the following, we will start
from the identity in Eq. (8.2) to derive the identity satisfied by the $\pi
-\rho $ three-line vertex without recourse to the ghost equation because the
ghost equation will not simplify our discussion. For this purpose, we need
the following one-particle irreducible decompositions: 
\begin{equation}
\bar G_{\;\;\mu }^{abc}(x.y.z)=\int d^4x^{\prime }d^4y^{\prime }d^4z^{\prime
}i\Delta _\pi ^{aa^{\prime }}(x-x^{\prime })\Gamma ^{a^{\prime }b^{\prime
}c^{\prime },\mu ^{\prime }}(x^{\prime },y^{\prime },z^{\prime })i\Delta
_\pi ^{b^{\prime }b}(y^{\prime }-y)iD_{\mu ^{\prime }\mu }^{c^{\prime
}c}(z^{\prime }-z)  \eqnum{8.5}
\end{equation}
where 
\begin{equation}
i\Delta _\pi ^{aa^{\prime }}(x-x^{\prime })=\left\langle 0^{+}\left| T[\hat 
\pi ^a(x)\hat \pi ^{a^{\prime }}(x^{\prime })]\right| 0^{-}\right\rangle 
\eqnum{8.6}
\end{equation}
is the pion propagator and 
\begin{equation}
\bar \Gamma _{\;\;\;\;\mu }^{a^{\prime }b^{\prime }c^{\prime }}(x^{\prime
},y^{\prime },z^{\prime })=i\frac{\delta ^3\Gamma }{\delta \pi ^{a^{\prime
}}(x^{\prime })\delta \pi ^{b^{\prime }}(y^{\prime })\delta A^{c^{\prime
}\mu }(z^{\prime })}\mid _{J=0}  \eqnum{8.7}
\end{equation}
denotes the $\pi -\rho $ three-line proper vertex. And 
\begin{equation}
\bar G^{adec}(x,y;z)=\int d^4x^{\prime }d^4z^{\prime }\Delta _\pi
^{aa^{\prime }}(x-x^{\prime })\Delta ^{cc^{\prime }}(z-z^{\prime })\gamma
^{a^{\prime }c^{\prime }de}(x^{\prime },z^{\prime },y)  \eqnum{8.8}
\end{equation}
where $\Delta ^{cc^{\prime }}(z-z^{\prime })$ is the ghost particle
propagator defined in Eq. (4.7) and 
\begin{equation}
\gamma ^{a^{\prime }c^{\prime }de}(x^{\prime },z^{\prime },y)=\int
d^4y^{\prime }d^4u\Delta _\pi ^{dd^{\prime }}(y-y^{\prime })\Delta
^{ee^{\prime }}(y-u)\bar \Gamma ^{a^{\prime }d^{\prime }e^{\prime }c^{\prime
}}(x^{\prime },y^{\prime },u,z^{\prime })  \eqnum{8.9}
\end{equation}
in which the propagators $\Delta _\pi ^{dd^{\prime }}(y-y^{\prime })$ and $%
\Delta ^{ee^{\prime }}(y-u)$ have a common coordinate $y$ and 
\begin{equation}
\bar \Gamma ^{a^{\prime }d^{\prime }e^{\prime }c^{\prime }}(x^{\prime
},y^{\prime },u,z^{\prime })=\frac{\delta ^4\Gamma }{i\delta \pi ^{a^{\prime
}}(x^{\prime })\delta \pi ^{d^{\prime }}(y^{\prime })\delta \bar C%
^{e^{\prime }}(u)\delta C^{c^{\prime }}(z^{\prime })}\mid _{J-0} 
\eqnum{8.10}
\end{equation}
is the pion-ghost particle four-line vertex.

On substituting Eqs. (8.5) and (8.8) into Eq. (8.2) and then transforming
the equation thus obtained into the momentum space, it is easy to get 
\begin{equation}
k_3^\mu \bar \Gamma _{\;\;\mu }^{abc}(k_1,k_2,k_3)=-ig\varepsilon
^{bde}\Delta _\pi (k_2)^{-1}\chi (k_3)\gamma ^{acde}(k_1,k_2,k_3) 
\eqnum{8.11}
\end{equation}
where $\chi (k_3)$ was defined in Eq. (5.13), 
\begin{equation}
\gamma ^{acde}(k_1,k_2,k_3)=\int d^4l\Delta _\pi (l)\Delta (k_2-l)\bar \Gamma
^{acde}(k_1,l;k_2-l,k_3)  \eqnum{8.12}
\end{equation}
and 
\begin{equation}
\Delta _\pi ^{ab}(k)=\delta ^{ab}\Delta _\pi (k)  \eqnum{8.13}
\end{equation}
with 
\begin{equation}
\Delta _\pi (k)=\frac 1{k^2-m_\pi ^2-\Omega _\pi (k)+i\varepsilon } 
\eqnum{8.14}
\end{equation}
is the pion propagator given in the momentum space . In the above, the pion
self energy is defined by $-i\Omega _\pi ^{ab}(k)=-i\delta ^{ab}\Omega _\pi
(k)$. With the definitions: 
\begin{equation}
\begin{array}{c}
\bar \Gamma _{\;\;\mu }^{abc}(k_1,k_2,k_3)=(2\pi )^4\delta ^4(k_1+k_2+k_3)%
\bar \Lambda _{\;\;\mu }^{abc}(k,k_2,k_3), \\ 
\gamma ^{acde}(k_1,k_2,k_3)=(2\pi )^4\delta ^4(k_1+k_2+k_3)\hat \gamma
^{acde}(k_1,k_2,k_3),
\end{array}
\eqnum{8.15}
\end{equation}
the identity in Eq. (8.11) can be written in the form 
\begin{equation}
k_3^\mu \bar \Lambda _{\;\;\mu }^{abc}(k_1,k_2,k_3)=-ig\varepsilon
^{bde}\Delta _\pi (k_2)^{-1}\chi (k_3)\hat \gamma ^{acde}(k_1,k_2,k_3) 
\eqnum{8.16}
\end{equation}

Let us proceed to discuss the renormalization of the pion propagator and the 
$\pi -\rho $ three-line vertex. In accordance with the Lorentz-covariance,
the self-energy can be written in the form 
\begin{equation}
\Omega _\pi (k)=k^2\omega _1(k^2)+m_\pi ^2\omega _2(k^2).  \eqnum{8.17}
\end{equation}
The divergence in the function $\omega _i(k^2)$ ($i=1,2$) can be subtracted
at the renormalization point $\mu $, $\omega _i(k^2)=\omega _i(\mu
^2)+\omega _i^c(k^2)$ where $\omega _i(\mu ^2)$ and $\omega _i^c(k^2)$ are
the divergent and finite parts of the function $\omega _i(k^2)$
respectively. Defining the renormalization constant $\bar Z_{3\text{ }}$of
the pion propagator as 
\begin{equation}
\bar Z_3^{-1}=1-\omega _1(\mu ^2)  \eqnum{8.18}
\end{equation}
the pion propagator will be renormalized as 
\begin{equation}
\Delta _\pi (k)=\bar Z_3\Delta _\pi ^R(k)  \eqnum{8.19}
\end{equation}
where 
\begin{equation}
\Delta _\pi ^R(k)=\frac 1{k^2-m_\pi ^{R2}-\Omega _\pi ^R(k)+i\varepsilon } 
\eqnum{8.20}
\end{equation}
in which $m_\pi ^R=Z_{m_\pi }^{-1}m_\pi $ is the renormalized pion mass with

\begin{equation}
Z_{m_\pi }=\{\bar Z_3[1+\omega _2(\mu ^2)]\}^{-1/2}  \eqnum{8.21}
\end{equation}
being the pion mass renormalization constant and $\Omega _\pi ^R(k)=\bar Z%
_3[k^2\omega _1^c(k^2)$ $+m_\pi ^2\omega _2^c(k^2)]$ represents the finite
correction to the propagator.

From the definition in Eq. (8.6) and the relation in Eq. (8.19), we see 
\begin{equation}
\pi ^a(x)=\bar Z_3^{1/2}\pi _R^a(x)  \eqnum{8.22}
\end{equation}
where $\pi _R^a(x)$ stands for the renormalized pion field function.
According to the definitions in Eq. (8.7) and (8.15) and the relations in
Eqs. (5.18) and (8.22), it is clear to see that the vertex $\Lambda
_{\;\;\mu }^{abc}(k_1,k_2,k_3)$ is renormalized as 
\begin{equation}
\bar \Lambda _{\;\;\mu }^{abc}(k_1,k_2,k_3)=\bar Z_3^{-1}Z_3^{-1/2}\bar 
\Lambda _{R\;\mu }^{abc}(k_1,k_2,k_3).  \eqnum{8.23}
\end{equation}
Analogous to the analysis mentioned in Eqs. (7.22)-(7.26), it can be proved
that the vertex $\hat \gamma ^{acde}(k_1,k_2,k_3)$ is renormalized in such a
fashion 
\begin{equation}
\hat \gamma ^{acde}(k_1,k_2,k_3)=\tilde Z_1^{-1}\hat \gamma
_R^{acde}(k_1,k_2,k_3).  \eqnum{8.24}
\end{equation}
This means that the vertex $\hat \gamma ^{acde}(k_1,k_2,k_3)$, as the vertex 
$\tilde \gamma _i^a(k_1,k_2,k_3)$ defined in Eq. (7.15), is renormalized in
the same way as the ghost vertex $\Lambda _{\;\;\lambda }^{abc}(k_1,k_2,k_3)$
because the divergence occurring in the vertex $\hat \gamma
^{acde}(k_1,k_2,k_3),$ as the vertex $\tilde \gamma ^a(k_1,k_2,k_3),$ comes
from the $\rho -$meson-ghost particle vertex which is included in the vertex 
$\hat \gamma ^{acde}(k_1,k_2,k_3)$. This fact can also be seen from a
perturbative analysis of the vertex. When Eqs. (8.19), (8.23), (8.24) and
(5.22) are inserted into Eq. (8.16), one may obtain a renormalized version
of the identity in Eq. (8.16) 
\begin{equation}
k_3^\mu \bar \Lambda _{R\;\mu }^{abc}(k_2,k_2,k_3)=-i\varepsilon
^{bde}g_R\Delta _\pi ^R(k_2)^{-1}\chi _R(k_3)\hat \gamma
_R^{acde}(k_1,k_2,k_3).  \eqnum{8.25}
\end{equation}
According to the usual discussion of vertex renormalization, we introduce a
vertex denoted by $\hat \Lambda _{\;\;\mu }^{abc}(k_1,k_2,k_3)$ which is
defined from $\bar \Lambda _{\;\;\mu }^{abc}(k_1,k_2,k_3)$ by extracting a
coupling constant. This vertex is renormalized as 
\begin{equation}
\hat \Lambda _{\;\;\mu }^{abc}(k_1,k_2,k_3)=\bar Z_1^{-1}\hat \Lambda
_{R\;\mu }^{abc}(k_1,k_2,k_3)  \eqnum{8.26}
\end{equation}
where $\bar Z_1$ is the renormalization constant of the $\pi -\rho $ vertex.
Clearly, the W-T identity for the vertex $\hat \Lambda _{\;\;\mu
}^{abc}(k_1,k_2,k_3)$ can be written from Eq. (8.16) by taking out a
coupling constant on the right hand side of Eq. (8.16), 
\begin{equation}
k_3^\mu \hat \Lambda _{\;\;\mu }^{abc}(k_1,k_2,k_3)=-i\varepsilon
^{bde}\Delta _\pi (k_2)^{-1}\chi (k_3)\hat \gamma ^{acde}(k_1,k_2,k_3) 
\eqnum{8.27}
\end{equation}
Upon substituting Eqs. (8.19), (8.24), (8.26) and (5.22) into Eq. (8.27) and
then multiplying Eq. (8.27) by a renormalized coupling constant, noticing $%
\bar \Lambda _{R\;\mu }^{abc}(k_1,k_2,k_3)=g_R\hat \Lambda _{R\;\mu
}^{abc}(k_1,k_2,k_3)$, we have 
\begin{equation}
k_3^\mu \bar \Lambda _{R\;\mu }^{abc}(k_2,k_2,k_3)=-i\frac{\tilde Z_3\bar Z_1%
}{\tilde Z_1\bar Z_3}\varepsilon ^{bde}g_R\Delta _\pi ^R(k_2)^{-1}\chi
_R(k_3)\hat \gamma _R^{acde}(k_1,k_2,k_3).  \eqnum{8.28}
\end{equation}
In comparison of Eq. (8.28) with Eq. (8.25), we see, it must be 
\begin{equation}
\frac{\tilde Z_3\bar Z_1}{\tilde Z_1\bar Z_3}=1  \eqnum{8.29}
\end{equation}
which leads to the S-T identity such that 
\begin{equation}
\frac{\bar Z_1}{\bar Z_3}=\frac{\tilde Z_1}{\tilde Z_3}.  \eqnum{8.30}
\end{equation}

\section{Pion-$\rho -$meson four-line vertex}

In this section, we plan to derive the W-T identity satisfied by the $\pi
-\rho $ four-line proper vertex. For this purpose, we first derive the W-T
identity obeyed by the $\pi -\rho $ four-point Green function by starting
the identity shown in Eq. (8.1). Changing the variable and indices $z,c$ and 
$\mu $ to $u,d$ and $\nu $ in Eq. (8.1), then differentiating Eq. (8.1) with
respect to the sources $K^a(x),K^b(y)$ and $J^{c\mu }(z)$ and finally
setting all the sources to be zero, we obtain an identity which, written in
the operator form, is 
\begin{equation}
\begin{array}{c}
\frac 1\alpha \partial _u^\nu \left\langle 0^{+}\left| T[\hat \pi ^a(x)\hat 
\pi ^b(y)\hat A_\mu ^c(z)\hat A_\nu ^d(u)]\right| 0^{-}\right\rangle \\ 
=-\left\langle 0^{+}\left| T^{*}[\hat \pi ^a(x)\Delta \pi ^b(y)\hat A_\mu
^c(z)\widehat{\overline{C}}^d(u)]\right| 0^{-}\right\rangle \\ 
-\left\langle 0^{+}\left| T^{*}[\hat \pi ^a(x)\hat \pi ^b(y)\Delta \hat A%
_\mu ^c(z)\widehat{\overline{C}}^d(u)]\right| 0^{-}\right\rangle
\end{array}
\eqnum{9.1}
\end{equation}
This identity may be simplified with the help of a ghost equation which can
be derived from the ghost equation in Eq. (3.18). Changing the variable $x$
and the index $a$ in Eq. (3.18) to $z$ and $c$, then differentiating Eq.
(3.18) successively with respect to the sources $\xi ^d(u)$, $K^b(y)$ and $%
K^a(x)$ and finally setting all the sources to vanish, one can obtain the
ghost equation. Written in the operator form, it reads 
\begin{equation}
\partial _z^\mu \left\langle 0^{+}\left| T^{*}[\hat \pi ^a(x)\hat \pi
^b(y)\Delta \hat A_\mu ^c(z)\widehat{\overline{C}}^d(u)]\right|
0^{-}\right\rangle =-\sigma ^2\left\langle 0^{+}\left| T[\hat \pi ^a(x)\hat 
\pi ^b(y)\hat C^c(z)\widehat{\overline{C}}^d(u)]\right| 0^{-}\right\rangle .
\eqnum{9.2}
\end{equation}
Differentiating Eq. (9.1) with respect to $z$, applying Eq. (9.2) and
noticing the definition of $\Delta \pi ^b(y)$ in Eq. (3.3), we arrive at 
\begin{equation}
\partial _{x_3}^\mu \partial _{x_4}^\nu \bar G_{\;\;\mu \nu
}^{abcd}(x_1,x_2,x_3,x_4)=-\alpha g\varepsilon ^{bij}\partial _{x_3}^\mu 
\bar G_{\;\;\;\mu }^{aijdc}(x_1,x_2,x_3,x_4)+\sigma ^2\bar G%
^{abcd}(x_1,x_2,x_3,x_4).  \eqnum{9.3}
\end{equation}
where we have changed the position variables for later convenience and the
Green functions are defined by 
\begin{equation}
\bar G_{\;\;\mu \nu }^{abcd}(x_1,x_2,x_3,x_4)=\left\langle 0^{+}\left| T[%
\hat \pi ^a(x_1)\hat \pi ^b(x_2)\hat A_\mu ^c(x_3)\hat A_\nu ^d(x_4)]\right|
0^{-}\right\rangle ,  \eqnum{9.4}
\end{equation}
\begin{equation}
\bar G_{\;\;\;\mu }^{aijdc}(x_1,x_2,x_3,x_4)=\left\langle 0^{+}\left| T[\hat 
\pi ^a(x_1)\hat \pi ^i(x_2)\hat A_\mu ^c(x_3)\hat C^j(x_2)\widehat{\overline{%
C}}^d(x_4)]\right| 0^{-}\right\rangle  \eqnum{9.5}
\end{equation}
and $\bar G^{abcd}(x_1,x_2,x_3,x_4)$ was defined in Eq. (8.4). These Green
functions are all connected.

The W-T identity for $\pi -\rho $ four-line vertex can be derived from Eq.
(9.3) with the aid of one-particle irreducible decompositions of the Green
functions which can easily be found by the standard procedure and are shown
below. For the Green function $\bar G_{\;\;\mu \nu }^{abcd}(x_1,x_2,x_3,x_4)$%
, we have the one-particle irreducible decompositions as follows 
\begin{equation}
\begin{array}{c}
\bar G_{\;\;\mu \nu }^{abcd}(x_1,x_2,x_3,x_4) \\ 
=\int \prod\limits_{i=1}^4d^4y_i\Delta _\pi (x_1-y_1)\Delta _\pi
(x_2-y_2)D_{\mu \rho }(x_3-y_3)D_{\nu \sigma }(x_4-y_4)\bar \Gamma
^{abcd,\rho \sigma }(y_1,y_2,y_3,y_4) \\ 
+i\int \prod\limits_{i=1}^3d^4y_id^4z_i\{\Delta _\pi (x_1-z_1)\Delta _\pi
(x_2-y_2)D_{\mu \rho }(x_3-y_3)D_{\nu \sigma }(x_4-z_3)\bar \Gamma
^{aed,\sigma }(z_1,z_2,z_3) \\ 
\times \Delta _\pi (z_2-y_1)\bar \Gamma ^{ebc,\rho }(y_1,y_2,y_3)+\Delta
_\pi (x_1-y_1)\Delta _\pi (x_2-z_1)D_{\mu \rho }(x_3-y_3)D_{\nu \sigma
}(x_4-z_3) \\ 
\times \bar \Gamma ^{aec,\rho }(y_1,y_2,y_3)\Delta _\pi (y_2-z_2)\bar \Gamma
^{bed,\sigma }(z_1,z_2,z_3)+\Delta _\pi (x_1-y_1)\Delta _\pi (x_2-y_2)D_{\mu
\rho }(x_3-z_1) \\ 
\times D_{\nu \sigma }(x_4-z_3)\bar \Gamma ^{abe,\lambda
}(y_1,y_2,y_3)D_{\lambda \tau }(y_3-z_2)\Gamma _{ced}^{\rho \tau \sigma
}(z_1,z_2,z_3)\}
\end{array}
\eqnum{9.10}
\end{equation}
where 
\begin{equation}
\bar \Gamma _{\;\;\mu \nu }^{abcd}(y_1,y_2,y_3,y_4)=i\frac{\delta ^4\Gamma }{%
\delta \pi ^a(y_1)\delta \pi ^b(y_2)\delta A^{c\mu }(y_3)\delta A^{d\nu
}(y_4)}\mid _{J=0}  \eqnum{9.11}
\end{equation}
is the $\pi -\rho $ four-line proper vertex, while the $\pi -\rho $
three-line vertex and the $\rho -$meson three-line vertex were already
defined in Eqs. (8.7) and (5.9) respectively. For the Green function $%
G_{\;\;\;\mu }^{aijdc}(x_1,x_2,x_3,x_4)$, its irreducible decomposition is

\begin{equation}
\bar G_{\;\;\;\mu }^{aijdc}(x_1,x_2,x_3;x_2,x_4)=-i\int
\prod\limits_{i=1}^3d^4y_i\Delta _\pi (x_1-y_1)\Delta (x_4-y_2)D_{\mu \nu
}(x_3-y_3)\sum\limits_{\alpha =1}^5\gamma _\alpha ^{aijdc,\nu
}(x_2,y_1,y_2,y_3)  \eqnum{9.12}
\end{equation}
where 
\begin{equation}
\gamma _1^{aijdc,\nu }(x_2,y_1,y_2,y_3)=\int d^4z_1d^4z_2\Delta _\pi
(x_2-z_1)\Delta (x_2-z_2)\bar \Gamma ^{aijdc,\nu }(y_1,y_2,y_3;z_1,z_2), 
\eqnum{9.13}
\end{equation}
\begin{equation}
\gamma _2^{aijdc,\nu }(x_2,y_1,y_2,y_3)=i\int
\prod\limits_{k=1}^4d^4z_k\Delta _\pi (x_2-z_2)\Delta (x_2-z_3)\Delta _\pi
(z_1-z_4)\bar \Gamma ^{aec,\nu }(y_1,y_3,z_4)\bar \Gamma
^{eijd}(z_1,z_2,z_3,y_2),  \eqnum{9.14}
\end{equation}
\begin{equation}
\gamma _3^{aijdc,\nu }(x_2,y_1,y_2,y_3)=i\int
\prod\limits_{k=1}^4d^4z_k\Delta _\pi (x_2-z_1)\Delta (x_2-z_4)\Delta _\pi
(z_2-z_3)\bar \Gamma ^{iec,\nu }(z_1,z_2,y_3)\bar \Gamma
^{aejd}(y_1,y_2;z_3,z_4),  \eqnum{9.15}
\end{equation}
\begin{equation}
\gamma _4^{aijdc,\nu }(x_2,y_1,y_2,y_3)=i\int
\prod\limits_{k=1}^4d^4z_k\Delta _\pi (x_2-z_3)\Delta (x_2-z_1)\Delta
(z_2-z_4)\Gamma ^{jec,\nu }(z_1,z_4,y_3)\bar \Gamma ^{aied}(y_1,y_2,z_2,z_3)
\eqnum{9.16}
\end{equation}
and 
\begin{equation}
\gamma _5^{aijdc,\nu }(x_2,y_1,y_2,y_3)=i\int
\prod\limits_{k=1}^4d^4z_k\Delta _\pi (x_2-z_1)\Delta (x_2-z_2)\Delta
(z_3-z_4)\Gamma ^{dec,\nu }(y_2,y_3,z_4)\bar \Gamma ^{aije}(y_1,z_1,z_2,z_3)
\eqnum{9.17}
\end{equation}
In the above, 
\begin{equation}
\Gamma ^{aijdc,\nu }(y_1,y_2,y_3;z_1,z_2)=\frac{\delta ^5\Gamma }{i\delta
\pi ^a(y_1)\delta \pi ^i(y_2)\delta \bar C^j(y_3)\delta C^d(z_1)\delta A_\nu
^c(z_2)}\mid _{J=0}  \eqnum{9.18}
\end{equation}
which contains a $\rho -$meson-ghost particle vertex in it, and the
three-line $\pi -\rho $ vertex, the three-line $\rho -$meson-ghost particle
vertex and the four-line pion-ghost particle vertex were respectively
defined Eqs. (8.7), (5.10) and (8.10). The typical feature of the functions $%
\gamma _\alpha ^{aijdc,\nu }(x_2,y_1,y_2,y_3)$ is that there are a pion
propagator and a ghost particle propagator in the functions which have a
common coordinate $x_2$. As for the Green function $%
G^{abcd}(x_1,x_2,x_3,x_4),$ its one-particle irreducible decomposition can
be found to be 
\begin{equation}
\bar G^{abcd}(x_1,x_2,x_3,x_4)=\int \prod\limits_{i=1}^3d^4y_i\Delta _\pi
(x_1-y_1)\Delta _\pi (x_4-y_2)\Delta (x_3-y_3)\Delta (x_4-y_4)\bar \Gamma
^{abcd}(y_1,y_2,y_3,y_4)  \eqnum{9.19}
\end{equation}
where the vertex $\bar \Gamma ^{abcd}(y_1,y_2,y_3,y_4)$ was defined in Eq.
(8.10). Upon substituting Eqs. (9.10), (9.12) and (9.19) into Eq. (9.3) and
then transforming the equation thus obtained into the momentum space, it is
not difficult to get the following W-T identity satisfied by the $\pi -\rho $
four-line vertex, 
\begin{equation}
\begin{array}{c}
k_3^\mu k_4^\nu \bar \Gamma _{\;\;\mu \nu }^{abcd}(k_1,k_2,k_3,k_4) \\ 
=-\int \frac{d^4q}{(2\pi )^4}\{k_3^\mu \bar \Gamma _{\;\;\mu
}^{ebc}(q,k_2,k_3)\Delta _\pi (q)k_4^\nu \bar \Gamma _{\;\;\nu
}^{aed}(k_1,-q,k_4) \\ 
+k_3^\mu \bar \Gamma _{\;\;\mu }^{aec}(k_1,-q,k_3)\Delta _\pi (q)k_4^\nu 
\bar \Gamma _{\;\;\nu }^{bed}(k_2,q,k_4) \\ 
+\bar \Gamma _{\;\;\lambda }^{abe}(k_1,k_2,-q)D^{\lambda \tau }(q)k_3^\mu
k_4^\nu \Gamma _{\mu \tau \nu }^{ced}(k_3,q,k_4)\} \\ 
-g\varepsilon ^{bij}\Delta _\pi ^{-1}(k_2)\chi (k_4)\sum\limits_{\alpha
=1}^5k_3^\mu \gamma _{\alpha \;\;\mu }^{aijdc}(k_1,k_2,k_3,k_4) \\ 
-\frac{\sigma ^2}\alpha \chi (k_3)\chi (k_4)\bar \Gamma
^{abcd}(k_1,k_2,k_3,k_4).
\end{array}
\eqnum{9.20}
\end{equation}
When we define 
\begin{equation}
\begin{array}{c}
\bar \Gamma _{\;\;\mu \nu }^{abcd}(k_1,k_2,k_3,k_4)=(2\pi )^4\delta
^4(\sum\limits_{i=1}^4k_i)\bar \Lambda _{\;\;\mu \nu
}^{abcd}(k_1,k_2,k_3,k_4), \\ 
\bar \Gamma _{\;\;\mu }^{abc}(k_1,k_2,k_3)=(2\pi )^4\delta
^4(\sum\limits_{i=1}^3k_i)\bar \Lambda _{\;\;\mu }^{abc}(k_1,k_2,k_3), \\ 
\gamma _{\alpha \;\;\mu }^{aijdc}(k_1,k_2,k_3,k_4)=(2\pi )^4\delta
^4(\sum\limits_{i=1}^4k_i)\tilde \gamma _{\alpha \;\;\mu
}^{aijdc}(k_1,k_2,k_3,k_4), \\ 
\bar \Gamma ^{abcd}(k_1,k_2,k_3,k_4)=(2\pi )^4\delta
^4(\sum\limits_{i=1}^4k_i)\bar \Lambda ^{abcd}(k_1,k_2,k_3,k_4),
\end{array}
\eqnum{9.21}
\end{equation}
the identity in Eq. (9.20) can be written as 
\begin{equation}
\begin{array}{c}
k_3^\mu k_4^\nu \bar \Lambda _{\;\;\mu \nu }^{abcd}(k_1,k_2,k_3,k_4) \\ 
=-k_3^\mu \bar \Lambda _{\;\;\mu }^{ebc}(k_2,k_3)\Delta _\pi
(k_1+k_4)k_4^\nu \bar \Lambda _{\;\;\nu }^{aed}(k_1,k_4) \\ 
-k_3^\mu \bar \Lambda _{\;\;\mu }^{aec}(k_1,k_3)\Delta _\pi (k_1+k_3)k_4^\nu 
\bar \Lambda _{\;\;\nu }^{bed}(k_2,k_4) \\ 
-\bar \Lambda _{\;\;\lambda }^{abe}(k_1,k_2,)D^{\lambda \tau
}(k_1+k_2)k_3^\mu k_4^\nu \Lambda _{\mu \tau \nu }^{ced}(k_3,k_4) \\ 
-g\varepsilon ^{bij}\Delta _\pi ^{-1}(k_2)\chi (k_4)\sum\limits_{\alpha
=1}^5k_3^\mu \tilde \gamma _{\alpha \;\;\mu }^{aijdc}(k_1,k_2,k_3,k_4) \\ 
-\frac{\sigma ^2}\alpha \chi (k_3)\chi (k_4)\bar \Lambda
^{abcd}(k_1,k_2,k_3,k_4).
\end{array}
\eqnum{9.22}
\end{equation}

We are now in a position to discuss the renormalized version of the above
identity. According to the definitions in Eqs. (9.11) and (8.10) and the
relations in Eqs. (5.18) and (8.22), it is easy to see 
\begin{equation}
\begin{array}{c}
\bar \Lambda _{\;\;\mu \nu }^{abcd}(k_1,k_2,k_3,k_4)=\bar Z_3^{-1}Z_3^{-1}%
\bar \Lambda _{R\;\mu \nu }^{abcd}(k_1,k_2,k_3,k_4) \\ 
\bar \Lambda ^{abcd}(k_1,k_2,k_3,k_4)=\bar Z_3^{-1}\tilde Z_3^{-1}\bar 
\Lambda _R^{abcd}(k_1,k_2,k_3,k_4) \\ 
\tilde \gamma _{\alpha \;\;\mu }^{aijdc}(k_1,k_2,k_3,k_4)=Z_3^{-1/2}\tilde Z%
_1^{-1}\tilde \gamma _{\alpha R\;\mu }^{aijdc}(k_1,k_2,k_3,k_4)
\end{array}
\eqnum{9.23}
\end{equation}
In the last equality, the factor $Z_3^{-1/2}$ is given by applying the
relations in Eq. (5.18) and (8.22) to all the propagators and vertices
contained in the functions $\gamma _\alpha ^{aijdc,\nu }(x_2,y_1,y_2,y_3)$
shown in Eqs. (9.13)-(9.17), while, the factor $\tilde Z_1^{-1}$ arises from
the consideration that the functions $\tilde \gamma _\alpha
^{aijdc}(k_1,k_2,k_3,k_4),$ as the function $\gamma ^{acde}(k_1,k_2,k_3,k_4)$%
, includes the contribution given by the ghost vertex which is contained in
functions $\tilde \gamma _\alpha ^{aijdc}(k_1,k_2,k_3,k_4)$ (more
explanations will be given soon later). Substituting Eqs. (9.23), (5.19),
(5.22), (8.19), (8.23) and (4.27) into Eq. (9.22), we obtain the
renormalized version of the identity 
\begin{equation}
\begin{array}{c}
k_3^\mu k_4^\nu \bar \Lambda _{R\;\mu \nu }^{abcd}(k_1,k_2,k_3,k_4) \\ 
=-k_3^\mu \bar \Lambda _{R\;\mu }^{ebc}(k_2,k_3)\Delta _\pi
^R(k_1+k_4)k_4^\nu \bar \Lambda _{R\;\nu }^{aed}(k_1,k_4) \\ 
-k_3^\mu \bar \Lambda _{R\;\mu }^{aec}(k_1,k_3)\Delta _\pi
^R(k_1+k_3)k_4^\nu \bar \Lambda _{R\;\nu }^{bed}(k_2,k_4) \\ 
-\bar \Lambda _{R\;\lambda }^{abe}(k_1,k_2,)D_R^{\lambda \tau
}(k_1+k_2)k_3^\mu k_4^\nu \Lambda _{R\mu \tau \nu }^{\;ced}(k_3,k_4) \\ 
-g_R\varepsilon ^{bij}\Delta _\pi ^{R-1}(k_2)\chi _R(k_4)\sum\limits_{\alpha
=1}^5k_3^\mu \tilde \gamma _{\alpha R\;\mu }^{aijdc}(k_1,k_2,k_3,k_4) \\ 
-\frac{\sigma _R^2}{\alpha _R}\chi _R(k_3)\chi _R(k_4)\bar \Lambda
_R^{abcd}(k_1,k_2,k_3,k_4).
\end{array}
\eqnum{9.24}
\end{equation}
where $g_R$, $\sigma _R$ and $\alpha _R$ were defined in Eqs. (7.25) and
(4.30).

As usual, we redefine the vertices in Eq. (9.22) by extracting a coupling
constant or its squared one as shown in the following 
\begin{equation}
\begin{array}{c}
\bar \Lambda _{\;\;\mu \nu }^{abcd}(k_1,k_2,k_3,k_4)=g^2\hat \Lambda
_{\;\;\mu \nu }^{abcd}(k_1,k_2,k_3,k_4), \\ 
\bar \Lambda _{\;\;\mu }^{abc}(k_1,k_2,k_3)=g\hat \Lambda _{\;\;\mu
}^{abc}(k_1,k_2,k_3), \\ 
\Lambda _{\mu \nu \lambda }^{abc}(k_1,k_2,k_3)=g\tilde \Lambda _{\mu \nu
\lambda }^{abc}(k_1,k_2,k_3), \\ 
\bar \Lambda ^{abcd}(k_1,k_2,k_3,k_4)=g^2\hat \Lambda
^{abcd}(k_1,k_2,k_3,k_4), \\ 
\tilde \gamma _{\alpha \;\;\mu }^{aijdc}(k_1,k_2,k_3,k_4)=g\hat \gamma
_{\alpha \;\;\mu }^{aijdc}(k_1,k_2,k_3,k_4)_{,}
\end{array}
\eqnum{9.25}
\end{equation}
With these definitions, the W-T identity in Eq. (9.22) will be replaced by 
\begin{equation}
\begin{array}{c}
k_3^\mu k_4^\nu \hat \Lambda _{\;\;\mu \nu }^{abcd}(k_1,k_2,k_3,k_4) \\ 
=-k_3^\mu \hat \Lambda _{\;\;\mu }^{ebc}(k_2,k_3)\Delta _\pi
(k_1+k_4)k_4^\nu \hat \Lambda _{\;\;\nu }^{aed}(k_1,k_4) \\ 
-k_3^\mu \hat \Lambda _{\;\;\mu }^{aec}(k_1,k_3)\Delta _\pi (k_1+k_3)k_4^\nu 
\hat \Lambda _{\;\;\nu }^{bed}(k_2,k_4) \\ 
-\hat \Lambda _{\;\;\lambda }^{abe}(k_1,k_2,)D^{\lambda \tau
}(k_1+k_2)k_3^\mu k_4^\nu \tilde \Lambda _{\mu \tau \nu }^{ced}(k_3,k_4) \\ 
-\varepsilon ^{bij}\Delta _\pi ^{-1}(k_2)\chi (k_4)\sum\limits_{\alpha
=1}^5k_3^\mu \hat \gamma _{\alpha \;\;\mu }^{aijdc}(k_1,k_2,k_3,k_4) \\ 
-\frac{\sigma ^2}\alpha \chi (k_3)\chi (k_4)\hat \Lambda
^{abcd}(k_1,k_2,k_3,k_4).
\end{array}
\eqnum{9.26}
\end{equation}
The vertices in Eq. (9 26) are renormalized in the manner as shown In Eqs.
(5.24), (8.24), (8.26) and in the following 
\begin{equation}
\hat \Lambda _{\;\;\mu \nu }^{abcd}(k_1,k_2,k_3,k_4)=\bar Z_4^{-1}\hat 
\Lambda _{R\;\mu \nu }^{abcd}(k_1,k_2,k_3,k_4)  \eqnum{9.27}
\end{equation}
\begin{equation}
\hat \Lambda ^{abcd}(k_1,k_2,k_3,k_4)=\bar Z_5^{-1}\hat \Lambda
_R^{abcd}(k_1,k_2,k_3,k_4)  \eqnum{9.28}
\end{equation}
\begin{equation}
\hat \gamma _{1\;\;\mu }^{aijdc}(k_1,k_2,k_3,k_4)=Z_5^{-1}\hat \gamma
_{1R\;\mu }^{aijdc}(k_1,k_2,k_3,k_4),  \eqnum{9.29}
\end{equation}
\begin{equation}
\hat \gamma _{\alpha \;\;\mu }^{aijdc}(k_1,k_2,k_3,k_4)=\bar Z_3\bar Z_1^{-1}%
\tilde Z_1^{-1}\hat \gamma _{\alpha R\;\mu }^{aijdc}(k_1,k_2,k_3,k_4),\text{
if }\alpha =2,3  \eqnum{9.30}
\end{equation}
and 
\begin{equation}
\hat \gamma _{\alpha \;\;\mu }^{aijdc}(k_1,k_2,k_3,k_4)=\tilde Z_3\tilde Z%
_1^{-2}\hat \gamma _{\alpha R\;\mu }^{aijdc}(k_1,k_2,k_3,k_4),\text{ if }%
\alpha =4,5.  \eqnum{9.31}
\end{equation}
It should be noted that the renormalization constant $\tilde Z_1^{-1}$ in
Eqs. (9.30) and (9.31) arises from the relation in Eq. (8.24) because the
functions $\hat \gamma _{\alpha \;\;\mu }^{aijdc}(k_1,k_2,k_3,k_4)$ with $%
\alpha =2,3,4$ and $5$ contain a function $\hat \gamma ^{abcd}(k_1,k_2,k_3)$%
. For example, by the Fourier transformation, one can obtain from Eq. (9.14)
that 
\begin{equation}
\hat \gamma _{2\;\;\mu }^{aijdc}(k_1,k_2,k_3,k_4)=i\int \frac{d^4q}{(2\pi )^4%
}\Delta _\pi (q)\hat \Lambda _{\;\;\mu }^{aec}(k_1,k_3,q)\hat \gamma
^{eijd}(k_2,k_4,q)  \eqnum{9.32}
\end{equation}
where $\hat \gamma ^{eijd}(k_2,k_4,q)$ has an expression similar to that as
shown in Eq. (8.12). It is easy to see that use of Eqs. (8.19), (8.24) and
(8.26) in Eq. (9.32) directly gives rise to Eq. (9.30). Similarly, from Eq.
(9.17), one can get 
\begin{equation}
\hat \gamma _{5\;\;\mu }^{aijdc}(k_1,k_2,k_3,k_4)=i\int \frac{d^4q}{(2\pi )^4%
}\Delta (q)\tilde \Lambda _{\;\;\mu }^{aec}(k_4,k_3,q)\hat \gamma
^{eijd}(k_1,k_2,q)  \eqnum{9.33}
\end{equation}
Apparently, substitution of Eqs. (4.26), (5.24) and (8.24) in Eq. (9.33)
leads to the relation in Eq. (9.31). The function $\hat \gamma _{3\;\;\mu
}^{aijdc}(k_1,k_2,k_3,k_4)$ is a kind of exchange term of $\hat \gamma
_{2\;\;\mu }^{aijdc}(k_1,k_2,k_3,k_4)$ and therefore the renormalization of $%
\hat \gamma _{3\;\;\mu }^{aijdc}(k_1,k_2,k_3,k_4)$ should be as the same as $%
\hat \gamma _{2\;\;\mu }^{aijdc}(k_1,k_2,k_3,k_4)$. The discussion for the
function $\hat \gamma _{4\;\;\mu }^{aijdc}(k_1,k_2,k_3,k_4)$ is the same
because the $\hat \gamma _{4\;\;\mu }^{aijdc}(k_1,k_2,k_3,k_4)$ is also a
kind of exchange term of $\hat \gamma _{5\;\;\mu }^{aijdc}(k_1,k_2,k_3,k_4)$.

When the relations in Eqs. (4.27), (5.22), (5.24), (8.19), (8.26) and
(9.27)-(9.31) are inserted into Eq. (9.26) and then multiplying Eq. (9.26)
by $g_R^2$, we arrive at 
\begin{equation}
\begin{array}{c}
k_3^\mu k_4^\nu \bar \Lambda _{R\;\mu \nu }^{abcd}(k_1,k_2,k_3,k_4) \\ 
=-\frac{\bar Z_4\bar Z_3}{\bar Z_1^2}[k_3^\mu \bar \Lambda _{R\;\mu
}^{ebc}(k_2,k_3)\Delta _\pi ^R(k_1+k_4)k_4^\nu \bar \Lambda _{R\;\nu
}^{aed}(k_1,k_4) \\ 
+k_3^\mu \bar \Lambda _{R\;\mu }^{aec}(k_1,k_3)\Delta _\pi
^R(k_1+k_3)k_4^\nu \bar \Lambda _{R\;\nu }^{bed}(k_2,k_4)] \\ 
-\frac{\bar Z_4Z_3}{Z_1\bar Z_1}\bar \Lambda _{R\;\lambda
}^{abe}(k_1,k_2,)D_R^{\lambda \tau }(k_1+k_2)k_3^\mu k_4^\nu \Lambda _{R\mu
\tau \nu }^{\;ced}(k_3,k_4) \\ 
-g_R\varepsilon ^{bij}\Delta _\pi ^{R-1}(k_2)\chi _R(k_4)k_3^\mu \{\frac{%
\bar Z_4\tilde Z_3}{\bar Z_3Z_5}\tilde \gamma _{1R\;\mu
}^{aijdc}(k_1,k_2,k_3,k_4) \\ 
+\frac{\bar Z_4\tilde Z_3}{\tilde Z_1\bar Z_1}[\tilde \gamma _{2R\;\mu
}^{aijdc}(k_1,k_2,k_3,k_4)+\tilde \gamma _{3R\;\mu
}^{aijdc}(k_1,k_2,k_3,k_4)] \\ 
\frac{\bar Z_4\tilde Z_3^2}{\bar Z_3\tilde Z_1^2}[\tilde \gamma _{4R\;\mu
}^{aijdc}(k_1,k_2,k_3,k_4)+\tilde \gamma _{5R\;\mu
}^{aijdc}(k_1,k_2,k_3,k_4)]\} \\ 
-\frac{\bar Z_4\tilde Z_3}{Z_3\bar Z_5}\frac{\sigma _R^2}{\alpha _R}\chi
_R(k_3)\chi _R(k_4)\Lambda _R^{abcd}(k_1,k_2,k_3,k_4)
\end{array}
\eqnum{9.34}
\end{equation}
where we have considered the relations in Eq. (9.25) which also hold for the
renormalized vertices. In comparison of Eq. (9.34) with Eq. (9.24), we have 
\begin{equation}
\frac{\bar Z_4\bar Z_3}{\bar Z_1^2}=1,\;\frac{\bar Z_4Z_3}{Z_1\bar Z_1}=1,\;%
\frac{\bar Z_4\tilde Z_3}{\tilde Z_1\bar Z_1}=1,\;\frac{\bar Z_4\tilde Z_3^2%
}{\bar Z_3\tilde Z_1^2}=1  \eqnum{9.35}
\end{equation}
and 
\begin{equation}
\frac{\bar Z_4\tilde Z_3}{\bar Z_3Z_5}=1,\;\frac{\bar Z_4\tilde Z_3}{Z_3\bar 
Z_5}=1.  \eqnum{9.36}
\end{equation}
From the four identities in Eq. (9.35), it is found that 
\begin{equation}
\frac{\bar Z_4}{\bar Z_1}=\frac{Z_1}{Z_3}=\frac{\tilde Z_1}{\tilde Z_3}=%
\frac{\bar Z_1}{\bar Z_3},  \eqnum{9.37}
\end{equation}
while, from the two identities in Eq. (9.36), one gets 
\begin{equation}
\frac{Z_3}{\bar Z_3}=\frac{Z_5}{\bar Z_5}.  \eqnum{9.38}
\end{equation}
Combining the identities given in Eqs. (5.28), (6.24), (7.32), (8.30),
(9.37) and (9.38), we finally obtain the S-T identities as follows 
\begin{equation}
\frac{Z_1^F}{Z_2}=\frac{Z_1}{Z_3}=\frac{\tilde Z_1}{\tilde Z_3}=\frac{\bar Z%
_1}{\bar Z_3}=\frac{Z_4}{Z_1}=\frac{\bar Z_4}{\bar Z_1}  \eqnum{9.39}
\end{equation}
and 
\begin{equation}
\frac{Z_1}{\tilde Z_1}=\frac{Z_3}{\tilde Z_3}=\frac{Z_4}{\tilde Z_4},\;\frac{%
Z_1}{\bar Z_1}=\frac{Z_3}{\tilde Z_3}=\frac{Z_5}{\bar Z_5}.  \eqnum{9.40}
\end{equation}

\section{Effective coupling constant}

In this section, we plan to perform the one-loop renormalization of the
SU(2)-symmetric hadrodynamics by using the renormalization group approach.
As argued in our previous paper [14, 25, 26], when the renormalization is
carried out in the mass-dependent momentum space subtraction scheme, the
solutions to the RGEs satisfied by renormalized propagators and vertices can
be uniquely determined by the boundary conditions of the renormalized
propagators and vertices. In this case, an exact S-matrix element can be
written in the form as given in the tree-diagram approximation provided that
the coupling constant and particle masses in the matrix element are replaced
by their effective (running) ones which are given by solving their
renormalization group equations. Therefore, the task of renormalization is
reduced to find the solutions of the RGEs for the renormalized coupling
constant and particle masses. Suppose $F_R$ is a renormalized quantity. In
the multiplicative renormalization, it is related to the unrenormalized one $%
F$ in such a way 
\begin{eqnarray}
F=Z_FF_R  \eqnum{10.1}
\end{eqnarray}
where $Z_F$ is the renormalization constant of $F$. The $Z_F$ and $F_R$ are
all functions of the renormalization point $\mu =\mu _0e^t$ where $\mu _0$
is a fixed renormalization point corresponding the zero value of the group
parameter $t$. Differentiating Eq. (10.1) with respect to $\mu $ and
noticing that the $F$ is independent of $\mu $, we immediately obtain a
renormalization group equation (RGE) satisfied by the function $F_R$ [21-23] 
\begin{eqnarray}
\mu \frac{dF_R}{d\mu }+\gamma _FF_R=0  \eqnum{10.2}
\end{eqnarray}
where $\gamma _F$ is the anomalous dimension defined by 
\begin{eqnarray}
\gamma _F=\mu \frac d{d\mu }\ln Z_F.  \eqnum{10.3}
\end{eqnarray}
Since the renormalization constant is dimensionless, the anomalous dimension
can only depend on the ratio ${\beta =\frac{m_R}\mu }${\ where }$m_R$
denotes a renormalized mass and ${\gamma }_F{=\gamma }_F{(g}_R{,\beta )}$ in
which $g_R$ is the renormalized coupling constant and depends on $\mu $.
Since the renormalization point is a momentum taken to subtract the
divergence, we may set $\mu =\mu _0\lambda $ where $\lambda =e^t$ which will
be taken to be the same as in the scaling transformation of momentum $%
p=p_0\lambda $. In the above, $\mu _0$ and $p_0$ are the fixed
renormalization point and momentum respectively. When we set $F$ to be the
coupling constant $g$ and noticing $\mu \frac d{d\mu }=\lambda \frac d{%
d\lambda }$, one can write from Eq. (10.2) the RGE for the renormalized
coupling constant 
\begin{equation}
\lambda \frac{dg_R(\lambda )}{d\lambda }+\gamma _g(\lambda )g_R(\lambda )=0 
\eqnum{10.4}
\end{equation}
with 
\begin{equation}
\gamma _g=\mu \frac d{d\mu }\ln Z_g.  \eqnum{10.5}
\end{equation}
According to the definition in Eq. (10.1) and the relation in Eq. (7.25), we
may take, 
\begin{equation}
Z_g=\frac{\widetilde{Z}_1}{\widetilde{Z_3}Z_3^{\frac 12}}  \eqnum{10.6}
\end{equation}
to calculate the anomalous dimension. As denoted in Eqs. (4.25) and (5.24),
the renormalization constants $Z_3$, $\widetilde{Z_3}$ and $\widetilde{Z}_1$
are determined by the $\rho -$meson self-energy, the ghost particle
self-energy and the ghost particle -$\rho -$meson vertex correction,
respectively. At one-loop level, the $\rho -$meson self-energy is depicted
in Figs. (1a)-(1f), the ghost particle self-energy is shown in Fig. (2) and
the ghost vertex correction is represented in Figs. (3a) and (3b). According
to the Feynman rules listed in Appendix and noticing that the symmetry
factors of the diagrams in Figs. (1a), (1c), (1e) and (1f) are $1/2$ and the
symmetry factors of the other diagrams are $1$, the expressions of the
self-energies and the vertex correction are easily written out. For the
gluon one-loop self-energy denoted by $-i\Pi _{\mu \nu }^{ab}(k)$, one can
write 
\begin{equation}
\Pi _{\mu \nu }^{ab}(k)=\sum_{i=1}^6\Pi _{\mu \nu }^{(i)ab}(k)  \eqnum{10.7}
\end{equation}
where $\Pi _{\mu \nu )}^{(1)ab}(k)-\Pi _{\mu \nu )}^{(6)ab}(k)$ represent
the self-energies given in turn by Figs.(1a)-(1f). They are separately
represented in the following: 
\begin{equation}
\begin{array}{c}
\Pi _{\mu \nu }^{(1)ab}(k)=i\delta ^{ab}g^2\int \frac{d^4l}{(2\pi )^4}\frac{%
g^{\lambda \lambda ^{\prime }}g^{\rho \rho ^{\prime }}}{[l^2-m_\rho
^2+i\epsilon ][(l+k)^2-m_\rho ^2+i\epsilon ]}[g_{\mu \lambda }(l+2k)_\rho
-g_{\lambda \rho }(2l+k)_\mu \\ 
+g_{\rho \mu }(l-k)_\lambda ][g_{\nu \rho ^{\prime }}(l-k)_{\lambda ^{\prime
}}-g_{\lambda ^{\prime }\rho ^{\prime }}(2l+k)_\nu +g_{\lambda ^{^{\prime
}}\nu }(l+2k)_{\rho ^{\prime }}],
\end{array}
\eqnum{10.8}
\end{equation}
\begin{equation}
\Pi _{\mu \nu }^{(2)ab}(k)=-i\delta ^{ab}2g^2\int \frac{d^4l}{(2\pi )^4}%
\frac{(l+k)_\mu l_\nu }{[(l+k)^2-m_\rho ^2+i\epsilon ][l^2-m_\rho
^2+i\epsilon ]},  \eqnum{10.9}
\end{equation}
\begin{equation}
\Pi _{\mu \nu }^{(3)ab}(k)=-i\delta ^{ab}2g^2\int \frac{d^4l}{(2\pi )^4}%
\frac{g^{\lambda \rho }}{(l^2-m_\rho ^2+i\epsilon )}(g_{\mu \nu }g_{\lambda
\rho }-g_{\mu \rho }g_{\lambda \nu }),  \eqnum{10.10}
\end{equation}
\begin{equation}
\begin{array}{c}
\Pi _{\mu \nu }^{(4)ab}(k)=-i\delta ^{ab}\frac 12g^2\int \frac{d^4l}{(2\pi
)^4}\frac 1{[(l-k)^2-M^2+i\epsilon ][l^2-M^2+i\epsilon ]} \\ 
\times Tr[\gamma _\mu ({\bf l}-{\bf k}+M)\gamma _\nu ({\bf l}+M)],
\end{array}
\eqnum{10.11}
\end{equation}
\begin{equation}
\Pi _{\mu \nu }^{(5)ab}(k)=i\delta ^{ab}g^2\int \frac{d^4l}{(2\pi )^4}\frac{%
(2l+k)_\mu (2l+k)_\nu }{(l^2-m_\pi ^2+i\epsilon )[(k+l)^2-m_\pi ^2+i\epsilon
)},  \eqnum{10.12}
\end{equation}
and 
\begin{equation}
\Pi _{\mu \nu }^{(6)ab}(k)=-i\delta ^{ab}2g^2\int \frac{d^4l}{(2\pi )^4}%
\frac{g_{\mu \nu }}{(l^2-m_\pi ^2+i\epsilon )}  \eqnum{10.13}
\end{equation}
where ${\bf l=}\gamma ^\lambda l_\lambda $, ${\bf k=}\gamma ^\lambda
k_\lambda $ with $\gamma ^\lambda $ are the $8\times 8$ block-diagonal $%
\gamma -$matrices. In the above, $\varepsilon ^{acd}\varepsilon
^{bcd}=2\delta ^{ab}$ and $Tr(T^aT^b)=\frac 12\delta ^{ab}$ have been
considered. It should be noted that in writing Eqs. (10.8)-(10.10), we
choose to work in the Feynman gauge for simplicity. This choice is based on
the fact that the SU(2)-symmetric model of hadrodynamics, as a non-Abelian
gauge field theory, has been proved to be an unitary theory [27], that is to
say, the S-matrix elements evaluated from the model are independent of gauge
parameter. Therefore, we are allowed to choose a convenient gauge in the
calculation. From Eqs. (10.8)-(10.13), it is clearly seen that 
\begin{equation}
\Pi _{\mu \nu }^{ab}(k)=\delta ^{ab}\Pi _{\mu \nu }(k)=\delta
^{ab}\sum_{i=1}^6\Pi _{\mu \nu }^{(i)}(k).  \eqnum{10.14}
\end{equation}
By the dimensional regularization approach [28-32], the divergent integrals
over $l$ in Eqs. (10.8)-(10.13) can be regularized in a $n$-dimensional
space and thus are easily calculated. The results are 
\begin{equation}
\begin{array}{c}
\Pi _{\mu \nu }^{(1)}(k)=-\frac{g^2}{(4\pi )^2}\int_0^1dx\frac 1{\varepsilon
[k^2x(x-1)+m_\rho ^2]^\varepsilon }\{g_{\mu \nu }[11x(x-1) \\ 
+5)k^2+9m_\rho ^2]+2[5x(x-1)-1]k_\mu k_\nu \},
\end{array}
\eqnum{10.15}
\end{equation}
\begin{equation}
\begin{array}{c}
\Pi _{\mu \nu }^{(2)}(k)=\frac{g^2}{(4\pi )^2}\int_0^1dx\frac 1{\varepsilon
[k^2x(x-1)+m_\rho ^2]^\varepsilon }\{[k^2x(x-1) \\ 
+m_\rho ^2]g_{\mu \nu }+2x(x-1)k_\mu k_\nu \},
\end{array}
\eqnum{10.16}
\end{equation}
\begin{equation}
\Pi _{\mu \nu }^{(3)}(k)=\frac{6g^2}{(4\pi )^2}\frac{m_\rho ^2}\varepsilon
g_{\mu \nu },  \eqnum{10.17}
\end{equation}
\begin{equation}
\Pi _{\mu \nu }^{(4)}(k)=-\frac{8g^2}{(4\pi )^2}\int_0^1dx\frac{k^2x(x-1)}{%
\varepsilon [k^2x(x-1)+M^2]^\varepsilon }[g_{\mu \nu }-\frac{k_\mu k_\nu }{%
k^2}],  \eqnum{10.18}
\end{equation}
\begin{equation}
\begin{array}{c}
\Pi _{\mu \nu }^{(5)}(k)=-\frac{g^2}{(4\pi )^2}\int_0^1dx\frac 1{\varepsilon
[k^2x(x-1)+m_\pi ^2]^\varepsilon }\{2[k^2x(x-1)+m_\pi ^2]g_{\mu \nu } \\ 
+(1-2x)^2k_\mu k_\nu \}
\end{array}
\eqnum{10.19}
\end{equation}
and 
\begin{equation}
\Pi _{\mu \nu }^{(6)}(k)=\frac{2g^2}{(4\pi )^2}\frac{m_\pi ^2}\varepsilon
g_{\mu \nu }  \eqnum{10.20}
\end{equation}
where $\varepsilon =2-\frac n2\rightarrow 0$ when $n\rightarrow 4$. In Eqs.
(10.15)-(10.20), except for the $\varepsilon $ in the factor $1/\varepsilon
[k^2x(x-1)+m^2]^\varepsilon $ where $m=M,m_{\rho \text{ }}$ or $m_\pi $, we
have set $\varepsilon \rightarrow 0$ in the other factors and terms by the
consideration that this operation does not affect the calculated result of
the anomalous dimension. According to the decomposition shown in Eqs. (4.15)
and (4.16) and noticing $g_{\mu \nu }={\cal P}_T^{\mu \nu }+{\cal P}_L^{\mu
\nu }$, it is easy to get the transverse part of $\Pi _{\mu \nu }(k)$ from
Eqs. (10.15)-(10.20) and furthermore, based on the decomposition denoted in
Eq. (4.20), the functions $\Pi _1(k^2)$ and $\Pi _2(k^2)$ can be written
out. The results are 
\begin{equation}
\begin{array}{c}
\Pi _1(k^2)=-\frac{g^2}{(4\pi )^2}\int_0^1dx\{\frac{5[2x(x-1)+1]}{%
\varepsilon [k^2x(x-1)+m_\rho ^2]^\varepsilon }+\frac{8x(x-1)}{\varepsilon
[k^2x(x-1)+M^2]^\varepsilon } \\ 
+\frac{2x(x-1)}{\varepsilon [k^2x(x-1)+m_\pi ^2]^\varepsilon }\}
\end{array}
\eqnum{10.21}
\end{equation}
and 
\begin{equation}
\begin{array}{c}
\Pi _2(k^2)=-\frac{g^2}{(4\pi )^2}\{\int_0^1dx\frac 8{\varepsilon
[k^2x(x-1)+m_\rho ^2]^\varepsilon }+\frac{2m_\pi ^2/m_\rho ^2}{\varepsilon
[k^2x(x-1)+m_\pi ^2]^\varepsilon } \\ 
-\frac{2m_\pi ^2}{\varepsilon m_\rho ^2}-\frac 9\varepsilon \}.
\end{array}
\eqnum{10.22}
\end{equation}
It is clear that the both functions $\Pi _1(k^2)$ and $\Pi _2(k^2)$ are
divergent in the four-dimensional space-time. When the divergences are
subtracted in the mass-dependent momentum space subtraction scheme [30-33],
in accordance with the definition in Eq. (4.25), we immediately obtain from
the expression in Eq. (10.21) the one-loop renormalization constant $Z_3$ as
follows 
\begin{equation}
\begin{array}{c}
Z_3=1-\Pi _1(\mu ^2) \\ 
=1+\frac{g^2}{(4\pi )^2}\int_0^1dx\{\frac{5[2x(x-1)+1]}{\varepsilon [\mu
^2x(x-1)+m_\rho ^2]^\varepsilon }+\frac{8x(x-1)}{\varepsilon [\mu
^2x(x-1)+M^2]^\varepsilon } \\ 
+\frac{2x(x-1)}{\varepsilon [\mu ^2x(x-1)+m_\pi ^2]^\varepsilon }\}
\end{array}
\eqnum{10.23}
\end{equation}

Next, we turn to the ghost particle one-loop self-energy denoted by $%
-i\Omega ^{ab}(q)$. From Fig. (2), in the Feynman gauge, one can write 
\begin{equation}
\Omega ^{ab}(q)=i\delta ^{ab}2g^2\int \frac{d^4l}{(2\pi )^4}\frac{q\cdot
(q-l)}{[(q-l)^2-m_\rho ^2+i\epsilon ][l^2-m_\rho ^2+i\epsilon ]} 
\eqnum{10.24}
\end{equation}
By the dimensional regularization, it is easy to get 
\begin{equation}
\Omega ^{ab}(q)=\delta ^{ab}q^2\hat \Omega (q^2)  \eqnum{10.25}
\end{equation}
where 
\begin{equation}
\hat \Omega (q^2)=\frac{g^2}{(4\pi )^2}\int_0^1dx\frac{2(x-1)}{\varepsilon
[q^2x(x-1)+m_\rho ^2]^\varepsilon }  \eqnum{10.26}
\end{equation}
According to the definition given in Eq. (4.25) and the above expression ,
the one-loop renormalization constant of ghost particle propagator is of the
form 
\begin{equation}
\widetilde{Z}_3=1-\hat \Omega (\mu ^2)=1-\frac{g^2}{(4\pi )^2}\int_0^1dx%
\frac{2(x-1)}{\varepsilon [\mu ^2x(x-1)+m_\rho ^2]^\varepsilon } 
\eqnum{10.27}
\end{equation}

Now, let us discuss the ghost vertex renormalization. In the one-loop
approximation. the vertex defined by extracting out a coupling constant is
expressed as 
\begin{equation}
\widetilde{\Lambda }_\lambda ^{abc}(p,q)=\varepsilon ^{abc}p_\lambda
+\Lambda _{1\lambda }^{abc}(p,q)+\Lambda _{2\lambda }^{abc}(p,q) 
\eqnum{10.28}
\end{equation}
where the first term is the bare vertex, the second and the third terms
stand for the one-loop vertex corrections shown in Figs. (3a) and (3b)
respectively. In the Feynman gauge, the vertex corrections are expressed as 
\begin{equation}
\Lambda _{1\lambda }^{abc}(p,q)=-i\varepsilon ^{abc}g^2\int \frac{d^4l}{%
(2\pi )^4}\frac{p\cdot (q-l)(p-l)_\lambda }{[l^2-m_\rho ^2+i\epsilon
][(p-l)^2-m_\rho ^2+i\epsilon ][(q-l)^2-m_\rho ^2+i\epsilon ]}  \eqnum{10.29}
\end{equation}
and 
\begin{equation}
\Lambda _{2\lambda }^{abc}(p,q)=i\varepsilon ^{abc}g^2\int \frac{d^4l}{(2\pi
)^4}\frac{l\cdot (p-q-l)p_\lambda -p\cdot lq_\lambda +p\cdot
(2q-p+l)l_\lambda }{[l^2-m_\rho ^2+i\epsilon ][(p-l)^2-m_\rho ^2+i\epsilon
][(q-l)^2-m_\rho ^2+i\epsilon ]}  \eqnum{10.30}
\end{equation}
where $\varepsilon ^{acd}\varepsilon ^{ebf}\varepsilon ^{dfc}=-\varepsilon
^{abc}$ has been noted. By employing the dimensional regularization to
compute the above integrals, it is not difficult to get 
\begin{equation}
\Lambda _{1\lambda }^{abc}(p,q)=\varepsilon ^{abc}\frac{g^2}{(4\pi )^2}%
\int_0^1dx\int_0^1dy\{\frac{\frac 12yp_\lambda }{\varepsilon \Theta
_{xy}^\varepsilon }-\frac 1{\Theta _{xy}}[p_\lambda A_1(p,q)+q_\lambda
B_1(p,q)]-\frac 18p_\lambda \}  \eqnum{10.31}
\end{equation}
where 
\begin{equation}
\begin{array}{c}
\Theta _{xy}=p^2xy(xy-1)+q^2[(x-1)^2y+(x-1)]y-2p\cdot qx(x-1)y^2+m_\rho ^2,
\\ 
A_1(p,q)=\{p\cdot q[1+(x-1)y]-p^2xy\}(1-xy)y, \\ 
B_1(p,q)=\{p\cdot q[1+(x-1)y]-p^2xy\}(x-1)y^2
\end{array}
\eqnum{10.32}
\end{equation}
and 
\begin{equation}
\Lambda _{2\lambda }^{abc}(p,q)=\varepsilon ^{abc}\frac{g^2}{(4\pi )^2}%
\int_0^1dx\int_0^1dy\{\frac{\frac 32yp_\lambda }{\varepsilon \Theta
_{xy}^\varepsilon }+\frac 1{\Theta _{xy}}[p_\lambda A_2(p,q)+q_\lambda
B_2(p,q)]-\frac 38p_\lambda \}  \eqnum{10.33}
\end{equation}
where 
\begin{equation}
\begin{array}{c}
A_2(p,q)=\{p^2(2xy-x^2y^2-1)-q^2[(x-1)y-1](x-1)y \\ 
+p\cdot q[2-(3x-2)y+2x(x-1)y^2]\}y, \\ 
B_2(p,q)=[p\cdot q(x-1)-p^2x]y^2.
\end{array}
\eqnum{10.34}
\end{equation}
The divergences in the both vertices $\Lambda _{1\lambda }^{abc}(p,q)$ and $%
\Lambda _{2\lambda }^{abc}(p,q)$ may be subtracted at the renormalization
point $p^2=q^2=\mu ^2$ which implies $k=p-q=0$, being consistent with the
momentum conservation held at the vertices. Upon substituting Eqs. (10.31)
and (10.33) in Eq. (10.28), at the renormalization point, one can get 
\begin{equation}
\widetilde{\Lambda }_\lambda ^{abc}(p,q)\mid _{p^2=q^2=\mu ^2}=\varepsilon
^{abc}p_\lambda (1+\widetilde{L}_1)=\widetilde{Z}_1^{-1}\varepsilon
^{abc}p_\lambda  \eqnum{10.35}
\end{equation}
where 
\begin{equation}
\widetilde{Z}_1=1-\widetilde{L}_1=1-\frac{g^2}{(4\pi )^2}\int_0^1dx\{\frac{2x%
}{\varepsilon [\mu ^2x(x-1)+m_\rho ^2]^\varepsilon }-\frac{2x^2(x-1)\mu ^2}{%
\mu ^2x(x-1)+m_\rho ^2}-\frac 12\}  \eqnum{10.36}
\end{equation}
which is the one-loop renormalization constant of the ghost vertex.

Now we are ready to calculate the anomalous dimension $\gamma _g(\lambda )$.
Substituting the expressions in Eqs. (10.6), (10.23), (10.27) and (10.36)
into Eq. (10.5), it is easy to find an analytical expression of the
anomalous dimension $\gamma _g(\lambda )$. When we set $\frac{m_\rho }\mu =%
\frac \rho \lambda $ , $\frac{m_\pi }\mu =\frac \sigma \lambda $ and $\frac M%
\mu =\frac \beta \lambda $ with defining $\rho =\frac{m_\rho }\Lambda $, $%
\sigma =\frac{m_\pi }\Lambda $ and $\beta =\frac M\Lambda $ (here we have
set $\mu _0\equiv \Lambda $), the expression of $\gamma _g(\lambda )$, in
the approximation of order $g^2$, is given by 
\begin{equation}
\gamma _g(\lambda )=\tilde \gamma _1(\lambda )-\tilde \gamma _3(\lambda )-%
\frac 12\gamma _3(\lambda )  \eqnum{10.37}
\end{equation}
where 
\begin{equation}
\tilde \gamma _1(\lambda )=\lim\limits_{\varepsilon \rightarrow 0}\mu \frac d%
{d\mu }\ln \widetilde{Z}_1=\frac{\alpha _R}{2\pi }[1+\frac{2\rho ^2}{\lambda
^2-4\rho ^2}-\frac{4\rho ^4}{\lambda (\lambda ^2-4\rho ^2)}I(\lambda ,\rho
)],  \eqnum{10.38}
\end{equation}
\begin{equation}
\tilde \gamma _3(\lambda )=\lim\limits_{\varepsilon \rightarrow 0}\mu \frac d%
{d\mu }\ln \widetilde{Z}_3=-\frac{\alpha _R}{2\pi }[1+\frac{2\rho ^2}\lambda
I(\lambda ,\rho )]  \eqnum{10.39}
\end{equation}
and 
\begin{equation}
\begin{array}{c}
\gamma _3(\lambda )=\lim\limits_{\varepsilon \rightarrow 0}\mu \frac d{d\mu }%
\ln Z_3 \\ 
=-\frac{\alpha _R}\pi \{\frac 53-\frac{5\rho ^2}{\lambda ^2}+\frac 5{\lambda
^3}\rho ^2(\lambda ^2-2\rho ^2)I(\lambda ,\rho )-\frac 16[1+\frac{6\sigma ^2%
}{\lambda ^2} \\ 
+\frac{12\sigma ^4}{\lambda ^3}I(\lambda ,\sigma )]-\frac 23[1+\frac{6\beta
^2}{\lambda ^2}+\frac{12\beta ^4}{\lambda ^3}I(\lambda ,\beta )]\}
\end{array}
\eqnum{10.40}
\end{equation}
here $\alpha _R=g_R^2/4\pi $ and 
\begin{equation}
\begin{array}{c}
I(\lambda ,a)=\frac 1{\sqrt{\lambda ^2-4a^2}}\ln \frac{\lambda +\sqrt{%
\lambda ^2-4a^2}}{\lambda -\sqrt{\lambda ^2-4a^2}} \\ 
=\{ 
\begin{array}{c}
\frac 2{\sqrt{4a^{2-}\lambda ^2}}\cot ^{-1}\frac \lambda {\sqrt{%
4a^{2-}\lambda ^2}},\text{ }if\text{ }\lambda \leq 2a \\ 
\frac 2{\sqrt{\lambda ^2-4a^2}}\coth ^{-1}\frac \lambda {\sqrt{\lambda
^2-4a^2}},\text{ }if\text{ }\lambda \geq 2a
\end{array}
\end{array}
\eqnum{10.41}
\end{equation}
with $a=\rho ,\sigma $ or $\beta $. With the expressions given in Eqs.
(10.38)-(10.40), the $\gamma _g(\lambda )$ can be represented as 
\begin{equation}
\begin{array}{c}
\gamma _g(\lambda )=\frac{\alpha _R}{4\pi }\{\frac{19}3-\frac{10\rho ^2}{%
\lambda ^2}+\frac{\lambda ^2}{(\lambda ^2-4\rho ^2)}+(8-\frac{10\rho ^2}{%
\lambda ^2}-\frac{\lambda ^2}{\lambda ^2-4\rho ^2})\frac{2\rho ^2}\lambda
I(\lambda ,\rho ) \\ 
-\frac 13[1+\frac{6\sigma ^2}{\lambda ^2}+\frac{12\sigma ^4}{\lambda ^3}%
I(\lambda ,\sigma )]-\frac 43[1+\frac{6\beta ^2}{\lambda ^2}+\frac{12\beta ^4%
}{\lambda ^3}I(\lambda ,\beta ]\}
\end{array}
\eqnum{10.42}
\end{equation}
We would like to note that the fixed renormalization point $\Lambda $ in $%
\rho $, $\sigma $ and $\beta $ can be taken arbitrarily. For example, the $%
\Lambda $ may be chosen to be the mass of nucleon . In this case, $\beta =1$%
, $\rho =m_\rho /M$ and $\sigma =m_\pi /M$. Apparently, $\beta =1$ implies $%
\lambda =\sqrt{p^2/M^2}.$ In practice, the $\Lambda $ will be treated as a
scaling parameter of renormalization.

With the $\gamma _g(\lambda )$ given above, the equation in Eq. (10.4) can
be solved to give the effective coupling constant as follows 
\begin{equation}
\alpha _R(\lambda )=\frac{\alpha _R}{1+\frac{\alpha _R}{2\pi }G(\lambda )} 
\eqnum{10.43}
\end{equation}
where $\alpha _R=\alpha _R(1)$ and 
\begin{equation}
\begin{array}{c}
G(\lambda )=\int_1^\lambda \frac{d\lambda }\lambda \frac{4\pi \gamma
_g(\lambda )}{\alpha _R}=\frac 23[\varphi _1(\lambda ,\rho )-\varphi
_1(1,\rho )] \\ 
-\frac 13[\varphi _2(\lambda ,\sigma )-\varphi _2(1,\sigma )]-\frac 43%
[\varphi _2(\lambda ,\beta )-\varphi _2(1,\beta )]
\end{array}
\eqnum{10.44}
\end{equation}
in which 
\begin{equation}
\varphi _1(\lambda ,\rho )=\frac{5\rho ^2}{\lambda ^2}+[(\frac{19}2-\frac{%
5\rho ^2}{\lambda ^2})\frac{(\lambda ^2-4\rho ^2)}{2\lambda }+\frac{3\lambda 
}4]I(\lambda ,\rho ),  \eqnum{10.45}
\end{equation}
\begin{equation}
\varphi _2(\lambda ,\sigma )=-\frac{2\sigma ^2}{\lambda ^2}+(1+\frac{2\sigma
^2}{\lambda ^2})\frac{(\lambda ^2-4\sigma ^2)}{2\lambda }I(\lambda ,\sigma )
\eqnum{10.46}
\end{equation}
and 
\begin{equation}
\varphi _2(\lambda ,\beta )=-\frac{2\beta ^2}{\lambda ^2}+(1+\frac{2\beta ^2%
}{\lambda ^2})\frac{(\lambda ^2-4\beta ^2)}{2\lambda }I(\lambda ,\beta ). 
\eqnum{10.47}
\end{equation}
here $\varphi _1(\lambda ,\rho )$ arises from the $\rho -$meson
self-interaction, $\varphi _2(\lambda ,\sigma )$ and $\varphi _2(\lambda
,\beta )$ come from the interaction between pion and $\rho -$meson and the
one between nucleon and $\rho $-meson, respectively.

In the large momentum limit ($\lambda \rightarrow \infty $), we have 
\begin{equation}
G(\lambda )=\frac{17}3\ln \lambda .  \eqnum{10.48}
\end{equation}
Therefore, in the limit mentioned above, we have $\alpha _R(\lambda
)\rightarrow 0$. This exhibits that the interaction given by the
SU(2)-symmetric model is also of the asymptotically free behavior. It should
be noted that the expressions in Eqs. (10.42) and (10.45)-(10.47) are
obtained at the timelike subtraction point where the $\lambda $ is a real
variable. We may also take spacelike momentum subtraction. For this kind of
subtraction, corresponding to $\mu \rightarrow i\mu $, the variable $\lambda 
$ in Eqs. (10.42), (10.45)- (10.47) should be replaced by $i\lambda $ where $%
\lambda $ is still a real variable. In this case, the function in Eq.
(10.41) will be replaced by 
\begin{equation}
\begin{array}{c}
\tilde I(\lambda ,a)=\frac 1{\sqrt{\lambda ^2+4a^2}}\ln \frac{\sqrt{\lambda
^2+4a^2}+\lambda }{\sqrt{\lambda ^2+4a^2}-\lambda } \\ 
=\frac 2{\sqrt{\lambda ^2+4a^2}}\tanh ^{-1}\frac \lambda {\sqrt{\lambda
^2+4a^2}}
\end{array}
\eqnum{10.39}
\end{equation}
It is easy to see that the function in Eq. (10.48) is the same for the both
subtractions.

The behavior of the function $\alpha _R(\lambda )$ is graphically described
in Fig. (4). In our test, we find that the behavior of the $\alpha
_R(\lambda )$ sensitively depends on the choice of the constant $\alpha _R$
and the scaling parameter $\Lambda .$ In this paper, we only take $\alpha
_R=0.5$ and $\Lambda =M$ as an illustration. Figs. (4a) and (4b) represent
respectively the effective coupling constants obtained at the timelike
subtraction point and the spacelike subtraction point. To exhibit the
effects of the $\rho $-meson self-interaction, the pion-$\rho $-meson
interaction and the nucleon-$\rho $-meson interaction on the effective
coupling constant, in each figure, besides the total effective coupling
constant, we also separately show the effective coupling constants given by
the functions $\varphi _1(\lambda ,\rho )$, $\varphi _2(\lambda ,\sigma )$
and $\varphi _2(\lambda ,\beta )$. These effective coupling constants are
represented by the solid, dotted and dashed lines respectively in the
subfigures within Figs. (4a) and (4b). Fig. (4a) shows that the effective
coupling constant given by the timelike momentum subtraction has a peak with
the maximum value $\alpha _R(\lambda )_{\max }=1.49222\alpha _R$ at $\lambda
=1.5303$. When $\lambda \rightarrow 0$, the $\alpha _R(\lambda )$ abruptly
falls down to zero, while, when $\lambda $ goes to infinity, the $\alpha
_R(\lambda )$ rather smoothly decreases from its maximum and tends to zero.
Fig. (4b) tells us that the effective coupling constant given by the
spacelike momentum subtraction has a different behavior in the low-energy
region. This coupling constant keeps almost a constant near the value of $%
\alpha _R$ in the region [0,1] of $\lambda $ and then decreases and tends to
zero with the growth of $\lambda $. From the subfigures, it is clearly seen
that at one-loop level, only the $\rho -$meson self-interaction is of the
behavior of asymptotic freedom, while the interactions between nucleon and $%
\rho -$meson and between pion and $\rho -$meson have no such a behavior. The
asymptotically free behavior of the total effective coupling implies that
the $\rho -$meson self-interaction plays an overwhelming role for the
one-loop interaction.

\section{Effective meson masses}

In this section, we proceed to derive the one-loop effective $\rho -$meson
and pion masses. First, we derive the $\rho $-meson effective mass. Setting $%
F_R=m_\rho ^R$ in Eq. (10.2), we have the RGE for the renormalized $\rho $%
-meson mass 
\begin{equation}
\lambda \frac{dm_\rho ^R(\lambda )}{d\lambda }+\gamma _\rho (\lambda )m_\rho
^R(\lambda )=0  \eqnum{11.1}
\end{equation}
where 
\begin{equation}
\gamma _{m_\rho }(\lambda )=\mu \frac d{d\mu }\ln Z_{m_\rho }.  \eqnum{11.2}
\end{equation}
From the last equality in Eq. (4.25) and Eqs. (10.21) and (10.22), in the
approximation of order $g^2$, we can write 
\begin{equation}
\begin{array}{c}
Z_{m_\rho }=1+\frac 12[\Pi _1(\mu ^2)+\Pi _2(\mu ^2)] \\ 
=1-\frac{g^2}{(4\pi )^2}\int_0^1dx\{\frac{5x(x-1)+13/2}{4\varepsilon
[k^2x(x-1)+m_\rho ^2]^\varepsilon }+\frac{4x(x-1)}{\varepsilon
[k^2x(x-1)+M^2]^\varepsilon } \\ 
+\frac{x(x-1)+m_\pi ^2/m_\rho ^2}{\varepsilon [k^2x(x-1)+m_\pi
^2]^\varepsilon }-\frac 1\varepsilon (m_\pi ^2/m_\rho ^2+\frac 92)\}.
\end{array}
\eqnum{11.3}
\end{equation}
On inserting Eq. (11.3) into Eq. (11.2) and completing the differentiation
with respect to $\mu $ and the integration over $x$, we find 
\begin{equation}
\begin{array}{c}
\gamma _{m_\rho }(\lambda )=\frac{g_R^2}{4\pi ^2}\{\frac{29}{12}+\frac{%
\sigma ^2}{2\rho ^2}-\frac 1{2\lambda ^2}(5\rho ^2+\sigma ^2+4\beta ^2) \\ 
+\frac{\rho ^2}{2\lambda ^3}(13\lambda ^2-10\rho ^2)I(\lambda ,\rho )+\frac{%
\sigma ^4}{\lambda ^3\rho ^2}(\lambda ^2-\rho ^2)I(\lambda ,\sigma ) \\ 
-\frac{4\beta ^4}{\lambda ^3}I(\lambda ,\beta )\}.
\end{array}
\eqnum{11.4}
\end{equation}
where the functions $I(\lambda ,\cdot \cdot \cdot )$ were defined in Eq.
(10.41). With this anomalous dimension, the RGE in Eq. (11.1) can be solved
to give an effective $\rho $-meson mass such that 
\begin{equation}
m_\rho ^R(\lambda )=m_\rho ^Re^{-S_\rho (\lambda )}  \eqnum{11.5}
\end{equation}
where $m_\rho ^R=m_\rho ^R(1)$ and 
\begin{equation}
S_\rho (\lambda )=\int_1^\lambda \frac{d\lambda }\lambda \gamma _{m_\rho
}(\lambda ).  \eqnum{11.6}
\end{equation}
In general, the coupling constant $g_R$ in Eq. (11.5) may be taken to be the
effective one. If the coupling constant is taken to be the constant $g_R$,
the function $S_\rho (\lambda )$ can be explicitly represented as 
\begin{equation}
\begin{array}{c}
S_\rho (\lambda )=\frac{\alpha _R}{6\pi }\{A_1(\lambda ,\rho )-A_1(1,\rho
)+A_2(\lambda ,\sigma )-A_2(1,\sigma ) \\ 
+A_3(\lambda ,\beta )-A_3(1,\beta )\}
\end{array}
\eqnum{11.7}
\end{equation}
where 
\begin{equation}
A_1(\lambda ,\rho )=\frac{5\rho ^2}{\lambda ^2}+(17-\frac{5\rho ^2}{\lambda
^2})\frac{(\lambda ^2-4\rho ^2)}{2\lambda }I(\lambda ,\rho ),  \eqnum{11.8}
\end{equation}
\begin{equation}
A_2(\lambda ,\sigma )=\frac{\sigma ^2}{\lambda ^2}-(1-\frac{6\sigma ^2}{\rho
^2}+\frac{2\sigma ^2}{\lambda ^2})\frac{(\lambda ^2-4\sigma ^2)}{4\lambda }%
I(\lambda ,\sigma )  \eqnum{11.9}
\end{equation}
and 
\begin{equation}
A_3(\lambda ,\beta )=\frac{4\beta ^2}{\lambda ^2}-(1+\frac{2\beta ^2}{%
\lambda ^2})\frac{(\lambda ^2-4\beta ^2)}\lambda I(\lambda ,\beta ). 
\eqnum{11.10}
\end{equation}
In the large momentum limit ($\lambda \rightarrow \infty $), we have 
\begin{equation}
S_\rho (\lambda )\rightarrow \frac{\alpha _R}\pi (\frac{29}2+3\frac{m_\pi ^2%
}{m_\rho ^2})\ln \lambda ,  \eqnum{11.11}
\end{equation}
therefore, 
\begin{equation}
\lim\limits_{\lambda \rightarrow \infty }m_\rho ^R(\lambda )=0. 
\eqnum{11.12}
\end{equation}
which exhibits the asymptotically free behavior.

The behavior of the effective $\rho -$meson mass $m_\rho ^R(\lambda )$ in
the whole range of momenta is displayed in Fig. (5) where the solid line
represents the effective mass given by the timelike momentum subtraction and
the dashed line represents the one obtained by the spacelike momentum
subtraction. In comparison of Fig. (5) with Fig. (4), one can see that the
behaviors of the effective masses are much similar to the behaviors of the
corresponding coupling constants. Saying concretely, the timelike momentum
effective mass has a peak with the maximum $m_\rho ^R(\lambda )_{\max
}=1.06674m_\rho ^R$ at $\lambda =1.16675$ and also abruptly falls down to
zero when $\lambda \rightarrow 0$ and rather smoothly decreases from its
maximum and tends to zero when $\lambda $ goes to infinity. While, the
spacelike momentum effective mass almost behaves as a constant near the
value $m_\rho ^R$ in the region [0,1] of $\lambda $ and then decreases and
tends to zero with the growth of $\lambda $.

Next, we turn to the effective pion mass. With setting $F_R=m_\pi ^R$ in Eq.
(10.2), we can write the RGE for the renormalized pion mass 
\begin{equation}
\lambda \frac{dm_\pi ^R(\lambda )}{d\lambda }+\gamma _{m_\pi }(\lambda
)m_\pi ^R(\lambda )=0  \eqnum{11.13}
\end{equation}
where 
\begin{equation}
\gamma _{m_\pi }(\lambda )=\mu \frac d{d\mu }\ln Z_{m_\pi }.  \eqnum{11.14}
\end{equation}
From Eq. (8.21), the one-loop renormalization constant $Z_{m_\pi }$ can be
written as 
\begin{equation}
Z_{m_\pi }=1-\frac 12[\omega _1(\mu ^2)+\omega _2(\mu ^2)]  \eqnum{11.15}
\end{equation}
where $\omega _1(\mu ^2)$ and $\omega _2(\mu ^2)$ are contributed from the
one-loop self-energies $-i\Omega _1^{ab}(k)$ and $-i\Omega _2^{ab}(k)$ as
depicted in Figs. (6a) and (6b) respectively. According to the Feynman rules
shown in Appendix, in the Feynman gauge, one can write 
\begin{equation}
\Omega _1^{ab}(k)=\delta ^{ab}\Omega _1(k)=-i\delta ^{ab}2g^2\int \frac{d^4l%
}{(2\pi )^4}\frac{(l-2k)^2}{[(k-l)^2-m_\pi ^2+i\epsilon ][l^2-m_\rho
^2+i\epsilon ]}  \eqnum{11.16}
\end{equation}
and 
\begin{equation}
\Omega _2^{ab}(k)=\delta ^{ab}\Omega _2(k)=-i\delta ^{ab}16g^2\int \frac{d^4l%
}{(2\pi )^4}\frac{l^2-k\cdot l-M^2}{[(k-l)^2-M^2+i\epsilon
][l^2-M^2+i\epsilon ]}  \eqnum{11.17}
\end{equation}
where $\varepsilon ^{acd}\varepsilon ^{bcd}=2\delta ^{ab}$, $Tr(\tau ^a\tau
^b)=2\delta ^{ab}$ and $Tr[\gamma _5(\not l-\not k+M)\gamma _5(\not
l+M)]=8(k\cdot l-l^2+M^2)$ have been used in writing the above expressions.
The integrals in Eqs. (11.16) and (11.17) are divergent. They can easily be
calculated in the dimensional regularization scheme. The results are 
\begin{equation}
\Omega _1(k)=\frac{2g^2}{(4\pi )^2}\int_0^1dx\frac{k^2(3x^2-6x+4)+2[(m_\pi
^2-m_\rho ^2)x+m_\rho ^2]}{\varepsilon [k^2x(x-1)+m_\pi ^2+m_\rho
^2(1-x)]^\varepsilon }  \eqnum{11.18}
\end{equation}
and 
\begin{equation}
\Omega _2(k)=\frac{16g^2}{(4\pi )^2}\int_0^1dx\frac{3k^2x(x-1)+M^2}{%
\varepsilon [k^2x(x-1)+M^2]^\varepsilon }  \eqnum{11.19}
\end{equation}
Substituting $\Omega _\pi (k)=\Omega _1(k)+\Omega _2(k)$ with $\Omega _1(k)$
and $\Omega _2(k)$ given above into Eq. (8.17), we find 
\begin{equation}
\begin{array}{c}
\omega _1(\mu ^2)=\frac{2g^2}{(4\pi )^2}\int_0^1dx\{\frac{3x^2-6x+4}{%
\varepsilon [\mu ^2x(x-1)+(m_\pi ^2-m_\rho ^2)x+m_\rho ^2]^\varepsilon } \\ 
+\frac{24x(x-1)}{\varepsilon [\mu ^2x(x-1)+M^2]^\varepsilon }\}
\end{array}
\eqnum{11.20}
\end{equation}
and 
\begin{equation}
\begin{array}{c}
\omega _2(\mu ^2)=\frac{2g^2}{(4\pi )^2}\int_0^1dx\{\frac{2[(1-m_\rho
^2/m_\pi ^2)x+m_\rho ^2/m_\pi ^2]}{\varepsilon [\mu ^2x(x-1)+(m_\pi
^2-m_\rho ^2)x+m_\rho ^2]^\varepsilon } \\ 
+\frac{8M^2/m_\pi ^2}{\varepsilon [\mu ^2x(x-1)+M^2]^\varepsilon }\}
\end{array}
\eqnum{11.21}
\end{equation}
where we have set $k^2=\mu ^2.$ On inserting Eqs. (11.20) and (11.21) into
Eq. (11.15), we obtain an explicit expression of the renormalization
constant $Z_{m_\pi }$. Substituting such a renormalization constant in Eq.
(11.14), through a lengthy derivation, we get 
\begin{equation}
\begin{array}{c}
\gamma _{m_\pi }(\lambda )=\frac{\alpha _R}{2\pi }\{\xi _1(\lambda )+\xi
_2(\lambda )\ln \frac \sigma \rho +\xi _3(\lambda )J(\lambda ;\rho ,\sigma )
\\ 
+\xi _4(\lambda )+\xi _5(\lambda )I(\lambda ,\beta )\}
\end{array}
\eqnum{11.22}
\end{equation}
where 
\begin{equation}
\xi _1(\lambda )=\frac 1{2\sigma ^2\lambda ^4}[6\sigma ^2(\rho ^2-\sigma
^2)^2-(4\rho ^4+\sigma ^4+\rho ^2\sigma ^2)\lambda ^2+2(\rho ^2+3\sigma
^2)\lambda ^4],  \eqnum{11.23}
\end{equation}
\begin{equation}
\begin{array}{c}
\xi _2(\lambda )=\frac 1{\sigma ^2\lambda ^6}[3\sigma ^2(\rho ^2-\sigma
^2)^3-2(\rho ^2-\sigma ^2)(\rho ^4 \\ 
+\sigma ^4+\rho ^2\sigma ^2)\lambda ^2+(2\rho ^4-3\sigma ^4+4\rho ^2\sigma
^2)\lambda ^4],
\end{array}
\eqnum{11.24}
\end{equation}
\begin{equation}
\begin{array}{c}
\xi _3(\lambda )=\frac 1{2\sigma ^2\lambda ^6}[3\sigma ^2(\rho ^2-\sigma
^2)^4-(\rho ^2-\sigma ^2)^2(2\rho ^4+5\sigma ^4+5\rho ^2\sigma ^2)\lambda ^2
\\ 
+(4\rho ^6+5\sigma ^6+6\rho ^4\sigma ^2-9\rho ^2\sigma ^4)\lambda ^4-(2\rho
^4+3\sigma ^4+4\rho ^2\sigma ^2)\lambda ^6],
\end{array}
\eqnum{11.25}
\end{equation}
\begin{equation}
\xi _4(\lambda )=\frac 4{\lambda ^2\sigma ^2}[(2\beta ^2-\sigma ^2)\lambda
^2-6\beta ^2\sigma ^2],  \eqnum{11.26}
\end{equation}
, 
\begin{equation}
\xi _5(\lambda )=\frac{16\beta ^4}{\lambda ^3\sigma ^2}(\lambda ^2-3\sigma
^2),  \eqnum{11.27}
\end{equation}
\begin{equation}
J(\lambda ;\rho ,\sigma )=\frac 1{\sqrt{K(\lambda ;\rho ,\sigma )}}\ln \frac{%
\lambda ^2-(\rho ^2+\sigma ^2)-\sqrt{K(\lambda ;\rho ,\sigma )}}{\lambda
^2-(\rho ^2+\sigma ^2)+\sqrt{K(\lambda ;\rho ,\sigma )}}  \eqnum{11.28}
\end{equation}
in which 
\begin{equation}
K(\lambda ;\rho ,\sigma )=\lambda ^4-2(\rho ^2+\sigma ^2)\lambda ^2+(\rho
^2-\sigma ^2)^2  \eqnum{11.29}
\end{equation}
and $I(\lambda ,\beta )$ was defined in Eq. (10.41).

With the anomalous dimension given above, the one-loop effective pion mass
will be obtained by solving the RGE in Eq. (11.13). The result is 
\begin{equation}
m_\pi ^R(\lambda )=m_\pi ^Re^{-S_\pi (\lambda )}  \eqnum{11.30}
\end{equation}
where 
\begin{equation}
S_\pi (\lambda )=\int_1^\lambda \frac{d\lambda }\lambda \gamma _{m_\pi
}(\lambda )  \eqnum{11.31}
\end{equation}
If the coupling constant in Eq. (11.22) is taken to be a constant, the
integral over $\lambda $ in Eq. (11.31) can partly be calculated
analytically, giving the result as follows 
\begin{equation}
\begin{array}{c}
S_\pi (\lambda )=\frac{\alpha _R}{2\pi }\{B_1(\lambda )-B_1(1)+B_2(\lambda
)-B_2(1) \\ 
+B_3(\lambda )-B_3(1)+B_4(\lambda )\}
\end{array}
\eqnum{11.32}
\end{equation}
where 
\begin{equation}
B_1(\lambda )=\frac{8\beta ^2}{\lambda ^2}-2(1-\frac{2\beta ^2}{\sigma ^2}+%
\frac{2\beta ^2}{\lambda ^2})\frac{(\lambda ^2-4\beta ^2)}\lambda I(\lambda
,\beta ),  \eqnum{11.33}
\end{equation}
\begin{equation}
\begin{array}{c}
B_2(\lambda )=\frac 1{4\sigma ^2}\{4(\rho ^2+3\sigma ^2)\ln \lambda +\frac 1{%
\lambda ^4}[(4\rho ^4 \\ 
+\sigma ^4+\rho ^2\sigma ^2)\lambda ^2-3\sigma ^2(\rho ^2-\sigma ^2)^2]\},
\end{array}
\eqnum{11.34}
\end{equation}
\begin{equation}
\begin{array}{c}
B_3(\lambda )=-\frac 1{2\sigma ^2\lambda ^6}\ln \frac \sigma \rho [(2\rho
^4-3\sigma ^4+4\rho ^2\sigma ^2)\lambda ^4 \\ 
-(\rho ^2-\sigma ^2)(\rho ^4+\sigma ^4+\rho ^2\sigma ^2)\lambda ^2+\sigma
^2(\rho ^2-\sigma ^2)^3]
\end{array}
\eqnum{11.35}
\end{equation}
and 
\begin{equation}
B_4(\lambda )=\int_1^\lambda \frac{d\lambda }\lambda \xi _3(\lambda
)J(\lambda ;\rho ,\sigma ).  \eqnum{11.36}
\end{equation}
The integral in Eq. (11.36) with the functions $\xi _3(\lambda )$ and $%
J(\lambda ;\rho ,\sigma )$ given respectively in Eqs. (11.25) and (11.28)
can only be evaluated numerically.

The effective pion mass $m_\pi ^R(\lambda )$ is graphically represented in
Fig. (7). In the figure, the solid line and the dashed one represent the
effective mass given by the timelike momentum subtraction and the spacelike
momentum subtraction, respectively. The figure indicates that the effective
mass given by the timelike momentum subtraction has a sharp peak around $%
\lambda =0.43057$ at which we have a maximum $m_\pi ^R(\lambda )_{\max
}=24.0411$. Departing from the maximum, the effective mass rapidly falls to
zero when either $\lambda \rightarrow 0$ or $\lambda \rightarrow \infty $.
In contrast, the effective mass given by the spacelike momentum subtraction
has a low peak around $\lambda =0.49384$ at which there is a maximum $m_\pi
^R(\lambda )_{\max }=1.49159.$ From the maximum, the effective mass smoothly
tends to zero.

\section{Effective nucleon mass}

Before deriving the one-loop effective nucleon mass, we need first to
discuss the subtraction of the nucleon one-loop self-energy on the basis of
the W-T identity represented in Eq. (7.16). For later convenience, the
identity in Eq. (7.16) will be given in another form. Introducing new vertex
functions $\hat \Lambda ^{a\mu }(p,q)$ and $\hat \gamma _i^a(p,q)$ defined
by 
\begin{equation}
\begin{array}{c}
\Gamma ^{a\mu }(p,q,k)=(2\pi )^4\delta ^4(p-q+k)ig\hat \Lambda ^{a\mu }(p,q)
\\ 
\gamma _i^a(p,q,k)=-(2\pi )^4\delta ^4(p-q+k)\hat \gamma _i^a(p,q)
\end{array}
\eqnum{12.1}
\end{equation}
where $i=1,2$ and $\hat \gamma _i^a(p,q)=-\widetilde{\gamma }_i^a(p,q)$ and
considering $k=q-p$, Eq. (7.16) can be rewritten as 
\begin{eqnarray}
(p-q)_\mu \hat \Lambda ^{a\mu }(p,q)=\chi (k^2)[S_F^{-1}(p)\hat \gamma
_2^a(p,q)-\hat \gamma _1^a(p,q)S_F^{-1}(q)].  \eqnum{12.2}
\end{eqnarray}
From the perturbative calculation, it can be found that in the lowest order
of perturbation, we have 
\begin{equation}
\begin{array}{c}
\hat \Lambda _\mu ^{(0)a}(p,q)=\gamma _\mu T^a, \\ 
\hat \gamma _1^{(0)a}(p,q)=\hat \gamma _2^{(0)a}(p,q)=T^a.
\end{array}
\eqnum{12.3}
\end{equation}
In the one-loop approximation, the nucleon-gluon vertex denoted by $\hat 
\Lambda _\mu ^{(1)a}(p,q)$ is of order $g^2$. The nucleon-ghost vertex
functions $\hat \gamma _i^{(1)a}(p,q)$ ($i=1,2$) are contributed from Figs.
(8a) and (8b) and can be represented as 
\begin{equation}
\hat \gamma _i^{(1)a}(p,q)=T^aK_i(p,q)  \eqnum{12.4}
\end{equation}
where 
\begin{equation}
K_1(p,q)=ig^2\int \frac{d^4l}{(2\pi )^4}\gamma ^\mu S_F(l)(q-l)^\nu D_{\mu
\nu }(p-l)\Delta (q-l)  \eqnum{12.5}
\end{equation}
and 
\begin{equation}
K_2(p,q)=ig^2\int \frac{d^4l}{(2\pi )^4}S_F(l)\gamma ^\mu D_{\mu \nu
}(q-l)(p-l)^\nu \Delta (p-l).  \eqnum{12.6}
\end{equation}
It is clear that the above functions are logarithmically divergent. In the
one-loop approximation, the function $\chi (k^2)$ can be written as $\chi
(k^2)=1-\hat \Omega ^{(1)}(k^2)$ where the one-loop ghost particle
self-energy $\hat \Omega ^{(1)}(k^2)$ was represented in Eq. (10.24). Thus,
up to the order of $g^2$, with setting $\hat \Lambda _\mu
^{(1)a}(p,q)=T^a\Lambda _\mu ^{(1)}(p,q)$, we can write 
\begin{equation}
\hat \Lambda _\mu ^a(p,q)=T^a[\gamma _\mu +\Lambda _\mu ^{(1)}(p,q)] 
\eqnum{12.7}
\end{equation}
and 
\begin{equation}
\chi (k^2)\hat \gamma _i^a(p,q)=T^a[1+I_i(p,q)]  \eqnum{12.8}
\end{equation}
where 
\begin{equation}
I_i(p,q)=K_i(p,q)-\Omega ^{(1)}(k^2).  \eqnum{12.9}
\end{equation}
Upon substituting Eqs. (12.7) and (12.8) and the inverse of the nucleon
propagator denoted in Eq. (7.17) into Eq. (12.2), then differentiating the
both sides of Eq. (12.2) with respect to $p^\mu $ and finally setting $q=p$,
in the order of $g^2$, we get 
\begin{equation}
\overline{\Lambda }_\mu (p,p)=-\frac{\partial \Sigma (p)}{\partial p^\mu } 
\eqnum{12.10}
\end{equation}
where 
\begin{equation}
\begin{array}{c}
\overline{\Lambda }_\mu (p,p)=\Lambda _\mu ^{(1)}(p,p)-\gamma _\mu I_2(p,p)-(%
{\bf p}-M)\frac{\partial I_2(p,q)}{\partial p^\mu }\mid _{q=p} \\ 
+\frac{\partial I_1(p,q)}{\partial p^\mu }\mid _{q=p}({\bf p}-M).
\end{array}
\eqnum{12.11}
\end{equation}
It is emphasized that at one-loop level, the both sides of Eq. (12.11) are
of the order of $g^2$. The terms of orders higher than $g^2$ have been
neglected in the derivation of Eq. (12.11). The identity in Eq. (12.10)
formally is the same as we met in QED. By the subtraction at ${\bf p}{=\mu }$%
, the vertex $\overline{\Lambda }_\mu (p,p)$ will be expressed in the form 
\begin{equation}
\overline{\Lambda }_\mu (p,p)=L\gamma _\mu +\overline{\Lambda }_\mu ^c(p) 
\eqnum{12.12}
\end{equation}
where $L$ is a divergent constant defined by 
\begin{equation}
L=\overline{\Lambda }_\mu (p,p)\mid _{{\bf p}=\mu }  \eqnum{12.13}
\end{equation}
and $\overline{\Lambda }_\mu ^c(p)$ is the finite part of $\overline{\Lambda 
}_\mu (p,p)$ satisfying the boundary condition 
\begin{equation}
\overline{\Lambda }_\mu ^c(p)\mid _{{\bf p}=\mu }=0.  \eqnum{12.14}
\end{equation}
On integrating the identity in Eq. (12.10) over the momentum $p_{\mu \text{ }%
}$and considering the expression in Eq. (12.12), we obtain 
\begin{equation}
\Sigma (p)=A+({\bf p}-\mu )[B-C(p^2)]  \eqnum{12.15}
\end{equation}
where

\begin{equation}
A=\Sigma (\mu ),  \eqnum{12.16}
\end{equation}

\begin{equation}
B=-L  \eqnum{12.17}
\end{equation}
and $C(p^2)$ is defined by 
\begin{equation}
\int_{p_0^\mu }^{p^\mu }dp^\mu \overline{\Lambda }_\mu ^c(p)=({\bf p}-\mu
)C(p^2).  \eqnum{12.18}
\end{equation}
Clearly, the expression in Eq. (12.15) gives the subtraction version of the
nucleon self-energy which is required by the W-T identity and correct at
least in the approximation of order $g^2$. With this subtraction, the
nucleon propagator in Eq. (7.17) will be renormalized as 
\begin{equation}
S_F(p)=\frac{Z_2}{{\bf p}-M_R-\Sigma _R(p)}  \eqnum{12.19}
\end{equation}
where $Z_2$ is the renormalization constant defined by 
\begin{equation}
Z_2^{-1}=1-B,  \eqnum{12.20}
\end{equation}
$M_R$ is the renormalized nucleon mass defined as 
\begin{equation}
M_R=Z_M^{-1}M  \eqnum{12.21}
\end{equation}
in which 
\begin{equation}
Z_M^{-1}=1+Z_2[AM^{-1}+(1-\mu M^{-1})B],  \eqnum{12.22}
\end{equation}
$Z_M$ is the nucleon mass renormalization constant and $\Sigma _R(p)$ is the
finite correction of the self-energy satisfying the boundary condition $%
\Sigma _R(p)_{\mid p^2=\mu ^2}=0.$

Now we are in a position to discuss the one-loop renormalization of nucleon
mass. The RGE for the renormalized nucleon mass can be written from Eq.
(10.2) by setting $F=M$, 
\begin{equation}
\lambda \frac{dM_R(\lambda )}{d\lambda }+\gamma _M(\lambda )M_R(\lambda )=0 
\eqnum{12.23}
\end{equation}
where 
\begin{equation}
\gamma _M(\lambda )=\mu \frac d{d\mu }\ln Z_M.  \eqnum{12.24}
\end{equation}
It is clear that to determine the one-loop renormalization constant $Z_M$,
we first need to determine the divergent constants $A$ and $B$ from the
nucleon one-loop self-energy which is represented in the form as shown in
Eq. (12.15). The one-loop self-energy denoted by $-i\Sigma (p)$ contains two
terms which can be written out from Figs. (9a) and (9b) respectively. In the
Feynman gauge, it is represented as

\begin{equation}
\Sigma (p)=\Sigma _1(p)+\Sigma _2(p)  \eqnum{12.25}
\end{equation}
where 
\begin{equation}
\begin{array}{c}
\Sigma _1(p)=-i\frac 34g^2\int \frac{d^4k}{(2\pi )^4}\frac{\gamma ^\mu ({\bf %
k}+{\bf p}+M)\gamma _\mu }{[(k+p)^2-M^2+i\varepsilon ](k^2-m_\rho
^2+i\varepsilon )}, \\ 
\Sigma _2(p)=i3g^2\int \frac{d^4k}{(2\pi )^4}\frac{({\bf k}+{\bf p}-M)}{%
[(k+p)^2-M^2+i\varepsilon ](k^2-m_\pi ^2+i\varepsilon )}
\end{array}
\eqnum{12.26}
\end{equation}
here ${\bf k=}\gamma ^\mu k_\mu $ and ${\bf p=}\gamma ^\mu p_\mu .$ By
making use of the dimensional regularization to calculate the above
integral, it is found that 
\begin{equation}
\begin{array}{c}
\Sigma _1(p)=\frac 32\frac{g^2}{(4\pi )^2}\int_0^1dx\frac{(x-1){\bf p+}2M}{%
\varepsilon [p^2x(x-1)+M^2x+m_\rho ^2(1-x)]^\varepsilon }, \\ 
\Sigma _2(p)=\frac{3g^2}{(4\pi )^2}\int_0^1dx\frac{(x-1){\bf p+}M}{%
\varepsilon [\mu ^2x(x-1)+M^2x+m_\pi ^2(1-x)]^\varepsilon }
\end{array}
\eqnum{12.27}
\end{equation}
According to Eqs. (12.16), (12.25) and (12.27), we have 
\begin{equation}
\begin{array}{c}
A=\Sigma (p)\mid _{{\bf p}=\mu }=\frac 32\frac{g^2}{(4\pi )^2}\int_0^1dx\{%
\frac{(x-1)\mu {\bf +}2M}{\varepsilon [\mu ^2x(x-1)+M^2x+m_\rho
^2(1-x)]^\varepsilon } \\ 
+\frac{2[(x-1)\mu {\bf +}M]}{\varepsilon [\mu ^2x(x-1)+M^2x+m_\pi
^2(1-x)]^\varepsilon }\}.
\end{array}
\eqnum{12.28}
\end{equation}
With the aid of the following formula 
\begin{equation}
\frac 1{a^\varepsilon }-\frac 1{b^\varepsilon }=\varepsilon \int_0^1dx\frac{%
b-a}{[ax+b(1-x)]^{1+\varepsilon }},  \eqnum{12.29}
\end{equation}
one can get from Eqs. (12.15) and (12.25)-(12.28) that 
\begin{equation}
\begin{array}{c}
B=[\Sigma (p)-A]({\bf p-}M{\bf )}^{-1}\mid _{{\bf p}=\mu } \\ 
=\frac 32\frac{g^2}{(4\pi )^2}\int_0^1dx\{\frac{(x-1)}{\varepsilon [\mu
^2x(x-1)+M^2x+m_\rho ^2(1-x)]^\varepsilon }-\frac{2x(x-1)[(x-1)\mu ^2+2M\mu ]%
}{\mu ^2x(x-1)+M^2x+m_\rho ^2(1-x)} \\ 
+\frac{2(x-1)}{\varepsilon [\mu ^2x(x-1)+M^2x+m_\pi ^2(1-x)]^\varepsilon }-%
\frac{4x(x-1)[(x-1)\mu ^2+M\mu ]}{\mu ^2x(x-1)+M^2x+m_\pi ^2(1-x)}\}
\end{array}
\eqnum{12.30}
\end{equation}
where $C(\mu ^2)=0$ has been considered. On inserting Eqs. (12.28) and
(12.30) into Eq. (12.22) and noting that in the approximation of order $g^2,$
$Z_2\simeq 1$ should be taken in Eq. (12.22), it can be found that 
\begin{equation}
\begin{array}{c}
Z_M=1-\frac AM-(1-\frac \mu M)B \\ 
=1-\frac{g^2}{(4\pi )^2}\frac 32\int_0^1dx\{\frac{(x+1)}{\varepsilon [\mu
^2x(x-1)+M^2x+m_\rho ^2(1-x)]^\varepsilon }+\frac{2x(x-1)[(x-1)(\mu /M-1)\mu
^2+2(\mu ^2-M\mu ]}{\mu ^2x(x-1)+M^2x+m_\rho ^2(1-x)} \\ 
+\frac{2x}{\varepsilon [\mu ^2x(x-1)+M^2x+m_\pi ^2(1-x)]^\varepsilon }+\frac{%
4x(x-1)[(x-1)(\mu /M-1)\mu ^2+(\mu ^2-M\mu ]}{\mu ^2x(x-1)+M^2x+m_\pi ^2(1-x)%
}\}
\end{array}
\eqnum{12.31}
\end{equation}
When substituting Eq. (12.31) in Eq. (12.24) and applying the familiar
integration formulas, through a lengthy calculation, we obtain 
\begin{equation}
\gamma _M(\lambda )=\gamma _M^{(1)}(\lambda )+\gamma _M^{(2)}(\lambda ) 
\eqnum{12.32}
\end{equation}
where $\gamma _M^{(1)}(\lambda )$ and $\gamma _M^{(2)}(\lambda )$ are
contributed from the self-energies depicted in Figs. (9a) and (9b)
respectively. They are represented as follows. 
\begin{equation}
\begin{array}{c}
\gamma _M^{(1)}(\lambda )=\frac{3\alpha _R}{4\pi }\{\eta _1(\lambda )+\eta
_2(\lambda )\ln \frac \rho \beta +\frac 2{\lambda ^2K(\lambda ;\beta ,\rho )}%
\eta _3(\lambda ) \\ 
+\frac 1{2\lambda ^4}[\eta _4(\lambda )+\frac 4{K(\lambda ;\beta ,\rho )}%
\eta _5(\lambda )]J(\lambda ;\beta ,\rho )\}
\end{array}
\eqnum{12.33}
\end{equation}
where 
\begin{equation}
\eta _1(\lambda )=\frac 1{\lambda ^2}[\frac{\lambda ^3}{2\beta }+\frac 32%
\lambda ^2+(\rho ^2-3\beta ^2)\frac \lambda \beta +\beta ^2-\rho ^2], 
\eqnum{12.34}
\end{equation}
\begin{equation}
\begin{array}{c}
\eta _2(\lambda )=\frac 1{\lambda ^4}[\frac{\rho ^2}\beta \lambda ^3+(6\beta
^2-7\rho ^2)\lambda ^2 \\ 
-\frac 3\beta (\beta ^2-\rho ^2)(3\beta ^2-\rho ^2)\lambda +3(\beta ^2-\rho
^2)^2],
\end{array}
\eqnum{12.35}
\end{equation}
\begin{equation}
\begin{array}{c}
\eta _3(\lambda )=\frac{\rho ^2}\beta \lambda ^5-(2\beta ^2+3\rho ^2)\lambda
^4+\frac 1\beta (3\beta ^4+3\beta ^2\rho ^2-2\rho ^4)\lambda ^3 \\ 
+(\beta ^2-\rho ^2)(\beta ^2-4\rho ^2)\lambda ^2-\frac 1\beta (\beta ^2-\rho
^2)^2(3\beta ^2-\rho ^2)\lambda +(\beta ^2-\rho ^2)^3,
\end{array}
\eqnum{12.36}
\end{equation}
\begin{equation}
\begin{array}{c}
\eta _4(\lambda )=\frac{\rho ^2}\beta \lambda ^5-(6\beta ^2+7\rho ^2)\lambda
^4+\frac 1\beta (9\beta ^4+11\rho ^2\beta ^2-8\rho ^2)\lambda ^3 \\ 
+11(\beta ^2-\rho ^2)(\beta ^2-2\rho ^2)\lambda ^2-\frac 7\beta (\beta
^2-\rho ^2)^2(3\beta ^2-\rho ^2)\lambda +7(\beta ^2-\rho ^2)^3,
\end{array}
\eqnum{12.37}
\end{equation}
\begin{equation}
\begin{array}{c}
\eta _5(\lambda )=\frac{\rho ^4}\beta \lambda ^7-(2\beta ^4+3\rho ^4)\lambda
^6+\frac 1\beta (3\beta ^6+4\beta ^2\rho ^4-3\rho ^6)\lambda ^5 \\ 
+(\beta ^2-\rho ^2)(3\beta ^4-\beta ^2\rho ^2-7\rho ^4)\lambda ^4-\frac 1%
\beta (\beta ^2-\rho ^2)^2(6\beta ^4+5\beta ^2\rho ^2-3\rho ^4)\lambda ^3 \\ 
+5\rho ^2(\beta ^2-\rho ^2)^3\lambda ^2+\frac 1\beta (\beta ^2-\rho
^2)^4(3\beta ^2-\rho ^2)\lambda -(\beta ^2-\rho ^2)^5,
\end{array}
\eqnum{12.38}
\end{equation}
$K(\lambda ;\beta ,\rho )$ and $J(\lambda ;\beta ,\rho )$ are the functions
defined in Eqs. (11.28) and (11.29) with $\rho $ and $\sigma $ being
replaced by $\beta $ and $\rho $. 
\begin{equation}
\begin{array}{c}
\gamma _M^{(2)}(\lambda )=\frac{3\alpha _R}{2\pi }\{\zeta _1(\lambda )+\zeta
_2(\lambda )\ln \frac \sigma \beta +\frac 2{\lambda ^2K(\lambda ;\beta
,\sigma )}\zeta _3(\lambda ) \\ 
+\frac 1{2\lambda ^4}[\zeta _4(\lambda )+\frac 4{K(\lambda ;\beta ,\sigma )}%
\zeta _5(\lambda )]J(\lambda ;\beta ,\sigma )\}
\end{array}
\eqnum{12.39}
\end{equation}
where 
\begin{equation}
\zeta _1(\lambda )=\frac 1{\lambda ^2}[\frac{\lambda ^3}{2\beta }+\frac{%
\lambda ^2}2+\frac \lambda \beta (\sigma ^2-2\beta ^2)+\beta ^2-\sigma ^2], 
\eqnum{12.40}
\end{equation}
\begin{equation}
\begin{array}{c}
\zeta _2(\lambda )=\frac 1{\lambda ^4}[\frac{\sigma ^2}\beta \lambda
^3+(3\beta ^2-4\sigma ^2)\lambda ^2 \\ 
-\frac 3\beta (\beta ^2-\sigma ^2)(2\beta ^2-\sigma ^2)\lambda +3(\beta
^2-\sigma ^2)^2],
\end{array}
\eqnum{12.41}
\end{equation}
\begin{equation}
\begin{array}{c}
\zeta _3(\lambda )=\frac{\sigma ^2}\beta \lambda ^5-(\beta ^2+2\sigma
^2)\lambda ^4+\frac 2\beta (\beta ^4+\beta ^2\sigma ^2-\sigma ^4)\lambda ^3
\\ 
-3\sigma ^2(\beta ^2-\sigma ^2)\lambda ^2-\frac 1\beta (\beta ^2-\sigma
^2)^2(2\beta ^2-\sigma ^2)\lambda +(\beta ^2-\sigma ^2)^3,
\end{array}
\eqnum{12.42}
\end{equation}
\begin{equation}
\begin{array}{c}
\zeta _4(\lambda )=\frac{\sigma ^2}\beta \lambda ^5-(3\beta ^2+4\sigma
^2)\lambda ^4+\frac 2\beta (3\beta ^4+4\beta ^2\sigma ^2-4\sigma ^2)\lambda
^3 \\ 
-(\beta ^2-\sigma ^2)(2\beta ^2+7\sigma ^2)\lambda ^2-\frac 7\beta (\beta
^2-\sigma ^2)^2(2\beta ^2-\sigma ^2)\lambda +7(\beta ^2-\sigma ^2)^3,
\end{array}
\eqnum{12.43}
\end{equation}
\begin{equation}
\begin{array}{c}
\zeta _5(\lambda )=\frac{\sigma ^4}\beta \lambda ^7-(\beta ^4+2\sigma
^4)\lambda ^6+\frac 1\beta (2\beta ^6+3\beta ^2\sigma ^4-3\sigma ^6)\lambda
^5 \\ 
+(\beta ^2-\sigma ^2)(\beta ^4-\beta ^2\sigma ^2-5\sigma ^4)\lambda ^4-\frac 
1\beta (\beta ^2-\sigma ^2)^2(4\beta ^4+3\beta ^2\sigma ^2-3\sigma
^4)\lambda ^3 \\ 
+(\beta ^2-\sigma ^2)^3(\beta ^2+4\sigma ^2)\lambda ^2+\frac 1\beta (\beta
^2-\sigma ^2)^4(2\beta ^2-\sigma ^2)\lambda -(\beta ^2-\sigma ^2)^5,
\end{array}
\eqnum{12.44}
\end{equation}
$K(\lambda ;\beta ,\sigma )$ and $J(\lambda ;\beta ,\sigma )$ are the those
defined in Eqs. (11.28) and (11.29) with $\rho $ being replaced by $\beta $.

With the anomalous dimension given above, the equation in Eq. (12.23) can be
solved and gives the nucleon effective mass as follows 
\begin{equation}
M_R(\lambda )=M_Re^{-S_M(\lambda )}  \eqnum{12.45}
\end{equation}
where 
\begin{equation}
S_M(\lambda )=\int_1^\lambda \frac{d\lambda }\lambda \gamma _M(\lambda ) 
\eqnum{12.46}
\end{equation}
This integral can only be calculated numerically when the coupling constant
in Eqs. (12.33) and (12.39) is taken to be the effective one. Even though
the coupling constant is taken to be a constant, due to that the last terms
in Eq. (12.33) and (12.39) involve the functions $J(\lambda ;\beta ,\rho )$
and $J(\lambda ;\beta ,\sigma )$, we are not able to give an explicit
expression of the integral through calculation.

Graphically, the effective nucleon masses are shown in Fig. (10). In the
figure, the solid line represents the effective mass given by the timelike
momentum subtraction with the coupling constant being taken to be a
constant. This effective mass is more clearly exhibited in the subfigure
within Fig. (10). From the subfigure, it is seen that in the regime defined
from $\lambda =0.2$ to $\lambda =1.34855$, the effective mass almost behaves
as a constant whose value is about the ordinary nucleon mass $M_R$. Beyond
the regime mentioned above, the $M_R(\lambda )$ rapidly falls to zero when $%
\lambda \rightarrow 0$, and from $\lambda =1.34855$, it suddenly decreases
and fast tends to zero when $\lambda \rightarrow \infty $, exhibiting the
asymptotically free behavior. The effective nucleon mass given by the
spacelike momentum subtraction can directly be written out from Eqs.
(12.32)- (12.46) by replacing the $\lambda $ in $\gamma _M(\lambda )$ with $%
i\lambda $. Clearly, the effective mass given in this way becomes a complex
one. the real part and imaginary part of the effective mass are represented
in Fig. (10) by the dashed and dashed-dotted lines respectively. The figure
shows that either the real part or the imaginary part behaves as an
oscillating function with a damping amplitude. It is noted that in the most
of practical applications to both of scattering and bound state problems,
only the effective nucleon mass given by the timelike momentum subtraction
is concerned.

\section{Conclusions and discussions}

In this paper, it has been shown that the SU(2)-symmetric model of
hadrodynamics, as a massive non-Abelian gauge field theory, can surely be
set up on the basis of gauge-invariance. This conclusion is achieved by the
consideration that the model built up Lorentz-covariantly actually describes
a constrained interacting system since the vectorial $\rho -$meson fields
included in the model contains redundant unphysical degrees of freedom.
Therefore, to establish a correct quantum theory of the model, it is
necessary to introduce appropriate constraint conditions to eliminate the
unphysical degrees of freedom, i.e., to introduce the Lorentz condition to
remove the unphysical longitudinal components of the $\rho -$meson vector
potentials and the ghost equation to constrain the residual gauge degrees of
freedom which exist in the physical subspace defined by the Lorentz
condition. As shown in Sec.2, under the constraint conditions, the model
action is exactly gauge-invariant. For the quantum theory, the
gauge-invariance is embodied in the fact that the effective action and the
generating functional are invariant with respect to a set of
BRST-transformations. From the BRST-invariance, we derived a set of W-T
identities satisfied by the generating functionals for full Green functions,
connected Green functions and proper vertex functions. Furthermore, from the
above identities, we derived the W-T identities respected by the $\rho -$%
meson propagator, the $\rho -$meson three-line and four-line proper
vertices, the pion-$\rho -$meson three-line and four-line proper vertices
and the nucleon-$\rho -$meson proper vertex. Based on these identities, we
discussed the renormalization of the propagators and the vertices. In
particular, from the renormalized forms of the W-T identities obeyed by
propagators and vertices, the S-T identity for the renormalization constants
is naturally deduced. This identity is helpful for the renormalization by
means of the renormalization group approach.

From the derivations given in this paper, it is clearly seen that there is
no any difficulty to appear in performing the renormalization of the
propagators and vertices as well as their W-T identities. This indicates
that the SU(2)-symmetric model of hadrodynamics is renormalizable.
Certainly, to give a complete proof of the renormalizability of the model,
we need to prove that all divergences occurring in perturbative calculations
can be removed by introducing a finite number of counterterms. For the
massive non-Abelian gauge field theory without Higgs mechanism in which all
the gauge bosons have the same masses such as the model under consideration,
this kind of proof has actually been given in Ref. [33]. As argued in Refs.
[14, 25], when we work in the renormalization group approach, an exact
renormalized S-matrix element can be given by writing out the expressions of
its tree diagrams provided that the coupling constant and particle masses in
the S-matrix element are replaced by their effective ones which are
determined by solving their RGEs. In this paper, to demonstrate the
renormalizability of the model, the one-loop renormalization is performed by
the renormalization group method. In this renormalization, the analytical
expressions of the one-loop effective coupling constant and the effective
particle masses have been derived. Since the renormalization was carried out
by employing the mass-dependent momentum space subtraction scheme and
exactly respecting the W-T identities, the results obtained are faithful and
allow us to discuss the physical behaviors of the effective coupling
constant and masses in the whole range of momentum (or distance), unlike the
results given in the minimal subtraction scheme which are only suitable in
the large momentum limit.

As shown in sections 10-12, in the mass-dependent renormalization, it is
necessary to distinguish the results given by the timelike momentum
subtraction from the corresponding ones obtained by the spacelike momentum
subtraction. For example, as one can see from Fig. (4), the effective
coupling constants given in the timelike and spacelike subtraction schemes
have different behaviors in the low and intermediate energy region although
in the large momentum limit, the difference between the both coupling
constants disappears. Obviously, the both results obtained in the timelike
and spacelike momentum subtraction schemes are meaningful and suitable for
different physical processes. For instance, when we study the
nucleon-nucleon scattering taking place in the t-channel, the transfer
momenta in the $\rho -$meson and pion propagators are spacelike. In this
case, it is suitable to take the effective coupling constant and the
effective $\rho -$meson and pion masses given by the spacelike momentum
subtraction. If we investigate the nucleon-antinucleon annihilation process
which takes place in the s-channel, since the transfer momenta are timelike,
the effective coupling constant and the effective $\rho -$meson and pion
masses given in the timelike momentum subtraction scheme should be used.

As mentioned in section 10-12, the one-loop effective coupling constant and
the effective masses tend to zero in the large momentum (or short distance)
limit. This shows that the interactions given by the SU(2)-symmetric
hadrodynamics. , as QCD and other non-Abelian gauge field theories, exhibits
an asymptotically free behavior at least in the one-loop approximation of
perturbation [34-36]. This behavior arises from the $\rho -$meson
self-interactions which seems to be stronger than both of the interaction
between $\rho -$mesons and nucleons and the interaction between pions and
nucleons. But, we are not sure whether the asymptotically free property is
still preserved beyond the one-loop approximation. To give a definite answer
to this question, it is necessary to perform a nonperturbative calculation.
Nuclear force is a complicated problem. At the level of hadrodynamics, it
can not be solved by using only the SU(2)-symmetric model. However, if one
attempts to solve the problem within the framework of gauge field theory,
besides the $\sigma -\omega $ model and some others, the SU(2)-symmetric
model, as an effective field theory, is necessarily to be taken into account.

\section{Acknowledgment}

This work was supported by National Natural Science Foundation of China.

\section{Appendix A: Feynman rules}

For the application of the SU(2)-symmetric model of hydrodynamics to
perturbative calculations, in this appendix, we list the Feynman rules of
the model which can easily be derived from the effective action given by the
effective Lagrangian in Eq. (2.26) with the Lagrangian ${\cal L}$ written in
Eqs. (2.1)-(2.9). In the momentum space, they represented as follows (Note:
In the following, the propagators and vertices in Eqs. (A1)-(A11) are in
turn represented graphically in Figs. 11a-11k, each figure should be put on
the right of the corresponding formula):

Nucleon propagator: 
\begin{equation}
iS_F(p)=\frac i{{\bf p}-M+i\varepsilon }.  \eqnum{A1}
\end{equation}

Pion propagator: 
\begin{equation}
i\Delta _\pi ^{ab}(k)=\frac{i\delta ^{ab}}{k^2-m_\pi ^2+i\varepsilon }. 
\eqnum{A2}
\end{equation}

$\rho -$meson propagator: 
\begin{equation}
iD_{\mu \nu }^{ab}(k)=\frac{-i\delta ^{ab}}{k^2-m_\rho ^2+i\varepsilon }%
[g_{\mu \nu }-(1-\alpha )\frac{k_\mu k_\nu }{k^2-\sigma ^2+i\varepsilon }] 
\eqnum{A3}
\end{equation}
where $\sigma =\sqrt{\alpha }m_\rho .$

Ghost particle propagator:

\begin{equation}
i\Delta ^{ab}(k)=\frac{-i\delta ^{ab}}{k^2-\sigma ^2+i\varepsilon }. 
\eqnum{A4}
\end{equation}

Nucleon-pion vertex: 
\begin{equation}
\Lambda ^a(p,q.k)=-g\gamma _5\tau ^a.  \eqnum{A5}
\end{equation}

Nucleon-$\rho -$meson vertex: 
\begin{equation}
\Lambda _\mu ^a(p,q.k)=ig\gamma _\mu T^a.  \eqnum{A6}
\end{equation}
where $T^a=\tau ^a/2.$

Pion-$\rho -$meson three-line vertex: 
\begin{equation}
\bar \Lambda _{\;\;\mu }^{abc}(k_1,k_2,k_3)=g\varepsilon ^{abc}(k_1-k_2)_\mu
.  \eqnum{A7}
\end{equation}
(corresponding to Fig. 11g)

Pion-$\rho -$meson four-line vertex: 
\begin{equation}
\bar \Lambda _{\;\;\mu \nu }^{abcd}(k_1,k_2,k_3,k_4)=ig^2(\varepsilon
^{ace}\varepsilon ^{bde}+\varepsilon ^{ade}\varepsilon ^{bce})g_{\mu \nu }. 
\eqnum{A8}
\end{equation}

$\rho -$meson three-line vertex: 
\begin{equation}
\Lambda _{\mu \nu \lambda }^{abc}(k_1,k_2,k_3)=-g\varepsilon ^{abc}[g_{\mu
\nu }(k_1-k_2)_\lambda +g_{\nu \lambda }(k_2-k_3)_\mu +g_{\lambda \mu
}(k_3-k_1)_\nu ].  \eqnum{A9}
\end{equation}

$\rho -$meson four-line vertex: 
\begin{equation}
\begin{array}{c}
\Lambda _{\mu \nu \rho \sigma }^{abcd}(k_1,k_2,k_3,k_4)=-ig^2[\varepsilon
^{abe}\varepsilon ^{cde}(g_{\mu \sigma }g_{\nu \rho }-g_{\mu \rho }g_{\nu
\sigma }) \\ 
+\varepsilon ^{ace}\varepsilon ^{dbe}(g_{\mu \rho }g_{\nu \sigma }-g_{\mu
\nu }g_{\rho \sigma })+\varepsilon ^{ade}\varepsilon ^{bce}(g_{\mu \nu
}g_{\rho \sigma }-g_{\mu \sigma }g_{\nu \rho })].
\end{array}
\eqnum{A10}
\end{equation}

Ghost vertex: 
\begin{equation}
\tilde \Lambda _{\;\;\mu }^{abc}(k_1,k_2,k_3)=g\varepsilon ^{abc}k_{2\mu }. 
\eqnum{A11}
\end{equation}

\section{References}

\begin{itemize}
\item[1]  J. J. Sakurai, Ann. Phys. {\bf 11}. 1 (1960).

\item[2]  D. Luri\'e, Paeticles and Fields, Interscieence Publishings, a
division of Jorn Wiley \& Sons, New York (1960).

\item[3]  C. N. Yang and R. L. Mills, Phys. Rev. {\bf 96,} 191 (1954).

\item[4.]  H. Umezawa and S. Kamefuchi, Nucl. Phys. {\bf 23}, 399 (1961).

\item[5]  A. Ionides, Nucl. Phys. {\bf 23,} 662 (1961).

\item[6]  A. Salam, Nucl. Phys. {\bf 18}, 681 (1960) ; Phys. Rev. {\bf 127,}
331 (1962).

\item[7]  D. G. Boulware, Ann. Phys. {\bf 56,} 140 (1970).

\item[8]  A. Salam and J. Strathdee, Phys. Rev. {\bf D 2,} 2869 (1970).

\item[9]  C. Itzykson and F-B. Zuber, Quantum Field Theory, McGraw-Hill, New
York (1980).

\item[10]  J. C. Su, IL Nuovo Cimento, {\bf 117 B,} 203 (2002).

\item[11]  J. C. Su and J. X. Chen, Phys. Rev.{\bf \ D 69}, 076002 (2004).

\item[12]  J. C. Su, Proceedings of Institute of Mathematics of NAS of
Ukraine, Vol. {\bf 50}, Part 2, 965 (2004).

\item[13]  L. D. Faddeev and V. N. Popov, Phys. Lett. B25, 29 (1967).

\item[14]  .J. C. Su and H. J. Wang, Phys. Rev. {\bf C 70}, 044003 (2004).

\item[15]  J. C. Ward, Phys. Rev. 77, 2931 (1950); Y. Takahashi, Nuovo
Cimento 6, 370 (1957).

\item[16]  L. D. Faddeev and A. A. Slavnov, Gauge Fields: Introduction to
Quantum Theory, The Benjamin Commings Publishing Company Inc. (1980).

\item[17]  E. S. Abers and B. W. Lee, Phys. Rep. {\bf C9,} 1 (1973); B. W.
Lee, In Methods in Field Theory (1976), ed. R. Balian and Zinn-Justin.

\item[18]  C. Becchi, A. Rouet and R. Stora, Phys. Lett. {\bf B 52,} 344
(1974); Commun. Math. Phys. {\bf 42}, 127 (1975); I. V. Tyutin, Lebedev
Preprint {\bf 39} (1975).

\item[19]  A. Slavnov, Theor. and Math. Phys. {\bf 10,} 99 (1972), (English
translation).

\item[20]  J. C. Taylor, Nucl. Phys. {\bf B 33,} 436 (1971).

\item[21]  C. G. Callan, Phys. Rev. {\bf D 2}, 1541 (1970) ; K. Symanzik,
Commun, Math. Phys. {\bf 18}, 227 (1970).

\item[22]  S. Weinberg, Phys. Rev. {\bf D }8, 3497 (1973).

\item[23]  J. C. Collins and A. J. Macfarlane, Phys. Rev. {\bf D 10}, 1201
(1974).

\item[24]  F. J. Dyson, Phys. Rev. {\bf 75,} 1736 (1949); J. Schwinger,
Proc. Nat. Acad. Sci. {\bf 37,} 452 (1951).

\item[25]  J. C. Su, X. X. Yi and Y.H. Cao, J. Phys. G: Nucl. Part. Phys. 
{\bf 25}, 2325 (1999).

\item[26]  J. C. Su, L. Shan and Y. H. Cao, Commun. Theor. Phys. {\bf 36},
665 (2000).

\item[27]  J. C. Su, hep-th/9805193 and hep-th/9805194, to be soon published
in Progress in Field Theory Research.

\item[28]  G. 't Hooft, Nucl. Phys. {\bf B} {\bf 61}, 455 (1973) .

\item[29]  W. A. Bardeen, A. J. Buras, D. W. Duke and T. Muta, Phys. Rev. 
{\bf D} {\bf 18}, 3998 (1978); W. A. Bardeen and R. A. J. Buras, Phys. Rev. 
{\bf D} {\bf 20}, 166 (1979).

\item[30]  W. Celmaster and R. J. Gonsalves, Phys. Rev. Lett. {\bf 42}, 1435
(1979); Phys. Rev. {\bf D} {\bf 20}, 1420 (1979); W. Celmaster and D.
Sivers, Phys. Rev. {\bf D} {\bf 23}, 227 (1981).

\item[31]  E. Braaten and J. P. Leveille, Phys. Rev. {\bf D} {\bf 24}, 1369
(1981).

\item[32]  S. N. Gupta and S. F. Radford, Phys. Rev. {\bf D} {\bf 25}, 2690
(1982); J. C. Collins and A. J. Macfarlane, Phys. Rev. {\bf D} {\bf 10},
1201 (1974).

\item[33]  J. C. Su, hep-th/9805193. Its revised version will soon appear in
Progress in Field Theory Research.

\item[34]  H. D. Politzer. Phys. Rev. Lett. {\bf 30}, 1346 (1973).

\item[35]  D. J. Gross and F. Wilczek, Phys. Rev. Lett. {\bf 30}, 1343
(1973); Phys. Rev.{\bf \ D} {\bf 8}, 3633 (1973).

\item[36]  H. Georgi and H. D. Politzer, Phys. Rev. {\bf D 14}, 1829 (1976).
\end{itemize}

\section{Figure captions}

Fig. (1): The $\rho -$meson one-loop self-energy. The solid, helical, wavy
and dashed lines represent the nucleon, $\rho -$meson, pion and ghost
particle free propagators, respectively.

Fig. (2): The one-loop ghost particle self-energy. The lines represent the
same as in Fig. (1).

Fig. (3): The one-loop ghost-$\rho -$meson vertices. The lines mark the same
as in Fig. (1).

Fig. (4): (a) The one-loop effective coupling constants ${\alpha _R(\lambda )%
}$ given by the timelike momentum space subtraction; (b) the one-loop
effective coupling constants ${\alpha _R(\lambda )}$ given by the spacelike
momentum space subtraction. In the subfigures, the solid, dashed and
dashed-dotted lines represent the coupling constants arising from the $\rho
- $meson self-interaction, the interaction between $\rho -$mesons and pions
and the interaction between $\rho -$mesons and nucleons, respectively.

Fig. (5): The one-loop effective $\rho -$meson masses $m_\rho ^R(\lambda )$.
The solid and the dashed lines represent the effective masses given in the
timelike and spacelike momentum subtractions respectively.

Fig. (6): The pion self-energy. The lines represnt the same as mentioned in
Fig. (1).

Fig. (7): The one-loop effective pion masses $m_\pi ^R(\lambda )$. The solid
and the dashed lines represent the effective masses given in the timelike
and spacelike momentum subtractions, respectively.

Fig. (8): The one-loop nucleon-ghost particle vertices. The lines represent
the same as in Fig. (1).

Fig. (9): The one-loop nucleon self-energy. The lines represent the same as
in Fig. (1).

Fig. (10): The one-loop effective nucleon masses $M_R(\lambda )$. The solid
line represents the effective mass given by the timelike momentum space
subtraction. The dashed and dashed-dotted lines respectively represent the
real part and the imaginary part of the effective mass given by the
spacelike momentum subtraction

Fig. (11): The figures for Feynman rules. There are 11 figures in Fig. (11).
These figures numbered as fig.11a-fig. 11k in turn correspond to the
formulas (A1)-(A11) in Appendix. Each figure should be put on the right of
the corresponding formula.

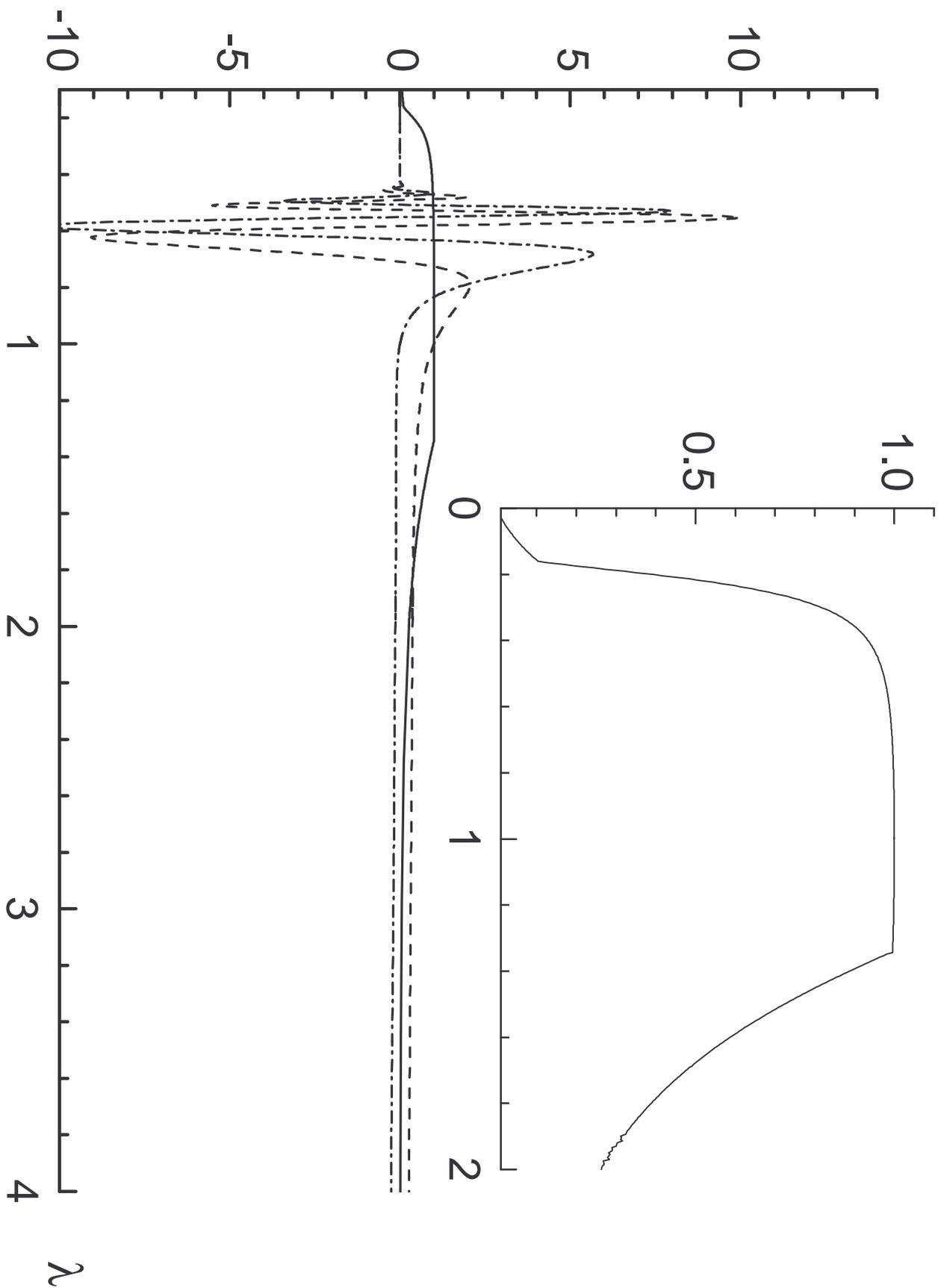
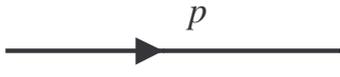
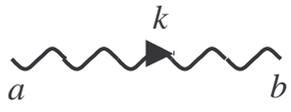
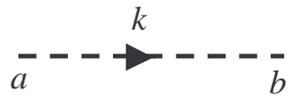
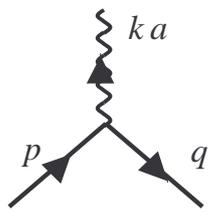
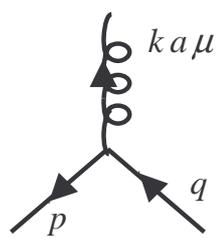
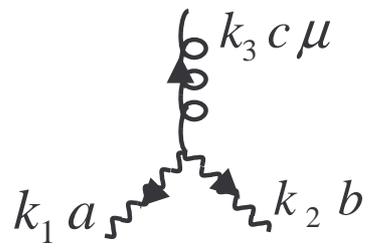
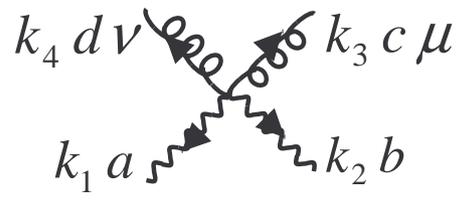
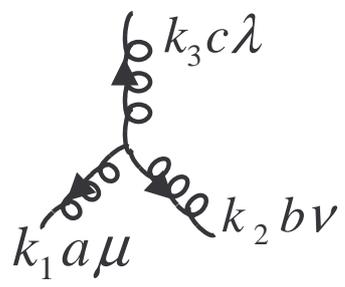
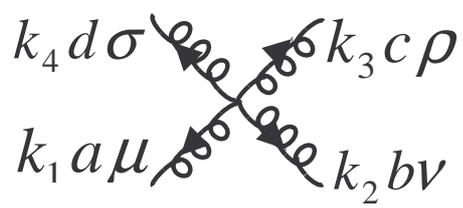
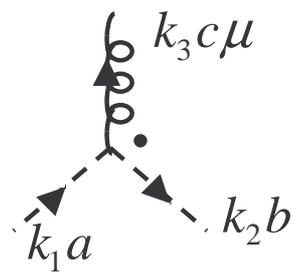
\end{document}